\documentclass{aa}
\usepackage{array}
\newcolumntype{L}{>{$}l<{$}} 
\newcolumntype{R}{>{$}r<{$}}
\newcolumntype{C}{>{$}c<{$}}
\usepackage{graphicx}
\graphicspath{{Images/}}
\usepackage{txfonts}
\usepackage{color}
\usepackage[dvipsnames]{xcolor}
\usepackage{soul}
\definecolor{linkcolor}{rgb}{0,0,0.6}
\usepackage[colorlinks=true,
linkcolor= linkcolor,
citecolor= linkcolor,
urlcolor= linkcolor]
{hyperref}
\usepackage{hyperref}
\makeatletter
\renewcommand*\aa@pageof{, page \thepage{} of \pageref*{LastPage}}
\makeatother
\def\lapprox{\;\rlap{\lower 2.5pt            
 \hbox{$\sim$}}\raise 1.5pt\hbox{$<$}\;} 

\begin{document}

   \title{Effects of radiative losses on the relativistic jets of high-mass microquasars}


   \author{A. Charlet\inst{1,}\inst{2},
          R. Walder\inst{1},
          A. Marcowith\inst{2},
          D. Folini\inst{1},
          J. M. Favre\inst{3},
          M. E. Dieckmann\inst{4}
          }
    \authorrunning{A. Charlet et al.}

   \institute{Univ Lyon, ENS de Lyon, Univ Lyon 1,  CNRS, Centre de Recherche Astrophysique de Lyon UMR5574\\
            F-69230, Saint-Genis-Laval, France
         \and
           Laboratoire Univers et Particules de Montpellier (LUPM), Universit\'e de Montpellier, CNRS/IN2P3, CC72, place Eug\`ene Bataillon, \\ F-34095 Montpellier Cedex 5, France
        \and
            Centre suisse de calcul scientifique (CSCS), Via Trevano 131, 6900 Lugano, Switzerland
        \and
            Department of Science and Technology (ITN),  Linköping  University, 60174  Norrköping, Sweden}

   \date{Received 7 July 2021 / Accepted 10 November 2021}

 
  \abstract
   {Relativistic jets are ubiquitous in astrophysics. High-mass microquasars (HMMQs) are useful laboratories for studying these jets because they are relatively close and evolve over observable timescales. The ambient medium into which the jet propagates, however, is far from homogeneous. Corresponding simulation studies to date consider various forms of a wind-shaped ambient medium, but typically neglect radiative cooling and relativistic effects.}
   {We investigate the dynamical and structural effects of radiative losses and system parameters on relativistic jets in HMMQs, from the jet launch to its propagation over several tens of orbital separations.}
   {We used 3D relativistic hydrodynamical simulations including parameterized radiative cooling derived from relativistic thermal plasma distribution to carry out parameter studies around two fiducial cases inspired by Cygnus X-1 and Cygnus X-3.}
   {Radiative losses are found to be more relevant in Cygnus X-3 than Cygnus X-1. Varying jet power, jet temperature, or the wind of the donor star tends to have a larger impact at early times, when the jet forms and instabilities initially develop, than at later times when the jet has reached a turbulent state.}
   {Radiative losses may be dynamically and structurally relevant at least for Cygnus X-3 and thus should be examined in more detail.}


   \maketitle
%

\section{Introduction}
Jets are an ubiquitous manifestation of the activity of compact objects that are at the origin of the microquasar phenomenon \citep{Romero2017}. High-mass microquasars (HMMQ) are a subclass of high-mass X-ray binaries (HMXRB) and are composed of a black hole (BH) and a massive star companion. HMMQ launch powerful collimated jets \citep[e.g.,][]{mirabel1999sources, gallo2005dark} at relativistic speeds either from the accretion disk of the compact object through the Blandford-Payne magneto-centrifugal ejection mechanism \citep{Blandford1982}, or the BH magnetosphere through the Blandford-Znajek mechanism \citep{Blandford1977}. Jets are of interest as integral parts of the astrophysical objects harboring them, but also because their impressive stability due to collimation allows them to extend orders of magnitude farther than their injection scale~\citep{migliori-et-al:17, gourgouliatos-komissarov-natast:18}, offering a very powerful way to study their environment or their contribution to observed thermal (radio), but also (nonthermal) emissions~\citep{malzac:14, Zdziarski2014, Rodriguez2015, molina-et-al:19, albert-et-al:21, motta-et-al:21}.

The HMMQ jets closely resemble to scaled-down jets from active galactic nuclei (AGN) in regard of the overall energy released by accretion. However, in HMMQs, the ambient medium is dominated by the powerful winds of the stellar companion, which are often the source of accretion for the compact object. This stellar wind dominates the environment in which the jet will be launched and evolve, which makes the jet propagation in HMMQ fundamentally different from jets in AGNs and low-mass microquasars.

The effects of stellar wind on jets were studied by \cite{perucho20103d}, who performed 3D simulations of relativistic hydrodynamical jets with a simulation box scale shorter than the orbital separation. They suggested that jets with a power of a few $10^{36}$ erg s$^{-1}$ can be disrupted by the wind through the effect of the Kelvin-Helmholtz instability (KHI). A companion paper~\citep{perucho2010interaction} highlighted the formation and evolution of the recollimation shock and its potential role in particle acceleration in HMMQs. A larger-scale nonrelativistic study was performed by \cite{yoon2015global} with a simulation box scale of $\sim15$ orbital distance, focusing on jet bending at larger scales and obtaining a simple analytical formula for the asymptotic bending angle. A follow-up study by~\citet{yoon2016formation} reconsidered the formation of a recollimation shock, emphasizing that such a shock is likely present in Cygnus X-1, while the situation in Cygnus X-3 is less clear. Several papers pointed to the fact that stellar winds are more clumpy than homogeneous and explored associated consequences for the jets~\citep{perucho20123d, Delacita2017}.

This paper adds to this picture with a set of 3D hydrodynamical simulations that distinguish themselves from existing work in that they are relativistic, include radiative cooling, and follow the jet evolution over comparatively large spatial distances of about 20 to 75 orbital separations. The comparatively large spatial domain allows us to follow jet evolution from an initial smooth phase through the nonlinear growth of instabilities to a turbulent, autosimilar state, thereby creating a larger-scale perspective for some of the results cited above. The parameter study we perform with this general setup is anchored at system parameters inspired by Cygnus X-1 \citep{orosz2011mass}, where cooling is moderate, and Cygnus X-3 \citep{zdziarski2013cyg}, where a strong cooling effect occurs due to the combination of a stronger stellar wind, magnetic field, and luminosity with an orbital separation that is one order of magnitude smaller than in Cygnus X-1.

We obtain typical timescales for the initial instability growth depending on the various parameters and the presence of cooling, highlighting the importance of the beam internal shocks in the growth of the KHI and therefore jet structure and dynamics. Cooling is found to play a role only in Cygnus X-3 on the scales covered by our simulations. Once cooling becomes dominant, the jet cocoon is immediately blown away by the stellar wind. Our simulations further suggest that the parameter sensitivities we explored somewhat diminish or are more difficult to clearly diagnose later on, when the jet has become fully turbulent. We confirm earlier findings that the jet is broken when the wind power is too strong. We find a strong instability developing at the jet beam - cocoon interface that destroys the beam. A turbulent expanding region develops subsequently that eventually expands away from the orbital plane, and the jet is recovered.

The layout of the article is as follows: in Sect. \ref{sect:phys_num}  we present the physics of hydrodynamical relativistic jets, our models for radiative losses, and the numerical setup and methods we used in our parameter study of jet outbreak. In Sect. \ref{sect:results} we discuss the jet propagation, destabilization, and structure with the cooling and parameters. In Sect. \ref{sect:discuss} we discuss the validity and limitations of our results before we summarize and conclude in Sect. \ref{sect:conclusion}.


\section{Physical scenario and numerical methods}\label{sect:phys_num}
\subsection{Relativistic equations}\label{sect:srhd}
The equations for special relativistic hydrodynamics (SRHD) can be written in the form
\begin{equation}
    \partial_t\mathcal{U} + \partial_i\mathcal{F}^i = 0,\label{eq:vecform}
\end{equation}
where $\mathcal{U}$ contains the "conservative" variables and $\mathcal{F}^i$ the corresponding fluxes in the $i$ direction, which are given by
\begin{equation}
    \mathcal{U} =
        \begin{bmatrix}D\\S^j\\\tau\end{bmatrix},\;\;
    \mathcal{F}^i =
        \begin{bmatrix}Dv^i\\S^jv^i+p\delta^{ij}\\S^i\end{bmatrix}.
\end{equation}
$D=\gamma\rho$, $S^i = \gamma^2\rho h v^i$ , and $\tau = \gamma^2\rho h - p$ are the conservatives variable, with $\gamma$ the Lorentz factor and $h$ the specific enthalphy. $\rho$, $v^i$ , and $p$ are the rest-mass density, velocity, and thermal pressure, respectively. They are called the "primitive" variables. The derivation of these equations can be found in textbooks such as \cite{LandauLifshitz1959} and \cite{mihalas2013foundations}. Additionally, a passive tracer $J$ distinguishing the jet material ($J=1$) from the ambient medium ($J=0$) is advected independently, following
\begin{equation}
    \partial_t(D J)+\partial_i(D J v^i) = 0.
\end{equation}
The system is closed by an adiabatic equation of state (EoS) with constant adiabatic index $\Gamma$ taken equal to 5/3 for both wind and injected material. A value of $\Gamma=4/3$ is better suited to model flows with high Lorentz factor. We chose the classical value of 5/3 for our mildly relativistic jets. A 1D comparison between these two values is given in Appendix \ref{sect:app_Gamma} . We find that changing the adiabatic index has an almost negligible impact on the jet head propagation speed, but jets with $\Gamma=5/3$ value present a more advanced front shock than jets with $\Gamma=4/3$. This may translate into a more extended cocoon in 2D and 3D jets. The adapted inversion method to recover the primitive from the conservative variables is taken from \cite{del2002efficient} and is detailed in Appendix \ref{sect:app_invscheme}. In the rest of this paper, quantities with subscript $b$, $ic,$ and $oc$ refer to the beam, inner cocoon, and outer cocoon, respectively. They are defined properly in Sect. \ref{sect:previous}. Values with subscript $j$ refer to injection parameters, and $w$ refers to the stellar wind.

\subsection{Jet propagation}
We generalize the model for the propagation of a relativistic jet as derived by  \cite{marti1997morphology} and \cite{mizuta2004propagation}, for example, where multidimensional effects are neglected and 1D momentum balance between the beam with velocity $v_b$ and the ambient gas is assumed in the rest frame of the contact discontinuity at the head of the jet, to the case of an ambient medium with its own (nonrelativistic) flow speed $v_w$. We obtain the following expression for the jet head velocity (details are given in Sect.~\ref{sect:propmodel}):
\begin{equation}
    v_h=\dfrac{\eta^*v_b-v_w-\sqrt{\eta^*}(v_b-v_w)}{\eta^*-1}\label{eq:mombalf},
\end{equation}
with $\eta^*$ the injected-to-ambient ratio of inertial density $\gamma^2\rho h=\gamma^2(\rho + \Gamma_1 p/c^2)$, $\Gamma_1\equiv\Gamma/(\Gamma-1)$. Parameters $\eta$ and $\eta^*$ are therefore linked by the relation
\begin{equation}
    \eta^* = \gamma_b^2\eta\frac{h_b}{h_w}.
      \label{Eq:Relate_Mu_MuStar}
\end{equation}
Equation \ref{eq:mombalf} recovers the equation derived in \cite{marti1997morphology} and \cite{mizuta2004propagation}, for instance, by taking $v_w=0$. This model is useful to describe the early jet evolution, but several effects such as the growth of instabilities limit its applicability to longer-term dynamics. We can also cite jet propagation models that take deceleration into account, such as the extended Begelman-Cioffi model from \cite{scheck2002does} and the decelerated momentum balance from \cite{mukherjee2020simulating}, but these models are not adapted to fit the dynamics of our HMMQ jets as they were developed in the context of AGN jets.

\subsection{Linear growth of the Kelvin-Helmholtz instability}
During the jet propagation, various hydrodynamical instabilities can be triggered and in time perturb the beam, reducing the effective beam speed at the front shock and decelerating the jet. An overview is given in Sect.~\ref{sect:instab}. Here we only mention the KHI at a relativistic flow interface. For the relativistic case we are interested in, we derive this dispersion equation from the resonance condition in \cite{hanasz1996kelvin} (for details, see again Sect.~\ref{sect:KHI}),
\begin{equation}
\begin{split}
&\left[(R-1)-n\pi\left(\frac{\omega^2}{\eta_c\Gamma} + (M_c-1)k_x^2 + 2\omega k_x\frac{M_c}{\sqrt{\eta_c\Gamma}} \right)^{1/2} \right]\\ &\quad\quad\quad\quad\times\left(\frac{\omega}{\sqrt{\eta_c\Gamma}} - M_ck_x \right) = 0,
\end{split}
\end{equation}
with $\eta_c\equiv\rho_b/\rho_{ic}$ the ratio of rest mass densities of the beam and inner cocoon, $R\equiv r_{ic}/r_b$ the radius ratio of the sheet and the core, $k_x$ the wave number in the jet propagation direction, and $n$ an integer number. To derive growth time from our simulations, densities are measured as a volume-averaged value over each jet zone, while beam and inner cocoon radius are derived from the respective measured volume and length in the propagating direction of the zone by approximating both as coaxial cylinders. This equation is solved using a Newton-Raphson method and leads to the calculation of the wavenumber $k_x$ , maximizing $\omega$ with the densities and radii as parameters for the first four modes ($n\leq4$). We then use the maximum corresponding $\omega$ to derive the linear growth time for the KHI. These timescales correspond to the linear growth times, whereas the observed growth rate in our simulations will be significantly higher due to nonlinear effects. They are still of interest when we compare them for different runs, as the relation between different linear growth time is the same as the observed runs.

\subsection{Radiative processes}\label{sect:radproc}
Following \cite{bodo2018recollimation}, radiative losses can be added in SRHD equations by introducing a source term in Eq. \ref{eq:vecform},
\begin{equation}
    \partial_t\mathcal{U} + \partial_i\mathcal{F}^i = \Psi.
    \label{Eq:BalanceLaw}
\end{equation}
Radiative losses occur by means of four main phenomena: inverse Compton (IC) scattering, free-free (or Bremsstrahlung) emission, synchrotron emission, and line and recombination cooling.

\begin{figure*}
\includegraphics[width=.5\textwidth]{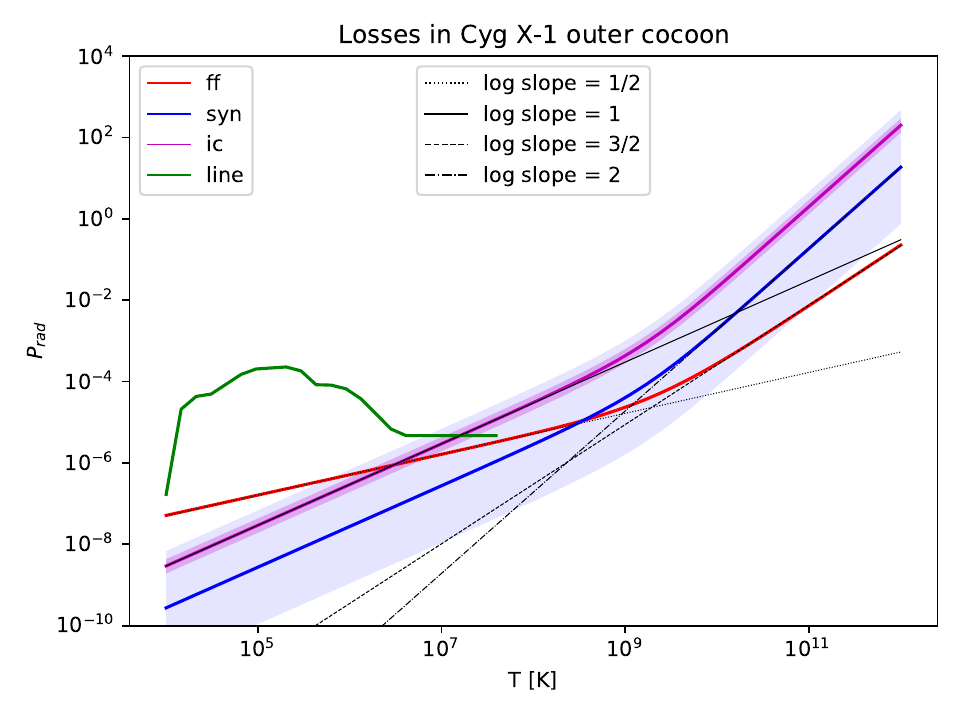}\includegraphics[width=.5\textwidth]{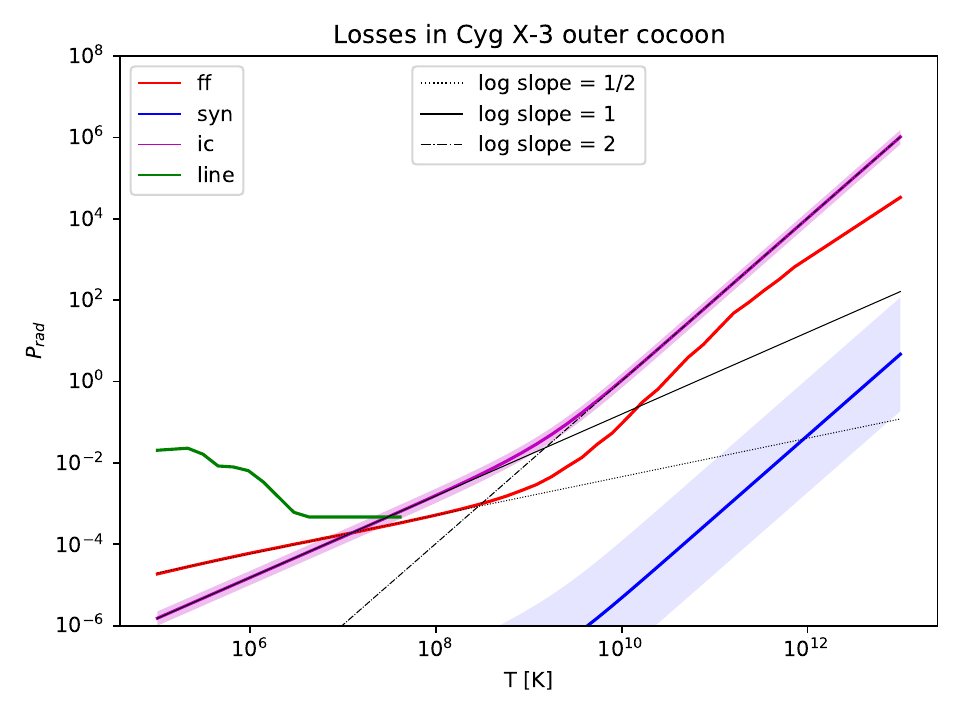}
\caption{Evolution of the loss processes with rest frame temperature in Cygnus X-1 (left) and Cygnus X-3 (right). The thin black lines shows the various temperature scalings detailed Table \ref{tab:powloss}. Line recombination losses ("line") are not drawn for $T>10^{7.7}$ as they are disabled above this temperature. Colored shading shows synchrotron losses ("syn") when the stellar magnetic field $B^\star$ is either multiplied or divided by 5, and the shading around inverse Compton losses ("ic") illustrates a $\pm10\%$ uncertainty on $T^\star$. The values for the physical quantities have been chosen in the outer cocoon leeward side in fiducial runs. They allow for a clear showcasing of the loss scaling, but are not representative of the losses over the simulation. We refer to Fig. \ref{fig:X1X3loss} to compare them over the whole jet. For Cygnus X-1: $\rho = 10^{-15}\text{ g cm}^{-3}$, $B^\star = 10$ G, $T^\star=3\cdot10^4$ K, $(x,y,z) = (1.5\cdot10^{12},4.05\cdot10^{13},4\cdot10^{13})$ cm, $(v_x,v_y,v_z) = (10^8,10^8,10^8)$ cm s$^{-1}$. For Cygnus X-3: $\rho=10^{-14}\text{ g cm}^{-3}$, $B^\star=100$ G, $T^\star=8\cdot10^4$ K, $(x,y,z) = (1.5\cdot10^{12},3.05\cdot10^{13},3\cdot10^{13})$ cm, $(v_x,v_y,v_z) = (10^8,5\cdot10^8,-10^3)$ cm s$^{-1}$.}
\label{fig:lossT}
\end{figure*}

When an external photon field acts as seeds for IC scattering, the emission pattern is anisotropic in the comoving frame of the emitting region, causing the jet to recoil. According to \cite{ghisellini2010compton}, however, this recoil can be neglected in the case of an ion-electron plasma because the majority of the jet momentum is transported by the ions. Both free-free emission and line cooling are dominated by the collisions between electrons and ions, which are isotropic in the fluid rest frame. This logic also applies to synchrotron losses because the pitch angle (between electron speed and magnetic field direction) distribution is isotropic, which results in no global momentum loss due to these radiative processes. We can then model the effect of the various radiative losses as a single energy-loss term,
\begin{equation*}
    \Psi = \begin{bmatrix}
    0\\\vec{0}\\P_{rad}
    \end{bmatrix},
\end{equation*}
where $P_{rad} = P_{IC} + P_{syn} +P_{ff} +  P_{line}$ is the volumic power losses due to inverse Compton scattering, synchrotron emission, Bremsstrahlung emission, and line and recombination cooling. Detailed expressions for each individual loss term including the derivation of the first two from a Maxwell-Jüttner distribution of the electrons and the necessary adaptations we made for them to be compatible with SRHD are given in Appendix~\ref{sect:radprocA}. The power-law exponents of each process are compiled in Table \ref{tab:powloss}. The evolution of the different loss terms with temperature for both types of runs is illustrated Fig. \ref{fig:lossT}.

The corresponding cooling time can be written as
\begin{equation}
    t_c = \gamma\left(t_{c,p}^{-1}+t_{c,\rho}^{-1}\right)^{-1},
\end{equation}
with $t_{c,p} = \gamma\rho/\Dot{\rho}$ the isobaric ($\Dot{p}=0$) cooling time and $t_{c,\rho} = \gamma p/\Dot{p}$ the isochoric ($\Dot{\rho}=0$) cooling time, see Appendix \ref{sect:tc} for the related derivation. Taking $\gamma\approx1$ as a first approximation, $t_{c,p} \approx 10^{21}\rho/P_{rad}$ and $t_{c,\rho} \approx 1.5 p/P_{rad}$. In the range of density and pressure observed in our simulations, the isochoric cooling time is consistently the shortest by about two orders of magnitude.

This work is limited to the case of thermal optically thin plasmas: photons produced in these processes can freely escape without interacting with the gas and thus carry all the energy away from the jet. This approximation is true for X-rays, but does not hold at every frequency. Verifying this hypothesis would require more specific investigations. We also applied our cooling model to the ambient medium, even though it is optically thick for optical and UV lines in Cygnus X-1 due to the properties of O-type star winds, and to optical and UV continuum in Cygnus X-3 because the companion star is a Wolf-Rayet (WR) star. We modeled this effect by setting up a temperature floor slightly lower than the surface temperature of the star to ensure that the medium did not cool to nonphysical values. This value is arbitrary and has no impact on the jet dynamics as it has close to no effect on the ambient inertial density, and the jet overpressure is such that modifying the ambient pressure has no effects.

The radiative processes have different scalings with rest mass density, temperature, and distance to the star in the orbital frame. Therefore the dominance of one or two processes over the others may vary with time and space. The power-law exponents of each process detailed Appendix \ref{sect:radprocA} are compiled in Table \ref{tab:powloss}. The evolution of the different loss terms with temperature for both types of runs is shown Fig. \ref{fig:lossT}.

\subsection{A-MaZe simulation toolkit}
We performed 3D simulations using the hydrodynamical module from the A-MaZe simulation toolkit \citep{walder2000maze, folini2003new, melzani2013apar}. It uses the method of lines, a semidiscretized finite-volume method: after discretizing in space, the resulting system of ordinary differential equations is solved with a forward Euler scheme. Fluxes are computed by the (stabilized) Lax-Friedrichs approximation using a second-order reconstruction based on min-mod limiters. The equations, the solution method, and a benchmark for the accuracy of this method are all reported in Appendix \ref{sect:det_num}. A relativistic solver was implemented by adding the recovery of primitive variables using the inversion method detailed Appendix \ref{sect:app_invscheme}. 

\begin{figure}
    \centering
    \includegraphics[width=\hsize]{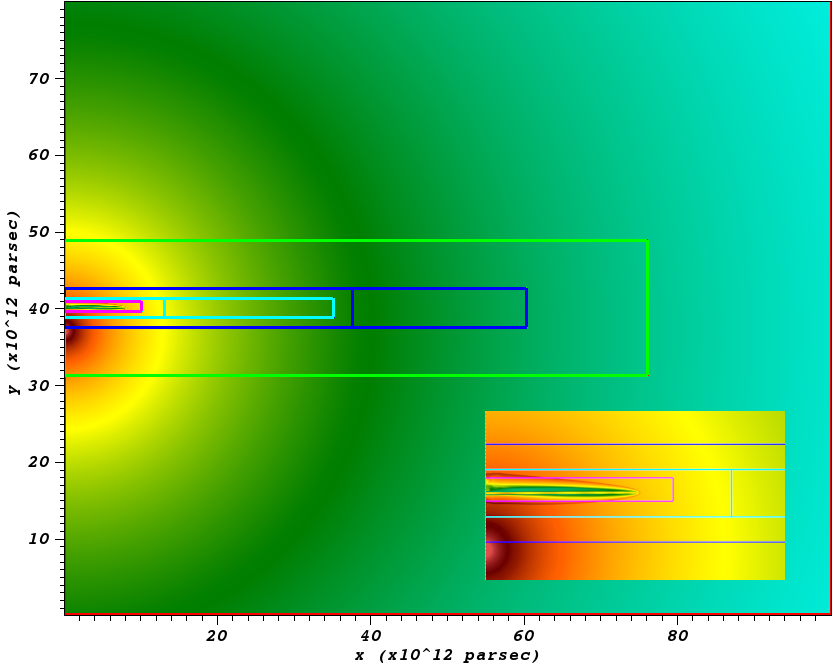}
    \caption{Structure of the computational grid over the whole domain, illustrated for a rest mass density slice along the plane containing the star and jet center. The density scale extends from $10^{-19}$ (deep blue) to $10^{-12}$ g cm$^{-3}$ (red), same as Fig. \ref{fig:CygX1_jetevol}. The grid contains five refinement levels: the whole domain is refined twice (green and blue interfaces) by a factor of 4 and then twice more (cyan and magenta interfaces) by a factor of 2 to attain a factor of 64 in the finest levels, which are better shown in the zoom into the jet injection at the orbital scale in the bottom right corner of the image.}
    \label{fig:grid}
\end{figure}

Our simulations were set in a static grid made of five refinement levels centered on the jet injection nozzle, as shown in Fig. \ref{fig:grid}. Cells from the coarse grid had a $4\cdot10^{11}$ cm edge, and the edge of the highest-level cells was 64 times lower for a maximum resolution of $6.25\cdot10^9$ cm. The number of coarse level grid cells was $250\times200\times200$ and $250\times150\times150$ for Cygnus X-1 and Cygnus X-3, respectively. The associated physical domain sizes are given in Table \ref{tab:X1vX3}. The cfl number was set to 0.15. The time step was refined along with the spatial grid. On the coarse grid, it was about 2s for Cygnus X-1 and 5s for Cygnus X-3. The jet was injected perpendicular to the orbital plane (y-z plane) by fixing $(\rho_j, \Vec{v}_j, T_j)$ on a few cells at x=0, always imposing at least 20 cells of the finest grid to fix the diameter of the beam. The environment was set by fixing the wind velocity and density at the stellar surface, resulting in an isotropic wind with constant speed modulus and density in $r^{-2}$. The boundary condition the at x=0 plane was reflective, while the other boundaries of the simulation grid had outflow conditions.

\subsection{Covered parameter space}
We defined runs CygX1 and CygX3 as our fiducial runs for Cygnus X-1 and Cygnus X-3, respectively. The main parameter values for these two runs are given in Table \ref{tab:X1vX3}. The choice of physical values was inspired by \cite{orosz2011mass} and \cite{yoon2015global} for Cygnus X-1 and by \cite{orosz2011mass} and \cite{dubus2010relativistic} for Cygnus X-3. The parameter choices for the various sensitivity studies are listed in Tables \ref{tab:CygX1} and \ref{tab:CygX3}.

Table \ref{tab:X1vX3} shows the value of the environment parameters relevant for jet radiative losses as well as the parameters of the respective fiducial runs. The characteristics of the Cygnus X-3 system mean the radiative losses will be stronger overall. We chose a stronger magnetic field base value for Cygnus X-3 to compensate for the addition of the distance scaling detailed in Appendix \ref{sect:domRad}. The luminosity of the two companion stars are similar because the companion star in Cygnus X-3 is hotter but smaller. Thus the smaller orbital distance implies stronger synchrotron and inverse Compton losses by a factor 100. The beam density $\rho_b$ was chosen to be ten times greater than in the Cygnus X-1 runs, implying stronger line and free-free losses by a factor 100 also. Both beams roughly have the same internal energy density, but will cool a $\sim100$ times faster in Cygnus X-3 case. The comparison of the position and speed diagrams of Figs. \ref{fig:CygX1_LvnoL_T} and \ref{fig:CygX3_LvnoL_T} shows that the first cooling effects are seen after 7600 s for Cygnus X-1, and the first cooling effects are visible after 100 s for Cygnus X-3.
\begin{table}
    \centering
    \begin{tabular}{|C||C|C|c|}
    \hline
     & \text{Cygnus X-1} & \text{Cygnus X-3} & unit \\\hline
    \rho_j & 1.3\cdot10^{-15} & 1.4\cdot10^{-14} & g cm$^{-3}$\\
    v_j & 10^{10} & 2.25\cdot10^{10} & cm s$^{-1}$\\
    T_j & 10^8 & 10^8 & K\\\hline
    d_{orb} & 3\cdot10^{12} & 2.6\cdot10^{11} & cm\\
    R^\star & 16.2 & 2.3 & $R\odot$\\
    T^\star & 3\cdot10^4 & 8\cdot10^4 & K\\
    B^\star & 10 & 100 & G\\
    \Dot{M}^\star & 3\cdot10^{-6} & 10^{-5} & $M_\odot$yr$^{-1}$\\
    v_\infty & 1000 & 1500 & km s$^{-1}$\\
    x_{max} & 10^{14} & 10^{14} & cm\\
    y_{max},\,z_{max} & 8\cdot10^{13} & 6\cdot10^{13} & cm\\\hline
    \end{tabular}
    \caption{Main parameters of fiducial runs.}
    \label{tab:X1vX3}
\end{table}

\subsection{Post-processing}\label{sect:postproc}
To perform a quantitative analysis of our simulations, we identified each computational cell according to the following rules for the various interfaces of the jet: the separation between ambient material and outer cocoon was made at $p = .01$ Ba \& $T = 10^7$ K, the working surface between inner and outer cocoon was defined where $J=0.05,$ following the definition for the mixing layer in \cite{perucho2004stabilityII}, and cells were considered part of the beam if $\zeta \equiv (v_x/v_j)J > 0.8$. This criterion was defined in \cite{yoon2015global}. We found that choosing 0.8 as the threshold value identified the beam up to the reverse shock with more success than a criterion that was purely based on $J$, especially in the later phases of the jet outbreak when beam and inner cocoon mix. The value $J=0.05$ was found to correctly separate the low-density high-temperature inner cocoon from the outer cocoon. These criteria are deemed correct in the sense that their limits correspond to the jumps in the various physical quantities between jet zones. Redundant cells between the different refinement levels were ignored to avoid errors. 

After identifying each computational cell as part of a zone, we measured various quantities relative to each zone in postprocessing, such as the length, the volume, the volume-averaged quantities, and the probability density functions over the jet. We also defined a proxy for the jet aspect ratio as $(\pi l^{3} / V)^{1/2}$ , where $l$ and $V$ are the total length and volume of the jet, respectively. This is equivalent to the ratio $l/R_{j,eff}$, where $R_{j,eff}$ is the radius of a cylindrical jet of length $l$ and volume $V.$ 

\section{Results}\label{sect:results}
The results we present below are, to the best of our knowledge, the first 3D simulations of jets in HMMQs that are relativistic and include radiative cooling in parameterized form. They cover the evolution of the jet from its launch over the onset of instabilities and radiative cooling to the turbulent phase at the end of our simulations. 

More specifically, we discuss the propagation of the jet through the stellar wind from its outburst close to the BH up to scales of about $6\cdot 10^{13}$~cm for Cygnus X-1. This corresponds to about 20 times the separation between the two stellar components $d_{orb}$, and $2\cdot 10^{13}$~cm $\approx75d_{orb}$ for Cygnus X-3 (the values for $d_{orb}$ are consigned in Table \ref{tab:X1vX3}).
A brief overview of previous studies is given in Sect. \ref{sect:previous}. Details about our fiducial simulations are given in Sect.~\ref{sect:fiducial}, with particular focus on the development of the KHI and its role in different phases of jet propagation (Sect. \ref{sect:growth}), as well as cocoon evolution (Sect. \ref{sect:coc_evol}). Then, the impact of radiative losses on jet structure and dynamics (Sect. \ref{sect:loss}) is investigated, before we perform a short parameter study (Sect. \ref{sect:paramsens}) of the jet temperature (Sect. \ref{sect:T}), kinetic power (Sect. \ref{sect:rho}), and stellar wind (Sect. \ref{sect:vw}).

\subsection{Previous studies}\label{sect:previous}
The interaction of a jet with its ambient medium typically results in a richly structured flow field. For any discussion of this complex flow field, it is useful to resort to its basic idealized morphology, which is characterized by three surfaces that separate the flow into four zones. The innermost zone, the jet beam or spine, consists of unprocessed jet material. A combination of discontinuities (including a reverse or terminal shock and potentially shear layers, or also reconfinement shocks) separates the jet beam from the inner cocoon, which is composed of shocked jet material (also called "shear layer", "jet sheath", or simply "cocoon"). A contact discontinuity or working surface marks the transition from inner to outer cocoon (or "cavity"), the latter containing shocked ambient medium. A third discontinuity, a bow or forward shock at the jet head, marks the transition to the ambient medium~\citep[good sketches of this structure can be found in other works, e.g.,][for a recent example]{matsumoto2019propagation}.

\cite{marti1997morphology} identified five parameters to completely specify a relativistic jet propagating into a homogeneous medium: the density ratio $\eta\equiv\rho_j/\rho_w$, the pressure ratio $K\equiv p_j/p_w$, the beam flow velocity $v_j$ (or its associated Lorentz number $\gamma_b$), the beam Mach number $M_j$,and the polytropic index $\Gamma$.

Quantities with subscript $j$ are relative to the values at injection, while the subscript $w$ denotes quantities of the (in our case, wind-dominated) ambient medium. As pointed out by \cite{marti1997morphology}, the propagation efficiency of a relativistic jet is mostly determined by the inertial mass density $\xi=\gamma^2\rho h$ introduced in Sect. \ref{sect:srhd}, and especially by the ratio $\eta^*$ between beam and ambient medium inertial mass densities because the latter determine the momentum balance at the contact discontinuity in the jet head.

Sufficiently light ($\eta<0.1$) supersonic jets, as considered in this work, display extended cocoons as the high pressure of the shocked gas drives a backflow toward the source. These jets also display a series of internal oblique shocks in the beam, whose strength and spacing are determined by the Mach number and the gradient in the pressure external to the beam \citep[see, e.g.,][and references therein]{gomez1995parsec, gomez1997hydrodynamical}: the higher these numbers, the stronger and closer to each other these oblique shocks. Increasing the Mach number also intensifies the expansion of the cocoon.

The structure of this cocoon is determined by the adiabatic index $\Gamma$: for models with $\Gamma = 5/3$, the cocoon is stable at first, but eventually evolves into vortices, producing turbulent structures and generating perturbations at the beam boundary. This enriches the internal structure of the beam. For models with $\Gamma = 4/3$, the first internal conical shock is strong enough for the resulting beam collimation to accelerate the flow. During this acceleration phase, the beam gas is less efficiently redirected to the cocoon downstream, accumulating at jet head. Once this acceleration is over, the continuous flow reestablishes itself, forming small turbulent vortices in the cocoon. The cocoon structure reflects this history, the parts of it formed before and after this beam acceleration phase presenting different morphologies. See \cite{marti1997morphology} for a more in-depth discussion.

\cite{bodo1994kelvin} observed that during its propagation, a jet will present different structures tightly linked to the evolution of the KHI modes, identifying three phases: in a first "linear phase" the various modes grow following the linear behaviour, and no shock is present in the beam. The apparition of biconical shocks centred on the beam axis marks the beginning of the "expansion phase", during which the strength of these shocks grow and the jet radius expands in the post-shock region. Finally, the evolution of the shocks leads to mixing between the jet and the external material, marking the start of the "mixing phase".

The stellar wind can influence the propagation of a relativistic jet greatly: jet disruption by a constant and perpendicular wind were observed in \cite{perucho2008interaction} and \cite{perucho20103d} even for moderate jet kinetic luminosities. In particular, the presence of a wind strengthens the initial recollimation shock in the beam which can also strengthen jet asymmetric KHI produced at the wind/jet contact discontinuity. These may in turn disrupt the jet.

\subsection{Cygnus X-1 and Cygnus X-3 fiducial cases}\label{sect:fiducial}
\begin{figure}
    \centering
    \includegraphics[width=\hsize]{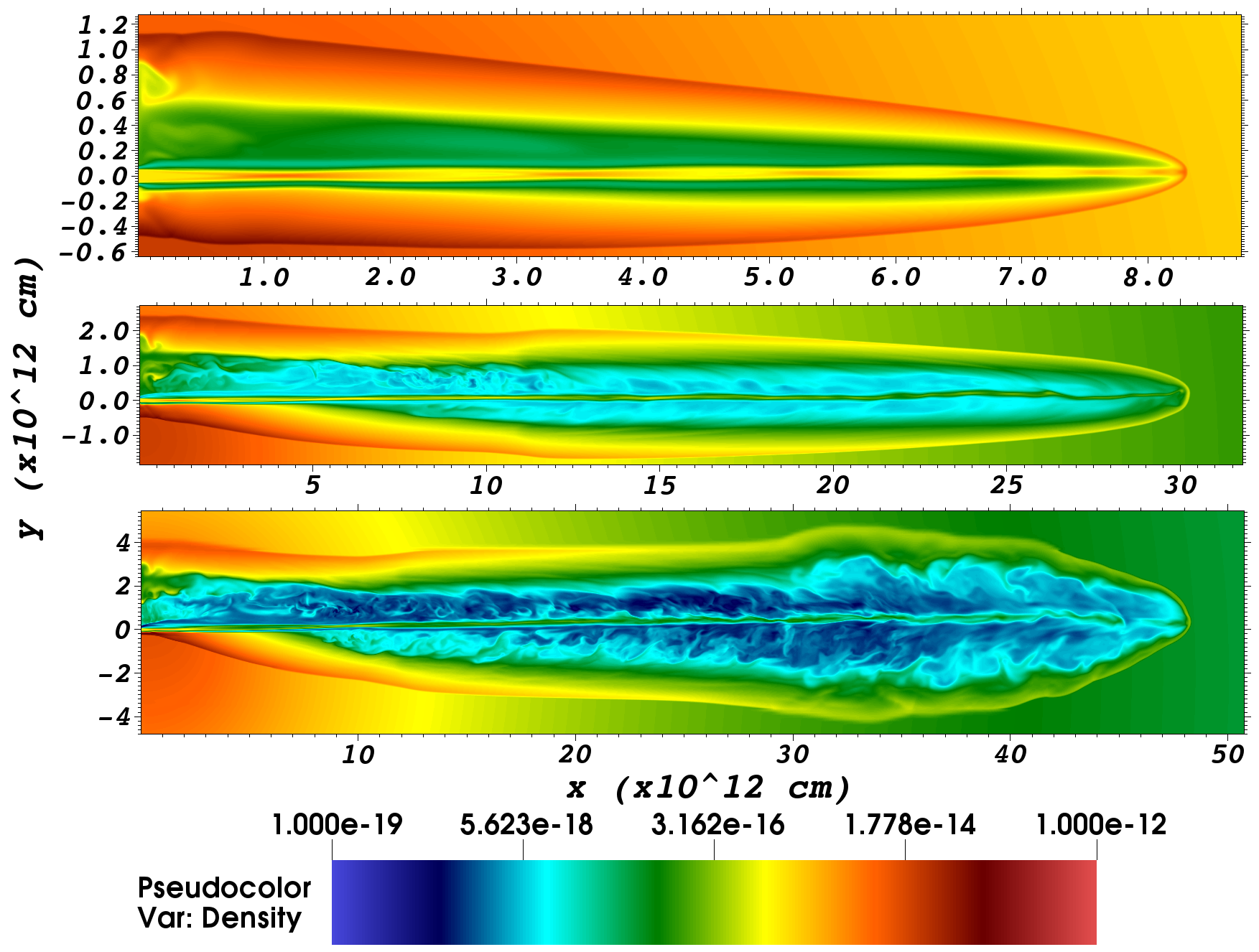}
    \caption{Rest-mass density slices of run CygX1 at times (top to bottom) t = 2000, 6000, and 12000 s, showing the three evolutionary phases detailed in the text in Sect. \ref{sect:fiducial}}
    \label{fig:CygX1_jetevol}
\end{figure}

\begin{figure*}
    \includegraphics[width=.5\textwidth]{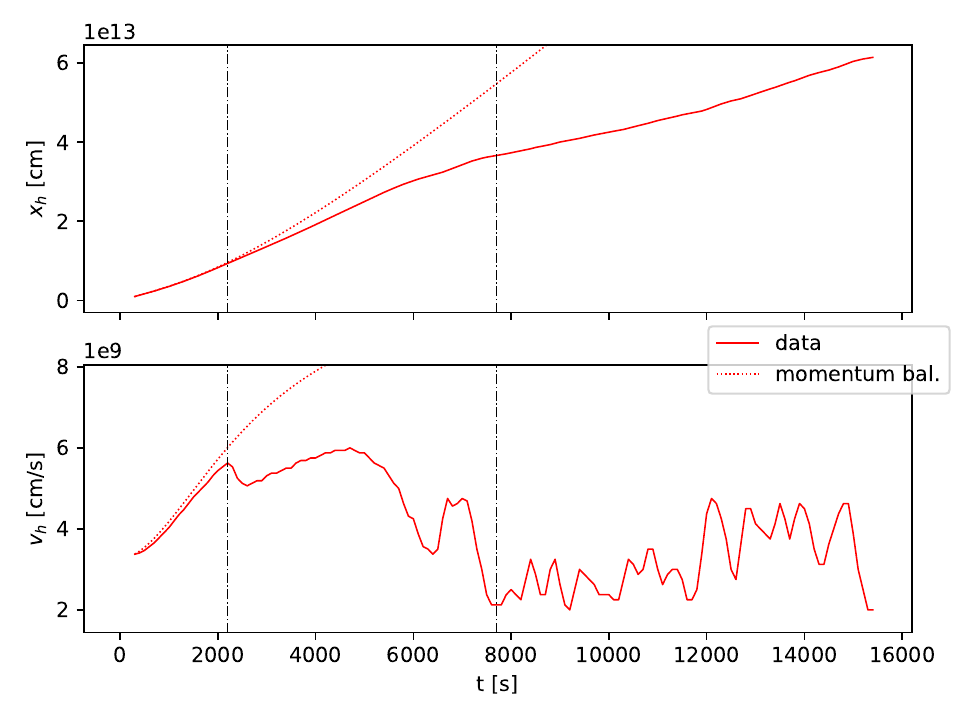}\includegraphics[width=.5\textwidth]{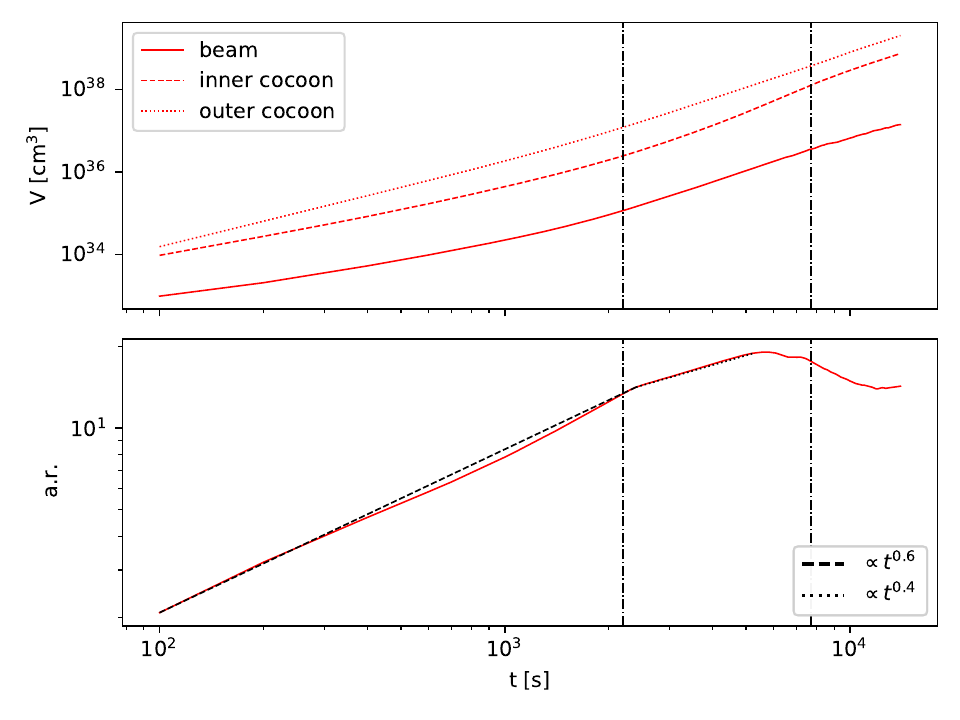}
    \caption{Dynamical and structural evolution of fiducial run CygX1. \textbf{Left:} Position and speed of the jet head for fiducial run CygX1. \textbf{Right:} Jet volume (beam and inner and outer cocoon) and aspect ratio. The limits of each evolutionary phase are marked by the vertical dash-dotted lines in the various panels.}
    \label{fig:CygX1_fid}
\end{figure*}

\begin{figure}
    \centering
    \includegraphics[width=\hsize]{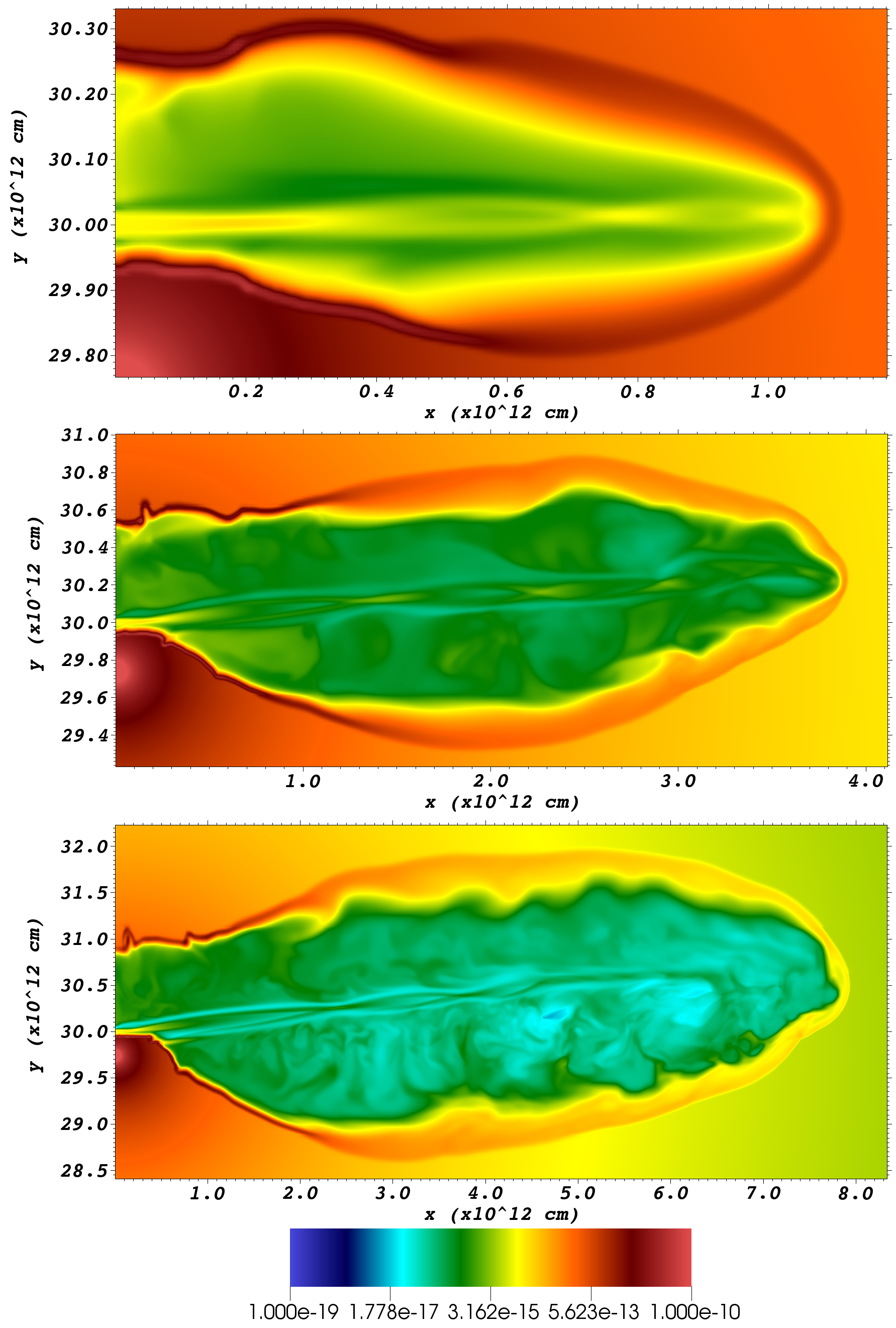}
    \caption{Rest-mass density slices of run CygX3 at times (top to bottom) t = 400, 1200, and 2500 s, showing the three evolutionary phases detailed in the text in Sect. \ref{sect:fiducial}}
    \label{fig:CygX3_jetevol}
\end{figure}

\begin{figure*}
    \includegraphics[width=.5\textwidth]{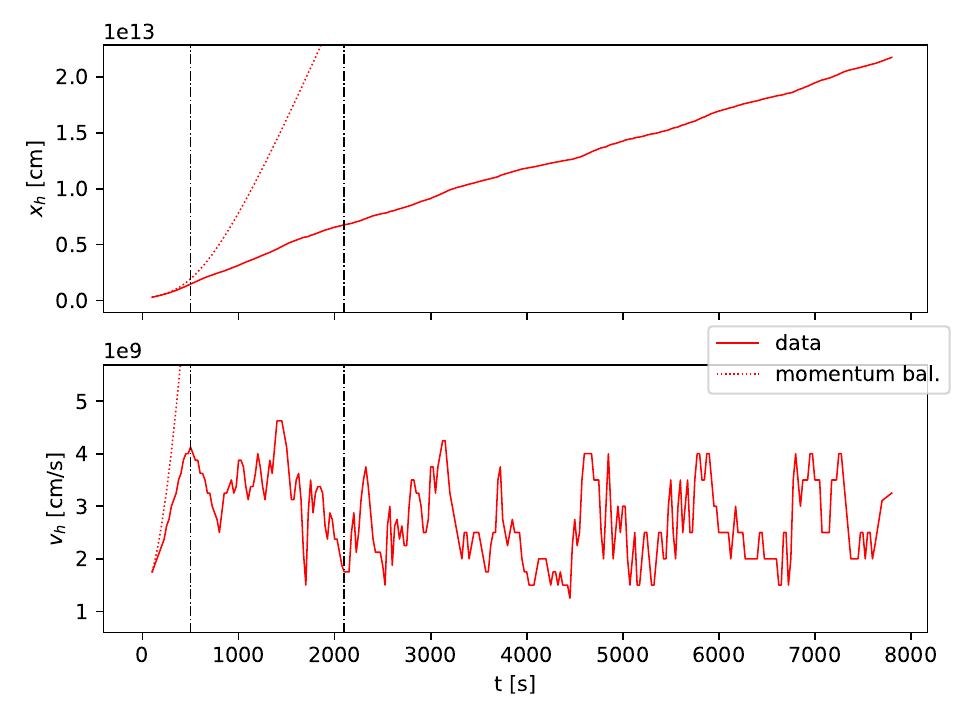}\includegraphics[width=.5\textwidth]{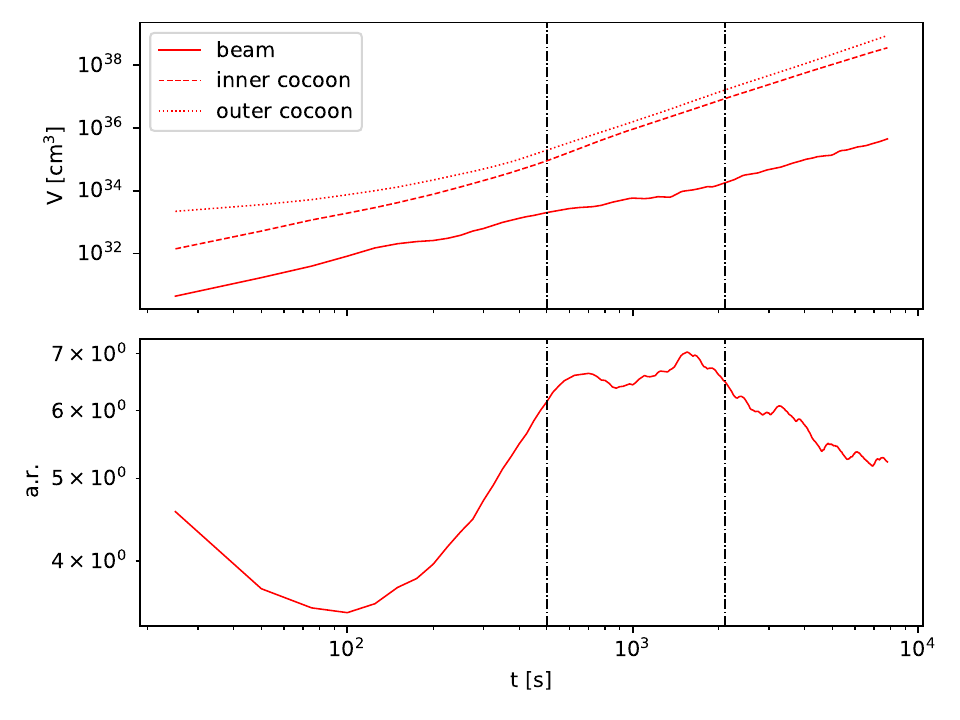}
    \caption{Dynamical and structural evolution of fiducial run CygX3. \textbf{Left:} Position and speed of the jet head for fiducial run CygX3. \textbf{Right:} Jet volume (beam and inner and outer cocoon) and aspect ratio. The limits of each evolutionary phase are marked by the vertical dash-dotted lines in the various panels.}
    \label{fig:CygX3_fid}
\end{figure*}

We start with a description of the fiducial cases for Cygnus X-1 and Cygnus X-3, respectively, against which all other sensitivity studies will be compared later on. Converting the numerical values of the parameters to the dimensionless quantities introduced in Sect. \ref{sect:previous} (Tables \ref{tab:CygX1params} and \ref{tab:CygX3params}) places our jets in the supersonic case with extended, turbulent cocoons and a beam with rich internal structure. Our fiducial runs indeed follow these expectations: the evolution of the jet is shown Figs. \ref{fig:CygX1_jetevol} and \ref{fig:CygX1_fid} for the CygX1 run, \ref{fig:CygX3_jetevol} and \ref{fig:CygX3_fid} for CygX3. Several features catch the eye, which we further elaborate on below. First, there is qualitative change in the appearance of the jet, from an early 'well ordered' state to a turbulent state later on. This change is also reflected in the propagation of the jet head and three phases of the jet evolution can be identified. Second, the aspect ratio of the jet is different for Cygnus X-1 and Cygnus X-3. Third, jet bending due to the lateral wind impact is observed in all simulations. Fourth, the jet is asymmetric due to the wind of the companion star.

\subsubsection{Instability growth and phases of jet propagation}\label{sect:growth}
The structure of our jets goes through three phases, common to both Cygnus X-1 and Cygnus X-3 runs. We refer to these three evolution stages as the "smooth", "instability growth" and "turbulent" phases respectively according to their inner structure. An illustration via rest-mass density slices is given in Figs.~\ref{fig:CygX1_jetevol} and \ref{fig:CygX3_jetevol}. The three phases also leave an imprint on the time-series data shown in Figs.~\ref{fig:CygX1_fid} and \ref{fig:CygX3_fid}.

During the first phase, the beam flow is surrounded by a smooth cocoon that is symmetrical at its head. A few internal shocks are present in the beam, starting with a strong recollimation shock situated at a few $\sim10^{12}$ cm downstream of the injection. Its existence and position are coherent with the criterion and analytical prediction from \cite{yoon2016formation} obtained by equating wind ram pressure and lateral ram pressure in the beam. The aspect ratio of the jet defined in Sect. \ref{sect:postproc} increases gradually. It roughly follows a power law in time in the early propagation phase with an exponent of $\sim$0.6. For Cygnus X-3, the aspect ratio in the same phase first decreases before it increases with what could be a power law with a similar exponent as for CygX1. This may be explained by the strong asymmetry of the cocoon during this phase due to the strong stellar winds. A small deviation of the CygX3 beam can already be observed at this point. The jet head position and velocity follow the theoretical 1D result from Sect.~\ref{sect:propmodel}. Deviations indicative of the transition from phase one (smooth) to phase two (instability growth) occur after roughly 2000 seconds in the case of CygX1 and much earlier, after a few hundred seconds, in the case of CygX3. In particular, the speed diagram for CygX3 breaks almost immediately from the theoretical profile. This may be a consequence of the already existing bending of the beam.

In the second phase, instabilities grow in the jet, which perturb the flow in both inner cocoon and beam head. While the jet volume tends to grow faster now than during the first phase, the growth of the aspect ratio slows down with a $\sim0.4$ exponent for CygX1 case, and the jet head velocity overall decreases while the position breaks from the theoretical values. In CygX1, the number of over- and underpressure regions in the beam (see Figs \ref{fig:CygX1_instgr_p} and \ref{fig:CygX3_instgr_p} in the appendix) stays approximately constant before increasing after about 5000 seconds. Ultimately, the growing instabilities cause oscillations of the beam head perpendicular to its propagation direction. For CygX3, these oscillations induce speed fluctuations even though the beam still retains its structure.

In the last phase, after about 6000 seconds in CygX1 and 2000 seconds in CygX3, the perturbations have reached the beam core. They modify the beam structure at the jet head severely, while the inner cocoon has become turbulent. This also marks a change in shape of both the cocoon and the jet head. The modification of the jet head shape can be linked to oscillations of the beam region that end in the reverse shock, as a beam head that is misaligned with the general jet propagation direction causes beam material to flow at higher speed and in same direction as the cocoon expansion, which deforming the jet. The jet head position evolves with an approximately constant mean velocity. Fluctuations up to roughly 30\% are visible in the speed plots, which is in line with the persisting motion of the jet head position perpendicular to the jet axis. The volume of the outer cocoon evolves roughly as a power law in time with an exponent of about three for CygX3 and half as much for CygX1. The volume of the beam features a similar time dependence in the case of CygX1, but a shallower dependence for CygX3, with a powe- law exponent of about two instead of three. The aspect ratio decreases somewhat before becoming constant, at least in the case of CygX1.

This classification can be compared to the classification from \cite{bodo1994kelvin} given in Sect. \ref{sect:previous}, but their "linear" phase escapes our data analysis because we dump data frames only every 100 seconds and 25 seconds for CygX1 and CygX3, respectively, while the estimated KHI linear growth timescale is typically on the order of a few tens of seconds for Cygnus X-1 and a few seconds for Cygnus X-3 (see Table \ref{tab:tKHI}). No value was found with this method for run CygX1\_mP, where the beam is heavily disrupted by the stellar wind and the approximations made are no longer valid. Our first two phases (smooth and instability growth) appear to be subdivisions of their "expansion" phase, while our turbulent and their "mixing" phase match.

\begin{table}[ht]
    \centering
    \begin{tabular}{|cc||cc|}
        \hline
        run name & $t_{KHI}$ (s) & run name & $t_{KHI}$ (s) \\\hline
        CygX1 & 71.4 & CygX3 & 0.80 \\
        CygX1\_noLoss & 68.0 & CygX3\_noLoss & 18.4 \\
        CygX1\_wind & 60.8 & CygX3\_mW & 2.7 \\
        CygX1\_mP & / & CygX3\_mP & 0.12 \\
        CygX1\_T7 & 207.0 & CygX3\_mPmW & 7.1 \\
        CygX1\_T9 & 26.8 & CygX3\_mPmmW & 10.0 \\\hline
    \end{tabular}
    \caption{Linear growth time of the KHI. The slash indicates no value could be found.}
    \label{tab:tKHI}
\end{table}

These phases are also visible in the speed diagram of the jet head that display the same trend for the Cygnus X-1 and Cygnus X-3 fiducial runs in Figs. \ref{fig:CygX1_fid} and \ref{fig:CygX3_fid}: we can link the smooth phase with the initial acceleration, the deceleration and the concave part with the instability growth, followed by the turbulent phase. The first two of these three phases are of interest in the context of dedicated studies on instability onset and growth. Although a large body of associated literature exists, we are not aware of any such studies for relativistic jets in HMMQs including radiative cooling.

The link between the internal structure and the dynamic was discussed in \cite{marti2016internal}, suggesting that the growth of the KHI is related to the strength of these oblique internal shocks inside the beam: the KHI grows as the sound wave travels back and forth between the beam surface and the contact discontinuity, therefore more and stronger internal shocks produce a greater number of reflections within a given time or distance. This ultimately accelerates the growth of the KHI. The ripple-like structures observed in the cocoon, similar to pressure perturbations in \cite{perucho2004stabilityI}, could be viewed as markers of such sound waves.

Returning to the density in Figs. \ref{fig:CygX1_jetevol} and \ref{fig:CygX3_jetevol}, although the beam and cocoon mix together at the jet head, the flow is not slowed down until the very end of the jet. The jets are bent away from the star almost as soon as the jet is established for the Cygnus X-3 runs, but also in Cygnus X-1 runs after enough lifetime of the jet. This bending angle $\psi$, defined in \cite{yoon2015global} as the angle between the local and the initial velocity vector, can be compared to the analytical value derived in the same paper. For run CygX1, we find $\psi=0.1$ rad for a beam end at $x = 6.8\cdot10^{13}$ cm, which is close to $\psi=0.09$ found analytically. For CygX3, we find $\psi=0.04$ rad for a beam end at $x=1.52\cdot10^{13}$ cm and $0.03$ analytically, showing good agreement of our runs with the analytical estimate.

\subsubsection{Cocoon evolution and radiative losses}\label{sect:coc_evol}
Over the course of the initial jet outburst, the outer cocoon expands in the directions perpendicular to the jet propagation due to its overpressure compared to the ambient medium. On the windward side closest to injection, the interface between cocoon and wind is a bow shock and its dynamic are determined at first order by the balance between wind ram pressure and internal thermal pressure of the cocoon: depending on this balance, the interface will move either away from or closer to the beam. Farther away from the plane of orbit, the wind ram pressure becomes negligible and the interface dynamics is driven by the balance between internal and external thermal pressure. On the leeward side, the cocoon expands in the same direction as the wind speed. The resulting asymmetry of the cocoon is apparent at early times in both CygX1 and CygX3. At later times, the difference between windward and leeward side diminishes as the wind speed is increasingly aligned with the propagation direction of the jet. (Figs.~\ref{fig:CygX1_jetevol} and \ref{fig:CygX3_jetevol}).

\begin{figure}
    \centering
    \includegraphics[width=\hsize]{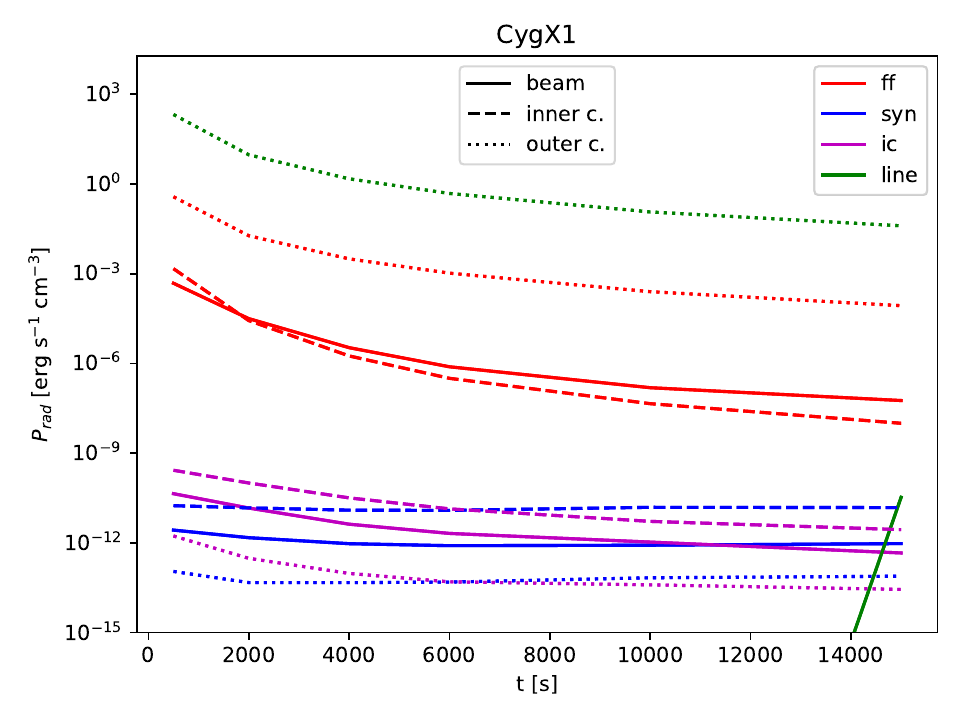}
    
    \includegraphics[width=\hsize]{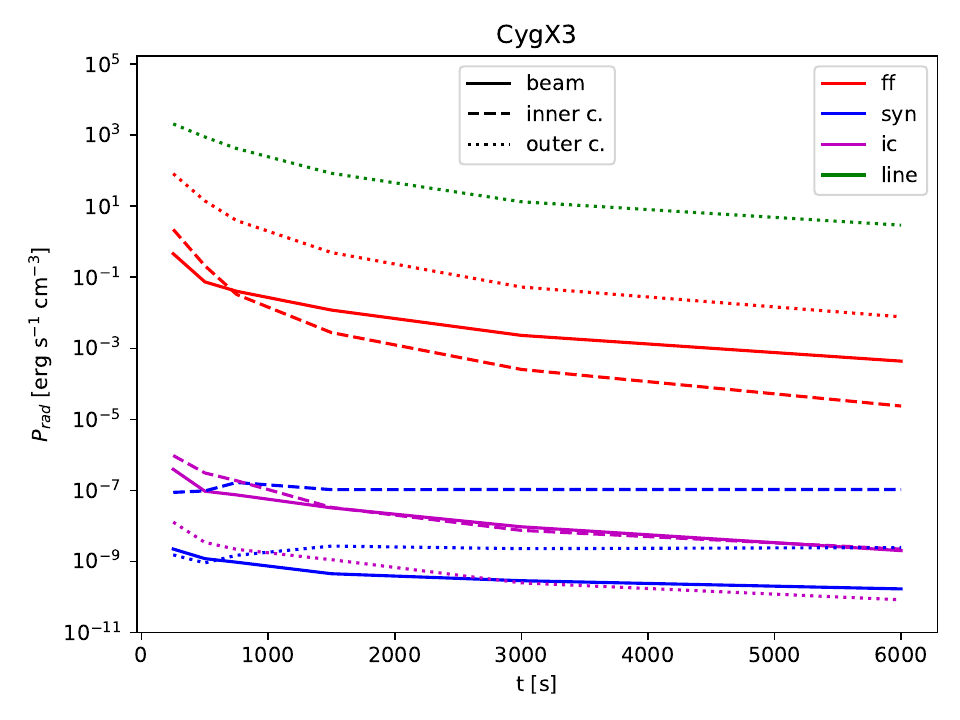}
    \caption{Time evolution of radiative losses for the fiducial runs CygX1 (top) and CygX3 (bottom), derived at each cell and summed per zone, at time steps t = (500, 2000, 4000, 6000, 10000, 15000) s for CygX1 and t = (250, 500, 750, 1500, 3000, 6000) s for CygX3 (two data points per evolutionary phase). In both cases, free-free losses dominate cooling in the inner cocoon and beam, while line cooling dominates the outer cocoon.}
    \label{fig:X1X3loss}
\end{figure}

As the cocoon cools down with time either adiabatically due to expansion and/or from radiative cooling, the thermal pressure of the cocoon diminishes, which increases the influence of the wind on its dynamics. Figure \ref{fig:X1X3loss} displays the volumic power losses per jet zone per process for the fiducial runs CygX1 and CygX3, measured over all the jet cells for two data points per evolutionary phase. In both cases, free-free losses dominate the cooling in the beam and inner cocoon, with a stronger cooling in the beam than in the inner cocoon. The colder outer cocoon is dominated by the very efficient line recombination cooling. This result holds true for all our simulated runs.

Moreover, the gas in the cocoon has a velocity component in the positive $x$-direction. Thus the cocoon moves outward of the system with the beam, but at a slower pace. Ultimately, no trace of the original cocoon is left in the innermost parts of the jet. A thin interface of shocked stellar wind only a few $r_b$ wide has instead formed between the wind and the beam (late times in Figs. \ref{fig:CygX1_jetevol} and \ref{fig:CygX3_jetevol}). This "naked beam" is of interest because it represents a (quasi-) stationary state structure studied in the literature (e.g., \citealt{wilson1987steady, komissarov2015stationary} for hydrodynamical jets, \citealt{marti2016internal, bodo2018recollimation} for MHD jets) and can be related to direct observations.

\subsection{Effects of losses on jet structure and dynamics}\label{sect:loss}
\begin{figure}
    \centering
    \includegraphics[width=\hsize]{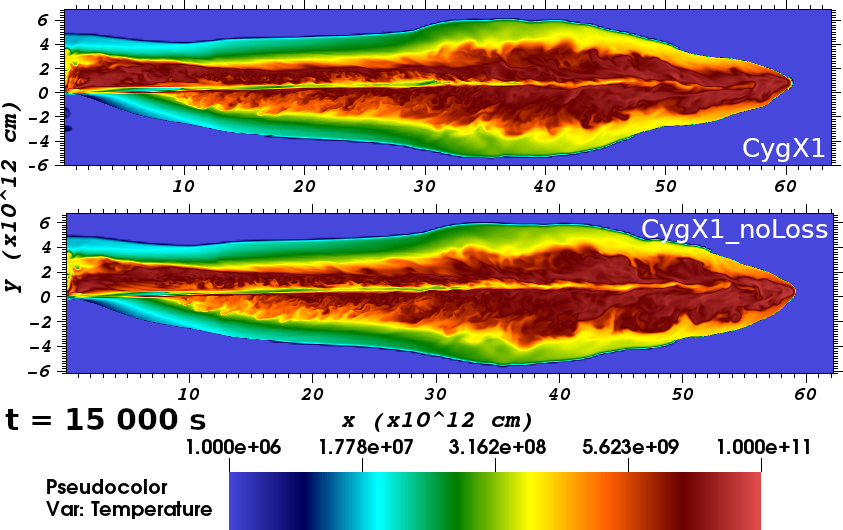}
    
\includegraphics[width=\hsize]{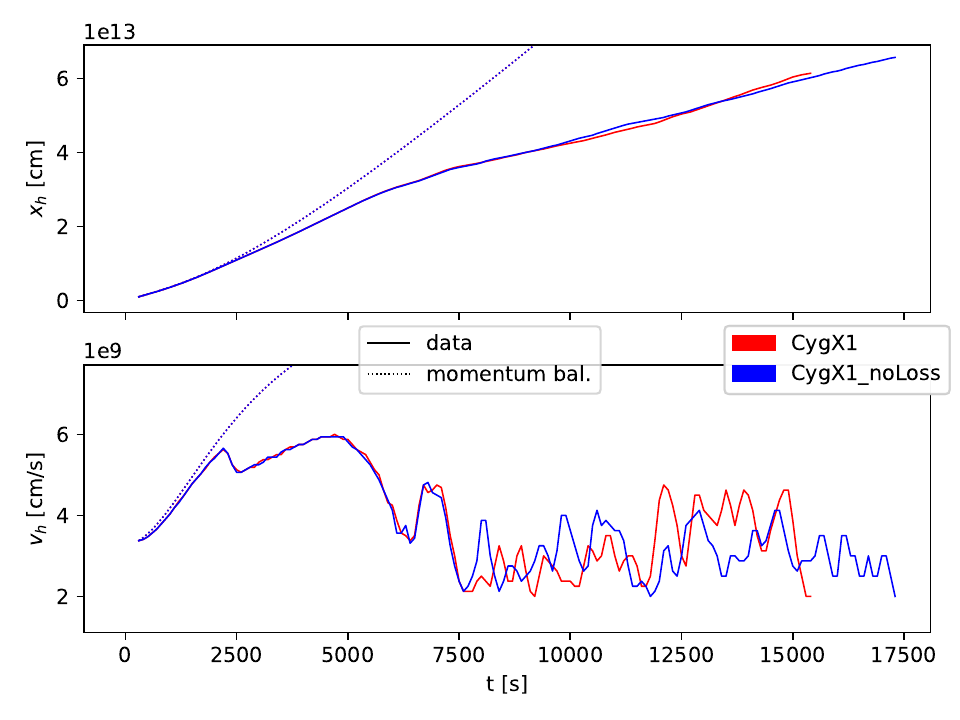}
    \caption{Effects of losses on run CygX1 structure and dynamics. \textbf{Top:} Temperature slices of runs CygX1 and CygX1\_noLoss at time t = 15 000 s. Both jets display similar structures, with the exception of a slightly larger outer cocoon at the head of the non-cooled jet. \textbf{Bottom:} Jet head propagation and speed of the same runs, CygX1 in red and CygX1\_noLoss in blue. The theoretical 1D propagation from Appendix \ref{sect:propmodel} is drawn as dotted lines following the same color-coding. The propagation is identical in both runs until the start of the turbulent phase, after which speed fluctuations differ, but the average propagation speed is identical in the two runs with almost no difference in the jet head position plot.}
    \label{fig:CygX1_LvnoL_T}
\end{figure}

\begin{figure}
    \centering
    \includegraphics[width=\hsize]{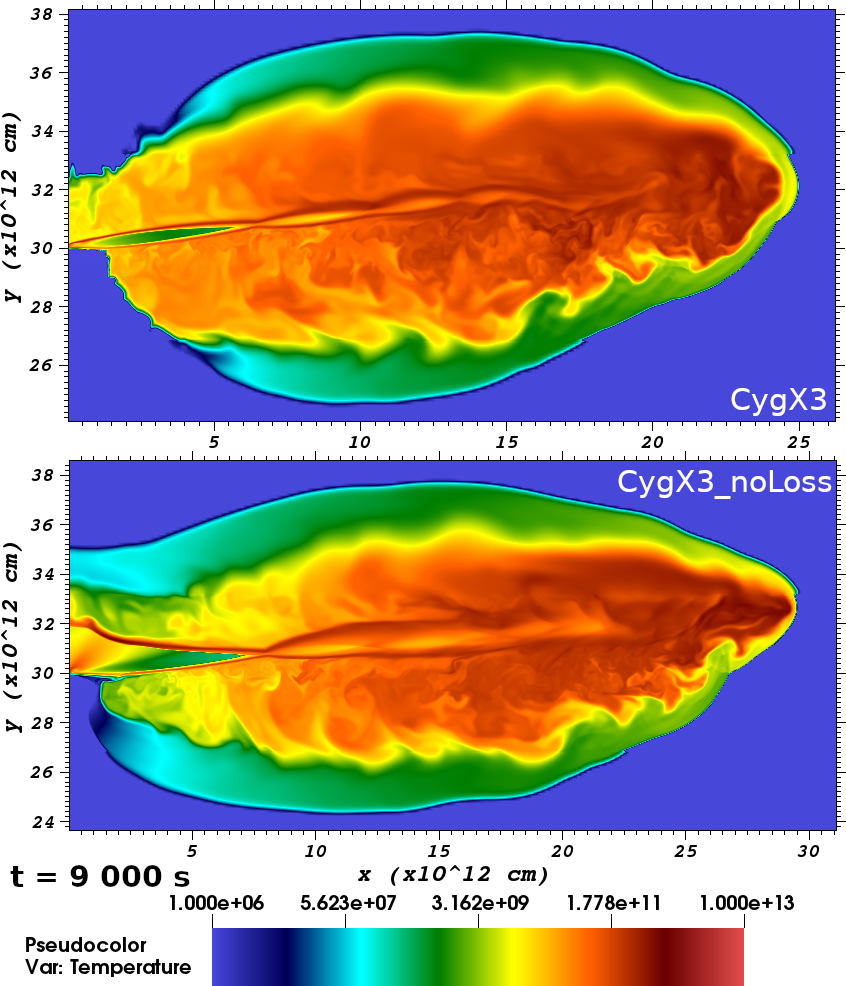}
    
\includegraphics[width=\hsize]{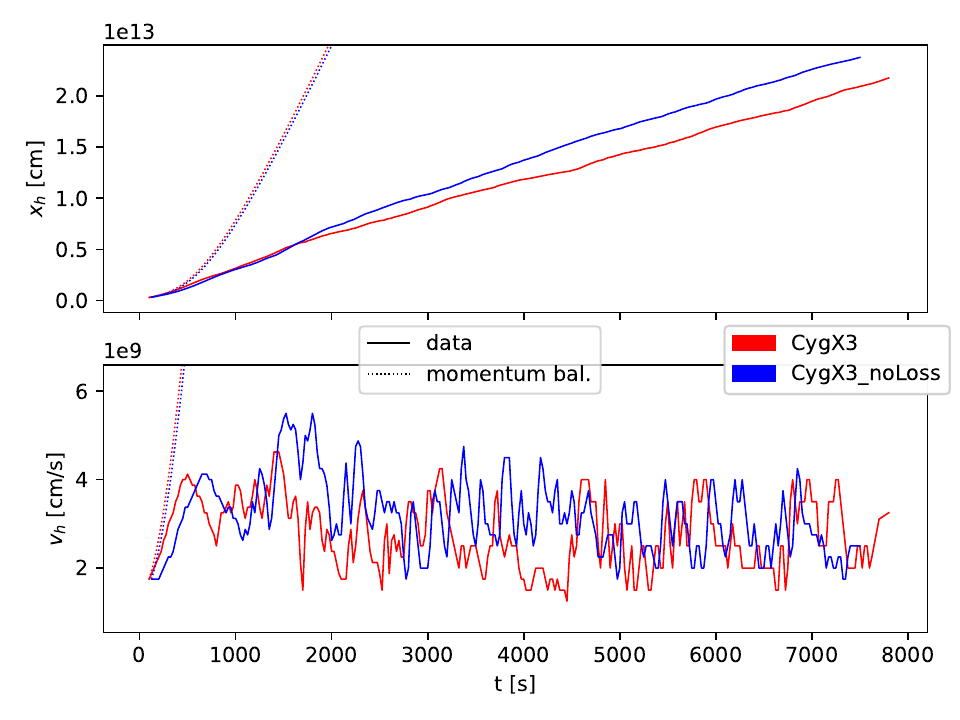}
    \caption{Effects of losses on run CygX3 structure and dynamics. \textbf{Top:} Temperature slices of runs CygX3 and CygX3\_noLoss at time t = 9 000 s. Two main differences appear: 1) The cooled beam is thinner. Its envelope closely follows the internal shocks structure in contrast to the non-cooled case. 2) The cocoon of the non-cooled jet expands farther at its basis, almost wrapping around the star, whereas in the cooled case, the cocoon has almost disappeared because ambient material has cooled enough to be blown back by the wind. \textbf{Bottom:} Same as Fig. \ref{fig:CygX1_LvnoL_T}. The cooled jet (in red) is initially faster, but leaves the smooth phase earlier, after which point it is slower on average, as seen in the propagation plot.}
    \label{fig:CygX3_LvnoL_T}
\end{figure}

Cooling times in our two fiducial cases are such that over the time covered by our simulations, radiative losses have no significant impact on CygX1, as shown Fig.~\ref{fig:CygX1_LvnoL_T}. Even the jet head velocity evolves remarkably similarly with and without radiative losses during the first two phases of jet evolution (Fig.~\ref{fig:CygX1_LvnoL_T}, bottom). In the case of CygX3, by contrast, radiative losses lead to a loss of much of the outer cocoon at distances of a few $10^{12}$~cm (Figure~\ref{fig:CygX3_LvnoL_T}, top) and slow the jet head dwon (Figure~\ref{fig:CygX3_LvnoL_T}, bottom). Therefore, we restrict the following discussion mostly to CygX3.

\subsubsection{Beam destabilizing effect through cocoon pressure}\label{sect:loss_unstab}
\begin{figure}
\includegraphics[width=\hsize]{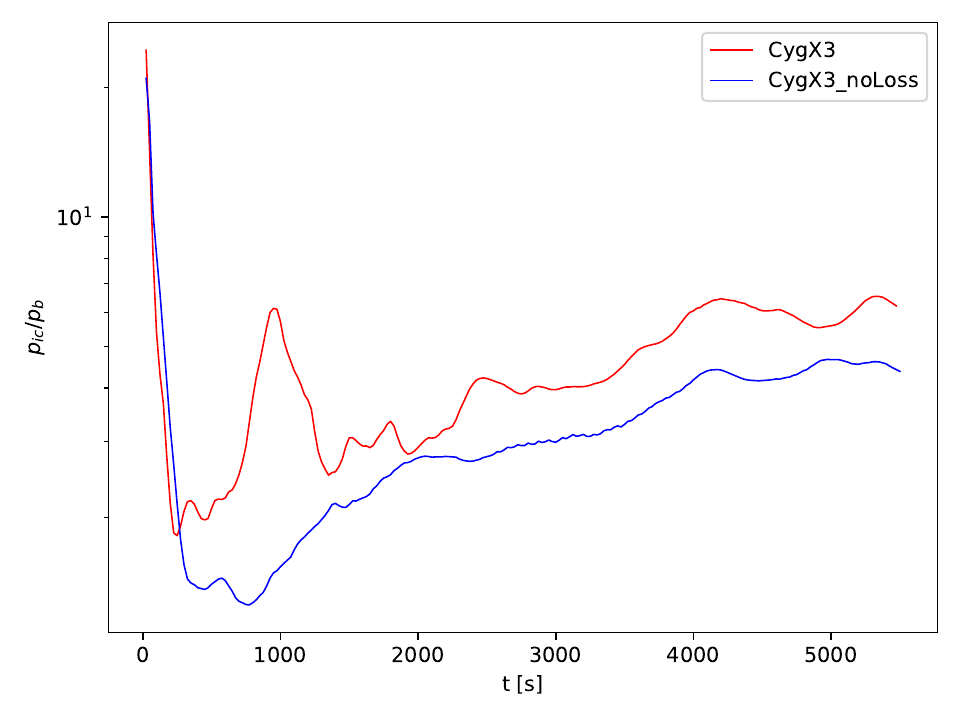}

\begin{center}
\includegraphics[width=\hsize]{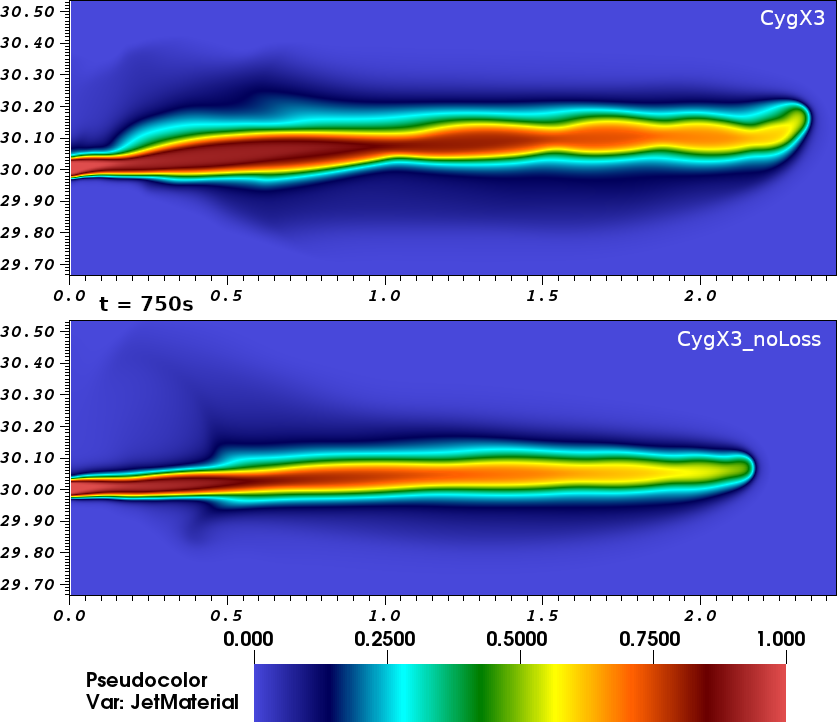}
\end{center}
\caption{Effect of radiative losses on the CygX3 beam structure. \textbf{Top:} Ratio of volume-averaged pressure of the inner cocoon and beam for runs CygX3 (red) and CygX3\_noLoss (blue). The overpressure is always higher in the cooled case. \textbf{Bottom:} Tracer density at t = 750 s for runs CygX3 and CygX3\_noLoss. The cooled jet features stronger oblique shocks.}
\label{fig:CygX3_LvnoL}
\end{figure}

The addition of the loss terms has a destabilizing effect on the beam through its interaction with the inner cocoon: free-free cooling, shown in Fig. \ref{fig:X1X3loss} to be the dominant process in both the beam and the inner cocoon, diminishes pressure in the jet with a different intensity depending on the jet zone: the beam cools faster than the inner cocoon, causing a stronger pressure gradient between inner cocoon and beam. This strengthens the oblique internal shocks (Figure~\ref{fig:CygX3_LvnoL}, left panels), which in turn accelerates the growth of KHI, as detailed Sect. \ref{sect:instab}. Thus KHI grows faster in the cooled case, changing the dynamical behavior of the cooled jet, as shown in Fig. \ref{fig:CygX3_LvnoL_T}. Pressure in each zone is derived from the mean rest mass density and temperature measured over the corresponding marked cells defined in Sect. \ref{sect:postproc}.

The cocoon-to-beam pressure ratio for runs CygX3 and CygX3\_noLoss is shown top panel of Fig. \ref{fig:CygX3_LvnoL}. It is to be noted that this ratio is greater than 1 at all time, ensuring a pressure collimation of the beam. In the non-cooled case, after the initial decrease that is caused by the settling in of the jet structure, the overpressure grows in two phases with a transition around $\sim$ 2000 s. In the cooled case, the growth rate seems constant from the start with the exception of a strong increase starting at t = 650 s, peaking at $\sim$ 1000 s, and joining the overall trend at 1350 s. This occurs as the jet transitions from the instability growth to the turbulent phase: t = 600 s indeed marks the apparition of strong oblique shocks in the cooled case. This stronger overpressure explains the difference in beam structure that is shown in the bottom panel of Fig. \ref{fig:CygX3_LvnoL}, displaying a longitudinal slice of the jet material tracer at t = 750 s including the plane containing the stellar center: a higher gradient of inner cocoon to beam pressure causes the stronger oblique shocks in CygX3 runs.

\subsubsection{Effects on the outer cocoon expansion}
\begin{figure}
    \centering
    \includegraphics[width=\hsize]{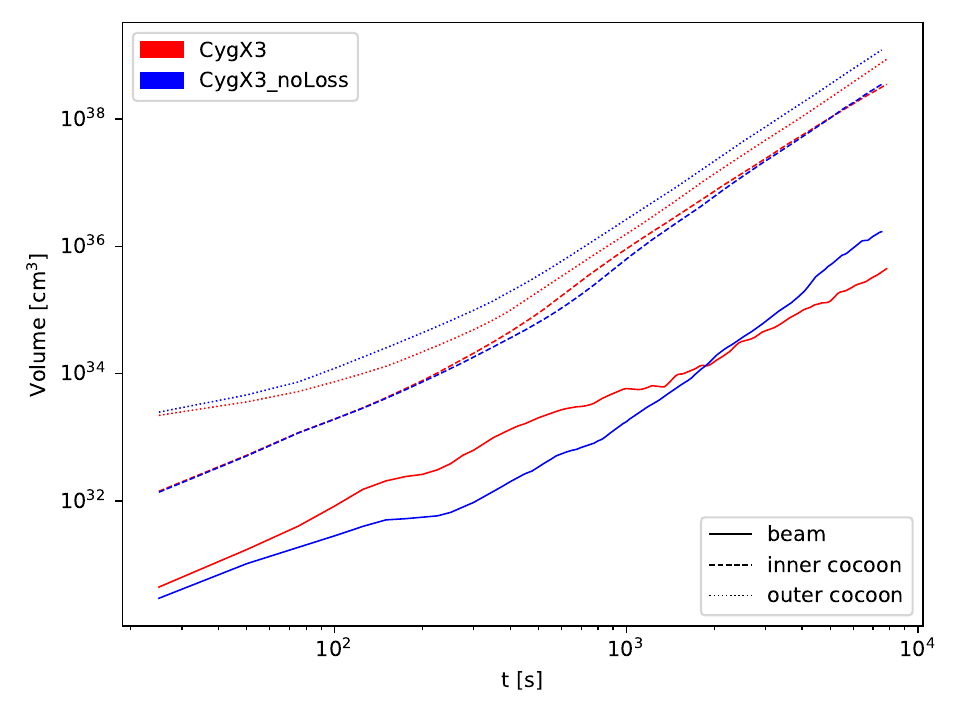}
    \caption{Volumes of the different jet zones as a function of time are affected differently by radiative cooling. The outer cocoon is larger without losses, while the volumes are comparable for the inner cocoon. The power-law dependence on time of the cocoon is robust for the cocoon, but not for the beam.}
    \label{fig:CygX3_LvnoL_logvol}
\end{figure}

Volumes of individual jet zones are affected by radiative cooling in different ways (Fig. \ref{fig:CygX3_LvnoL_logvol}). As explained in Sect. \ref{sect:coc_evol}, the dynamics of the wind-cocoon interface near injection zone is mostly controlled by the inner thermal pressure of the outer cocoon. Therefore, the more efficient the cooling, the smaller the outer cocoon and the faster the evolution of the cocoon up to the "naked beam" situation. In Cygnus X-3 case, the effect on the cocoon is shown in the volume diagram in Fig. \ref{fig:CygX3_LvnoL_logvol}: the outer cocoon very quickly evolves to be consistently of greater volume in the non-cooled case. In contrast, the volumes of the two inner cocoons are similar in the smooth and turbulent phases: in the first phase, the cooling effects on the inner cocoon are negligible because it is both the hottest and least dense part of the jet, while in the second phase, the instability-induced turbulences dominate the cocoon flow. The period during which the volume of the inner cocoon differs is likely due to the different starting time of the mixing phase between these runs, as explained in Sect. \ref{sect:loss_unstab}.

When the turbulent phase is reached, the inner and outer cocoon volumes show a similar power law dependence on time of roughly $t^{3}$, regardless of whether radiative losses are included. By contrast, the beam volume displays a different power-law dependence in the cooling and non-cooling case. The relative volume of outer to inner cocoon is much larger in the no-loss case than in the loss case. This may be an issue if radiative losses are diagnosed only during postprocessing from adiabatic solutions.

The cocoon form is also affected: The comparison of runs CygX3 and CygX3\_noLoss in the top panel of Fig. \ref{fig:CygX3_LvnoL_T} shows that the expansion is strong enough in the non-cooled case to cause the cocoon to almost wrap around the star before it is blown back as the cocoon pressure diminishes, while in the cooled case, the cocoon is almost immediately blown back to a thin shell by the strong stellar winds.

\subsection{Parameter sensitivities}\label{sect:paramsens}
We start with sensitivities to the assumed beam temperature (Sect.~\ref{sect:T}), which has been covered comparatively little in the literature and is thus somewhat more extensively dealt with here. Sensitivities to beam power and wind parameters follow (Sects.~\ref{sect:rho} and~\ref{sect:vw}).

\subsubsection{Effects of the jet temperature on instabilities growth}\label{sect:T}
An increase in the jet injection temperature $T_j$ lowers the beam Mach number. We expect the jet to display a smaller cocoon and to be less stable as the distance between internal shocks in the beam diminishes with it. This is true for the step from $10^8$ to $10^9$ K, but not for the step from $10^7$ to $10^8$ K. We ascribe this difference to the action of the first recollimation shock, which heats the beam of CygX1\_T7 to similar values as are found in CygX1. Increasing $T_j$ also results in a higher overpressure between the inner cocoon and the beam, which further strengthens the effects of oblique shocks we described above and leads to a faster destabilization of the beam, as explained in Sect. \ref{sect:growth}.

\begin{figure}
    \centering
    \includegraphics[width=\hsize]{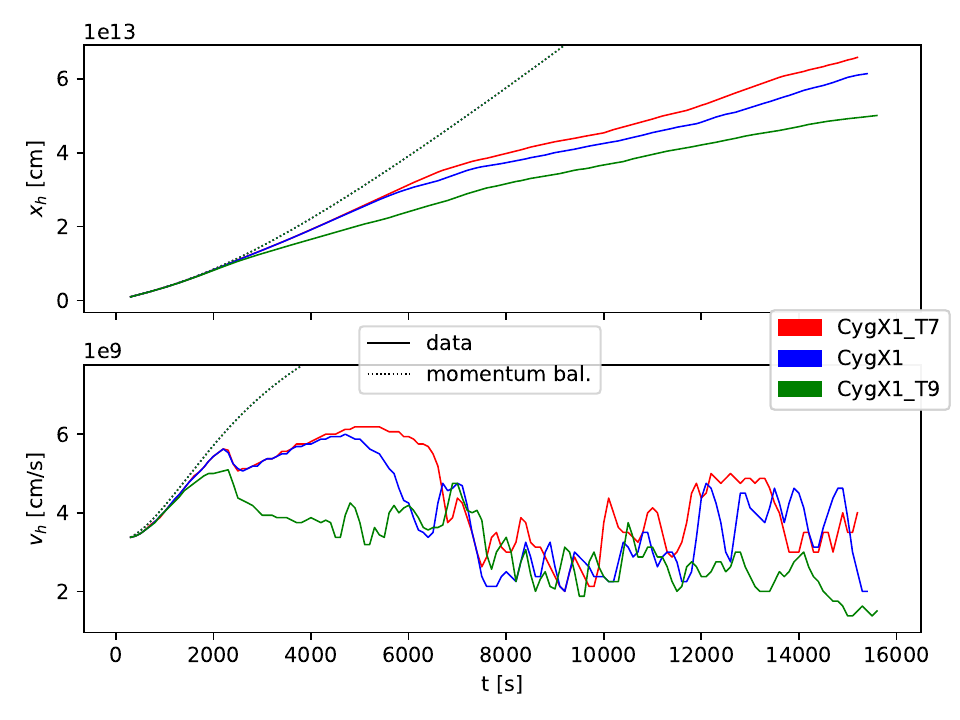}
    \caption{Jet head propagation and speed for runs CygX1\_T7 (red), CygX1 (blue) and CygX1\_T9. Run CygX1 slows down to the turbulent phase earlier than CygX1\_T7, but display the same average speed during the turbulent phase, as shown by the almost parallel propagation curves. Run CygX1\_T9 displays a different behaviour, decelerating to a plateau in the instability growth phase and with a lower average speed than the other two runs, which appears to decelerate after t = 12 000 s.}
    \label{fig:CygX1_T_xv}
\end{figure}

\begin{figure}
    \centering
    \includegraphics[width=\hsize]{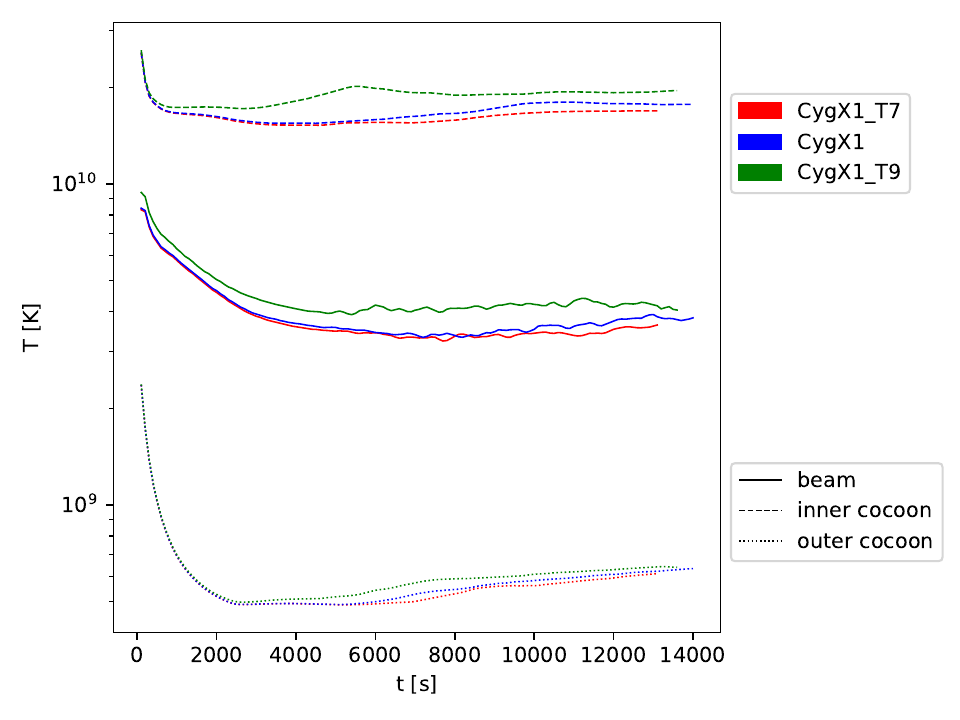}
    
\includegraphics[width=\hsize]{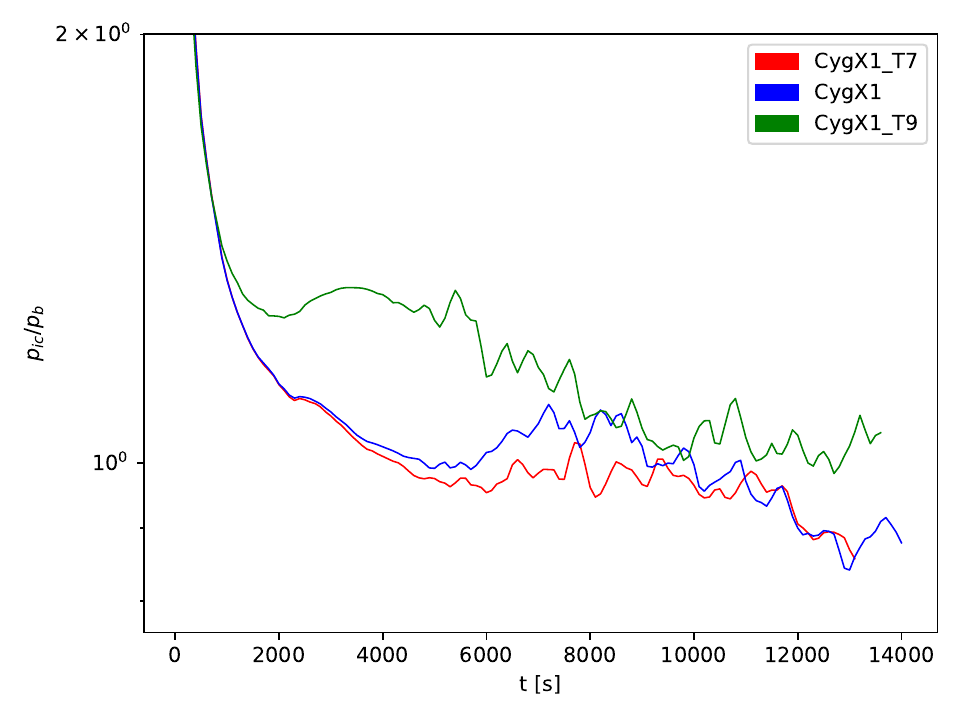}
    \caption{Comparison of temperature and overpressure of runs CygX1\_T7, CygX1 and CygX1\_T9, the color-coding is the same as in Fig. \ref{fig:CygX1_T_xv}. \textbf{Top:} Evolution of the zone-averaged temperature with simulation time. CygX1 and CygX1\_T7 differ only slightly at first, but then start to evolve differently after the 5000 s mark. CygX1\_T9 shows an overall higher temperature, with a small peak in inner cocoon temperature around t=6000 s. \textbf{Bottom:} Evolution of the ratio of the inner cocoon to beam pressure, as defined Sect. \ref{sect:loss_unstab}. CygX1 and CygX1\_T7 present similar values up to t $\sim$ 5 000 s, while CygX1\_T9 presents a stronger overpressure.}
    \label{fig:CygX1_T_Tndp}
\end{figure}

\begin{figure}
\includegraphics[width=\hsize]{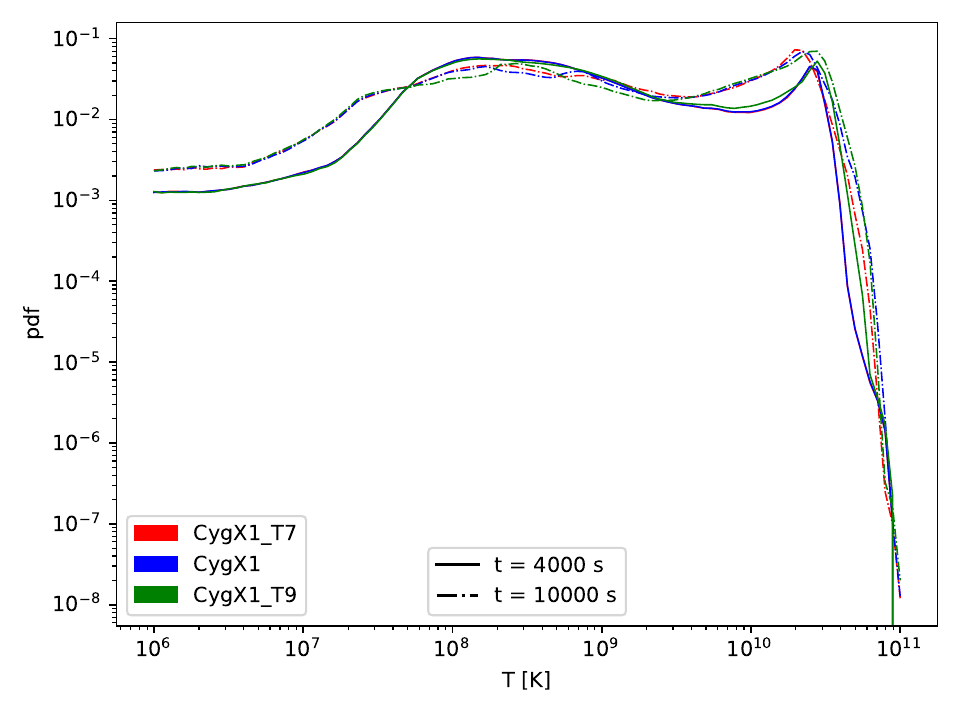}
\caption{Comparison between runs CygX1 (red), CygX1\_T7 (blue) and CygX1\_T9 (magenta). Probability density function of the temperature at t = 4000 s (full lines) and t = 10 000 s (dash-dotted lines). In the early stages, CygX1 and CygX1\_T7 are indistinguishable from each other, while CygX1\_T9 is slightly hotter and more of its volume is hotter than $\sim5\cdot10^9$ K. In the mixing phase, the temperature repartition is more in line with the injected temperature.}
\label{fig:CygX1_T_pdf}
\end{figure}

This destabilization is visible in the speed diagrams in Fig. \ref{fig:CygX1_T_xv}, showing colder jets to be more stable than hot jets: CygX1\_T7 shows similar dynamics as the fiducial run with a longer instability growth phase. In contrast, the run CygX1\_T9 displays different dynamics in this phase from the other two: the jet propagation speed slows down to a plateau instead of exhibiting a progressive acceleration. This difference in dynamical regime can be linked to the mean beam temperature in the top panel of Fig. \ref{fig:CygX1_T_Tndp}: beams associated with runs with $T_j=10^7$ and $10^8$K display almost the same temperature as a result of the heating at the initial recollimation shock, which raises them to a few $10^9$ K with very little differences. They begin to deviate from each other around the same time as the jet propagation speed does. With $T_j=10^9$ K, the temperature upstream of the shock is about the same as the downstream temperature, resulting in an higher effective beam temperature. This in turn means that the internal shocks that are closer to each other. This observation is confirmed by drawing the probability density function (PDF) of the temperature in Fig. \ref{fig:CygX1_T_pdf}. At t = 4000 s (full lines, during the instability growth phase), the PDFs for CygX1 and CygX1\_T7 are identical, while CygX1\_T9 differs for temperatures higher than $\sim4\cdot10^9$ K. At t = 10 000 s, the PDFs of the three runs differ by roughly the same amount, especially at the peak around $2\cdot10^{10}$ K. This can be interpreted as showing that the turbulence in the cocoon distributes the available thermal energy and therefore makes the difference in injected temperature visible.

The bottom panel of Fig. \ref{fig:CygX1_T_Tndp} shows the pressure ratio between inner cocoon and beam for the same three runs, where the pressure is derived from the mean temperature and from the rest mass density obtained by averaging over the marked cells. The pressure ratio in run CygX1\_T9 displays a different behavior from the other two runs, showing higher values as soon as t = 1000 s. This results in stronger internal shocks in the beam. The values for runs CygX1\_T8 and T7 are similar up to time t $\sim$ 6000 s, however, meaning that the internal structure of these two jets are similar during this period. These two effects both accelerate the growth of KHI modes, which is confirmed by the derived values of $t_{KHI}$ of 207, 71.4 and 26.8 s found in Table \ref{tab:tKHI} for runs CygX1\_T7, CygX1, and CygX1\_T9, respectively.

For the turbulent phase, the effect of the different beam temperatures is weaker. The velocity of the jet head is comparable to within its fluctuation range, except possibly for the very late time that is still covered by our simulations when the jet head velocity for the hottest jet seems to slow down slightly as compared to the two simulations with cooler jets. This would suggest that as soon as a more generic turbulent behavior takes over the dynamics, the beam injection temperature becomes less important.

\subsubsection{Effects of the injected power}\label{sect:rho}
A decrease in the jet kinetic power decreases its propagation efficiency as well as its stability. Starting from our fiducial test cases CygX1 and CygX3, we reduced the jet power by a factor of 10 (CygX1\_mP) and 2 (CygX3\_mP) via reducing the jet density at constant beam speed, as detailed in Tables \ref{tab:CygX1params} and \ref{tab:CygX3params} in the appendix. In these modified settings, the jets are expected to propagate more slowly and to be more prone to instabilities because the inertial mass density is lower as long as the jet is not disrupted by the stellar wind, as pointed out in \cite{perucho20103d}. This occurs for run CygX1\_mP, as shown in Fig. \ref{fig:CygX1mP_vx}. At constant beam speed $v_b$, the amplitude of the speed variations along the trend defined in the beginning of this section and the timescales at which they occur are controlled by the injected kinetic power, but the trend itself is not affected.

\begin{figure}
    \centering
    \includegraphics[width=\hsize]{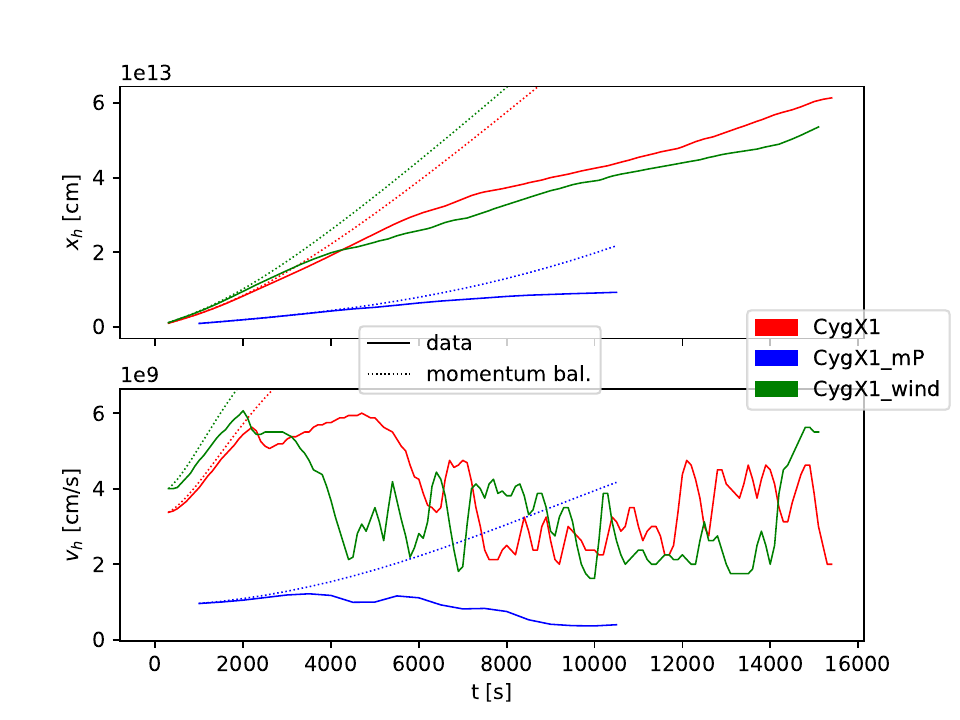}
    \caption{Sensitivity of the jet head position and speed to kinetic power and wind ram pressure for Cygnus X-1. The jet kinetic power is divided by 10 from CygX1 (red) to CygX1\_mP (blue). In the weak case, the jet is slower, but the position fits the theoretical evaluation for a longer time. The speed diagram shows no oscillations. The wind speed is increased by 50\% for run CygX1\_wind (green), which results in a higher starting speed, a shorter reacceleration phase, and a weaker bump in the second deceleration phase. The average speed in the turbulent phase is weaker. The speed and position plots differ faster from the theoretical 1D values in the strong wind case.}
    \label{fig:CygX1_params}
\end{figure}

\begin{figure}
    \centering
    \includegraphics[width=\hsize]{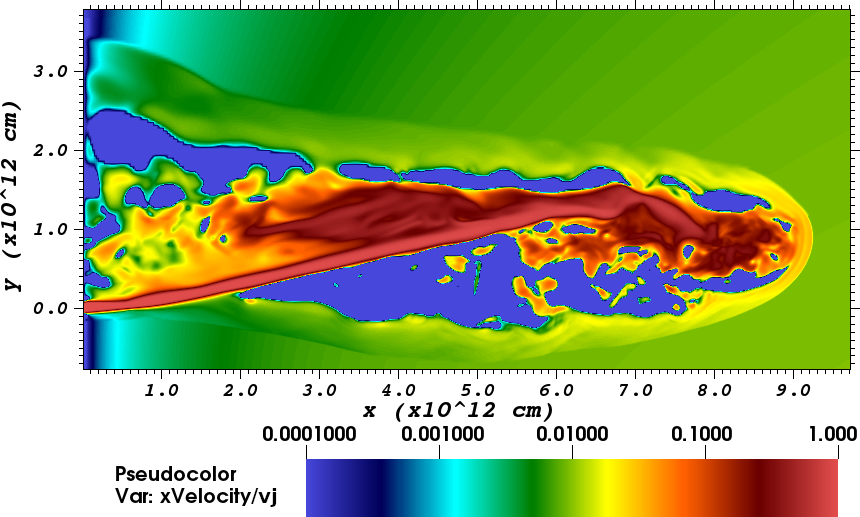}
    \caption{$v_x/v_j$ slice for run CygX1\_mP at time t = 10 000 s. The jet is heavily bent from the wind effects, and its velocity breaks down before it arrives at the head.}
    \label{fig:CygX1mP_vx}
\end{figure}

\begin{figure*}
\includegraphics[width=.5\textwidth]{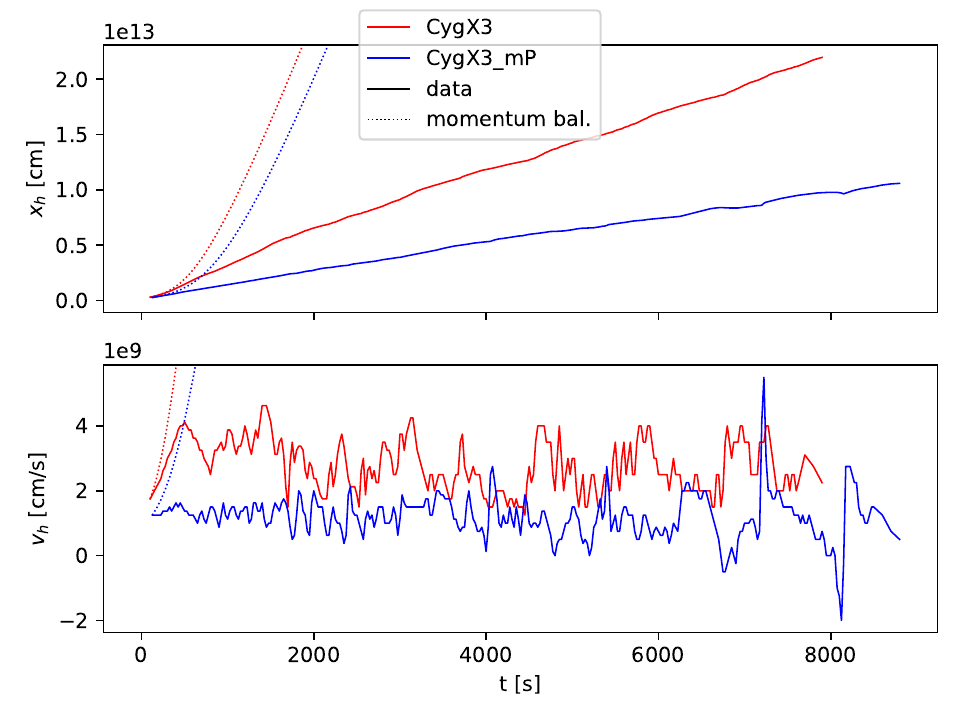}\includegraphics[width=.5\textwidth]{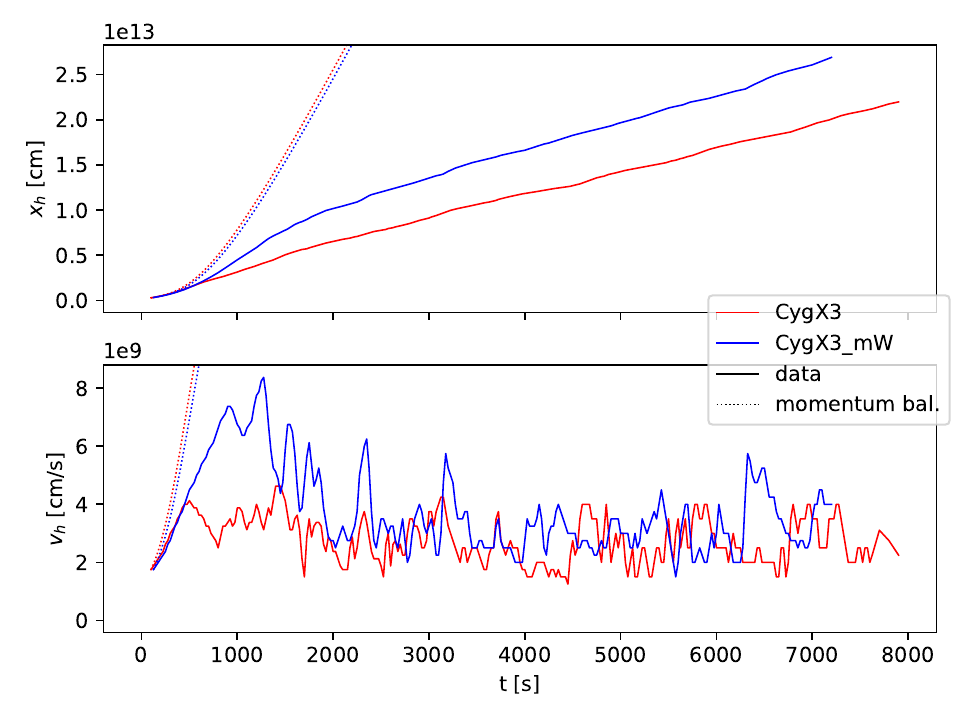}

\flushright
\includegraphics[width=.5\textwidth]{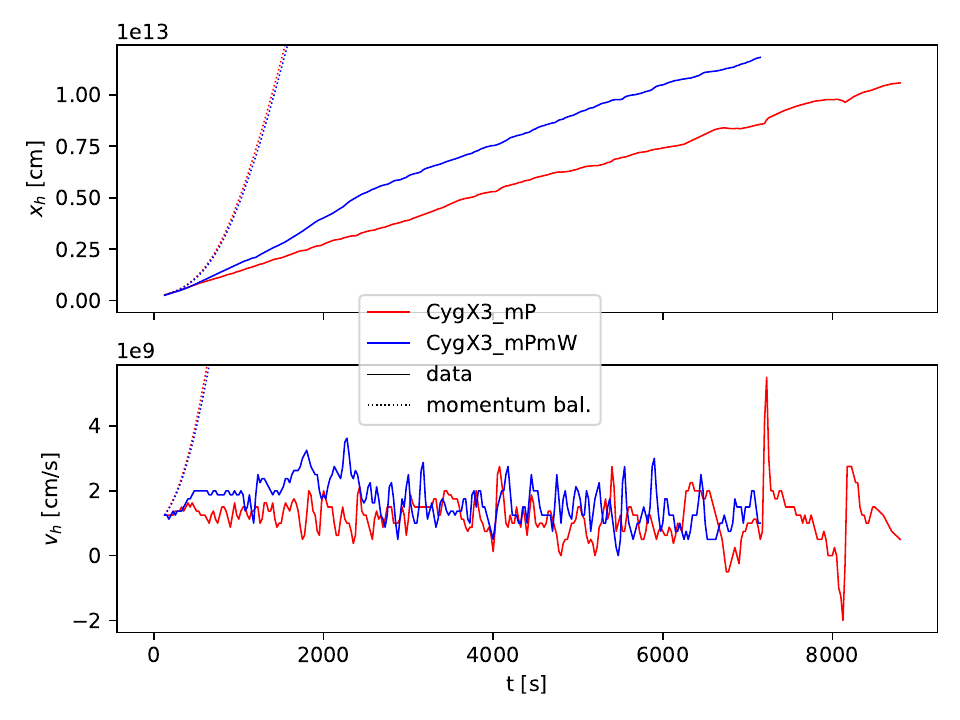}
\caption{Sensitivity of the jet head position and speed to kinetic power (left) and wind momentum (right) for the Cygnus X-3 runs. \textbf{Top left:} A division of the jet power by 2 decreases all dynamical values, starting from the mean propagation speed. In particular, the acceleration is far less efficient: it shows a gain of $\sim40$\% in speed between the starting point and the peak against a gain of $\sim120$\% in the CygX3 run. \textbf{Top right:} A reduction of $\Dot{M}^\star$ by 25\% and $v_w$ by 33\% (keeping $\eta$ constant) lengthens the smooth phase and strengthens the speed fluctuations in the following phases. The jets settle to the same median speed in the turbulent phase. \textbf{Bottom right:} Same reduction as in the top right panel, with half the jet power. In the CygX3\_mPmW case (right, blue), the initial acceleration phase is followed by a speed plateau.}
\label{fig:CygX3_params}
\end{figure*}

Figure \ref{fig:CygX1_params} shows propagation plots for runs CygX1 (red) and CygX1\_mP (blue), with the same parameters except for $\rho_b$ , which is ten times smaller in the second case. The weak jet displays a different behavior as the beam is strongly bent away by the wind and is then broken down by instabilities after t = 8000 s, as shown in red in Fig. \ref{fig:CygX1mP_vx}, drawn at t = 10 000 s. The beam mixes with the cocoon farther away from the contact discontinuity, and the momentum flowing from the reverse shock at beam head is partly dissipated in the cocoon. This smoothes the effects of beam head oscillations on the jet head dynamics and also slows the jet propagation down.

The propagation of runs CygX3 and CygX3\_mP is shown in the left panel of Fig. \ref{fig:CygX3_params}. $\rho_b$ is here divided by 2 between the two runs. In this case, the remark about the shape of the speed plot holds also: weaker jets show a similar evolution with smaller amplitudes in the global variations of the jet velocity. In contrast to the Cygnus X-1 case, however, a decrease in density destabilizes the jet: the first speed peak that occurs when the jet propagation breaks from the momentum balance model occurs earlier in the weak case, and the speed fluctuations begin earlier.

\subsubsection{Wind effects on the jet propagation}\label{sect:vw}
An increase in the wind ram pressure shortens the instability growth phase. This shows that this deceleration and the wind impact on the beam are linked. In the Cygnus X-1 runs, the impact of a stellar wind speed that is 50\% higher and therefore causes an increase of 50\% in the wind ram pressure at a constant mass-loss rate is shown in Fig. \ref{fig:CygX1_params} by comparing the CygX1 (red) and CygX1\_wind (green) runs: the stronger the wind speed, the shorter the instability growth phase. The beginning of this phase also starts slightly earlier: at 1800 s for CygX1\_wind versus 2200 s in the fiducial case. The ambient density, moreover, drops slightly from CygX1 to CygX1\_wind because $\Dot{M}^\star$ is constant between the two runs. This means a higher $\eta$ and therefore an easier jet propagation through the medium, and it explains the difference in starting propagation speed between the two runs. The theoretical 1D estimates for position and speed deviate from the measured values earlier in the strong wind case because the multidimensional effects are stronger. A higher wind speed induces a small plateau in jet speed at the very beginning of its propagation, which cannot be modeled by our 1D theoretical estimate.

For the Cygnus X-3 runs, the right panel of Fig. \ref{fig:CygX3_params} compares the fiducial run CygX3 (red) with run CygX3\_mW (blue), in which the wind is slower. All other parameters were kept constant, with the exception of mass-loss rate, to ensure same $\eta$ value between the runs and halving the wind ram pressure on the jet. The right panel shows the same modification using the weaker jet setup (runs CygX3\_mP and CygX3\_mPmW). In the first case, the initial accelerating phase is twice as long for run CygX3\_mW than run CygX3, and the second phase in run CygX3\_mW  consists only of a global deceleration because no reacceleration is observable. After this deceleration, run CygX3\_mW shows a slightly higher median propagation speed.

When we compare runs CygX3\_mP and CygX3\_mPmW in the bottom right panel of Fig. \ref{fig:CygX3_params}, a new effect arises: instead of a deceleration, the initial acceleration phase of run CygX3\_mPmW is followed by a plateau. The velocity fluctuations around the trend also appear later, but with a greater amplitude when the wind is weaker. After this speed plateau, the jet decelerates to a median value similar to the strong wind case, while the speed fluctuation timescale also decreases to a similar value as the strong wind case. This occurs after $t\sim3000$ s, when the jet has propagated to a distance of $\sim5\cdot10^{12}$ cm ($=20\, d_{orb}$). At this point, the wind is almost colinear with the jet propagation, and its lateral ram pressure on the jet is negligible.

\section{Discussion}\label{sect:discuss}
We have simulated jet outbreak and propagation in a large spatial and temporal frame based on the physical model developed in Sects. \ref{sect:srhd} and \ref{sect:radproc}. The simulation particularly included radiative cooling by different processes. We now critically discuss some underlying points of this model and the limitations of the results, and we describe the directions in which future work should be oriented. This point is of peculiar importance if we wish to compare simulation results to observations.

\subsection{Assumption of a hydrodynamical framework}\label{sect:disc_hydro}
High-energy nonthermal photons are a prominent feature of microquasar observations. They are present in the low and high state, but in different forms. They are thought to be produced by nonthermal processes that occur in the jet  \citep{2018A&A...618A.146M,2014SSRv..183...61P,2018MNRAS.480.2054M}, in which high-energy particles are accelerated to relativistic speeds and adopt a power-law spectrum. The mechanisms invoked to accelerate particles are stochastic acceleration (Fermi mechanism) at shocks, magnetic reconnection, or wave turbulence; see the recent reviews by \citet{2020LRCA....6....1M, 2020NewAR..8901543M}. All these acceleration mechanisms imperatively demand the plasma to be collisionless to a high degree. The Coulomb-logarithm of a typical jet beam in microquasars is approximately 15 and even higher in large regions of the cocoon. Consequently, kinetic timescales and kinetic inertial length scales are more than 10~orders of magnitude smaller than hydrodynamical scales. In short, the mean free path of thermal particles of a jet may easily be as long as the dynamical length scale of the jet and its cocoon. This poses the question to which degree jets can be understood on the basis of a hydrodynamical model.\\
Attempts to develop kinetical models of jets have been pursued; see~\citet{2021LRCA....7....1N} for a recent review. 
These models provide good insight into the various microphysical processes, and show how instabilities such as the Kelvin-Helmholtz, Rayleigh-Taylor, and kink instabilities translate into the collisionless regime. They also show that jet-like structures can develop on kinetic scales. However, it remains an open question whether the results found on kinetic scales can be scaled up to hydrodynamical scales and thus to scales on which the objects are seen in the skies. Some first steps toward the connection between kinetic and macroscopic scales have recently been made: \citet{2019A&A...621A.142D} showed that for even weak magnetic guide fields, kinetic jets can develop a structure resembling the hydrodynamical structure of a jet over scales much larger than the kinetic scales. An electromagnetic piston-like structure is coherently formed, acting like the contact discontinuity between jet and ambient material in a hydrodynamical jet \citep[see also][]{2017PhPl...24i4502D}. We note that in the same context, \citet{2019A&A...621A.142D} also showed that a leptonic jet propagating into an ambient ion-electron plasma can accelerate positrons to speeds of several hundred MeV. In this way, microquasars candidate sources can explain the positron population in the cosmic ray spectrum.

\subsection{Electron temperature and the single-fluid assumption}
In single-fluid simulations, the dynamics is dominated by the ions, while the electrons are responsible for most of the radiative losses. Studies such as \cite{vink2015electron, zhdankin2021production} have shown that processes such as shocks and radiative relativistic turbulences may create and/or maintain a difference between ion and electron temperature. This temperature difference has been observed in supernova remnants \citep{vink2003slow}. In particular, \cite{vink2015electron} suggested from thermodynamical consideration at shocks that the sonic Mach number of the upstream flow is the main parameter governing this temperature difference: at low Mach numbers ($M\leq2$), the shock will not induce a temperature difference, while at a high Mach number above $M\sim60,$ the ratio of the electron and ion temperature will be equal to the mass ratio $T_e/T_i = m_e/\langle m_i \rangle$. Between these extremal values, this ratio will roughly vary with $M^{-2}$.

In relativistic jets, at least two strong shocks present an upstream Mach number strong enough to induce a difference between ion and electron temperature downstream of it: the recollimation shock after injection, and the reverse shock at the beam end. To ensure that the calculated energy losses are valid, we need to determine the time taken by the electrons to thermalize to the flow temperature. Using formulas from \cite{trubiknov1965particle} found in \cite{huba2016nrl}, we obtained an estimate for the thermalization timescale considering the extremal case for a high Mach number. Table \ref{tab:thermal} presents the rest mass density and temperature values after the recollimation shock and reverse shock in CygX1 at t = 5000 s and CygX3 at t = 1000 s, the derived thermalization timescale, and the associated characteristic length using the local flow speed value. In all cases except downstream of the CygX3 reverse shock, the associated length scale is shorter than the grid resolution, meaning that the electrons would have thermalized to the flow temperature after a single simulation grid downstream of the shock. In the last case, we can estimate the volume whithin which an error is made using the flow temperature to derive losses as $V_{therm}=v_hc_s^2t_{therm}^3$, where $c_s$ is the local sound speed. With $v_h=3\cdot10^9$ cm s$^{-1}$ at t = 1000 s, we obtain $V_{therm}=3.3\cdot10^{33}$ cm$^3$, which accounts for 0.37\% of the inner cocoon volume at that time. We then consider our $T_e=T_i=T$ approximation to be valid in our hydrodynamical framework. In a more realistic framework in which the plasma is collisionless (see the discussion in Sect. \ref{sect:disc_hydro}), Coulomb interactions cannot equilibrate the electron and ion temperatures, but other processes such as plasma instabilities may fill that role.

\begin{table}[ht]
    \centering
    \begin{tabular}{|c||C|C|C|C|}
        \hline
        shock & \rho\text{ (g cm)}^{-3} & T \text{ (K)} & t_{therm} \text{ (s)} & l_{therm} \text{ (cm)} \\\hline
        X1 recoll. & 5\cdot10^{-15} & 5\cdot10^8 & 4.2\cdot10^{-4} & 4\cdot10^6\\
        X1 reverse & 10^{-15} & 3\cdot10^{10} & 0.6 & 2\cdot10^9\\
        X3 recoll. & 10^{-14} & 4\cdot10^{10} & 0.1 & 2.25\cdot10^9\\
        X3 reverse & 5\cdot10^{-15} & 10^{12} & 19 & 2\cdot10^{11}\\\hline
    \end{tabular}
    \caption{Thermalization of electrons downstream of a shock.}
    \label{tab:thermal}
\end{table}

\subsection{Thermal diffusion}
Another important process that in not included in our numerical model is thermal diffusion by relativistic thermal and nonthermal electrons or X-ray photons. This process is likely important as it creates hot shock-precursors, decreases peak temperatures, smoothes out contact interfaces, and enhances cooling. These features are suggested when looking at work which includes the process in the context of supernova remnants \citep{1975ApJ...198..355C, 2006MNRAS.371.1106T} and colliding winds in binary systems \citep{1998MNRAS.300..686M, 1999IAUS..193..378M}. Including heat transfer in a simulation is numerically demanding as it demands (semi-) implicite solvers due to the very stiffness of the process. Nevertheless, a few attempts have been made to develop performing solvers \citep{2008MNRAS.386..627B, 2009MNRAS.394.1727P, 2011A&A...531A..86V, 2012JCoPh.231.3561K,  2014A&A...563A..11C, 2013MNRAS.428...71B, 2016A&A...585A.138D, 2016A&A...586A.153V}. Future work may expand these attempts to jet simulations.

\subsection{Wind structure and jet bending}
In high-mass microquasars, jets are launched into winds originating from orbiting and rotating stars. This causes a circumbinary environment structured in Archimedean spirals in colliding-wind binaries \citep{1995IAUS..163..420W} and accreting binaries \citep{2008A&A...484L...9W, 2014ASPC..488..141W}. These spirals, also called corotating interaction regions (CIRs), are bound by shocks that confine high-density, high-temperature regions. These structures will likely have an impact on the jet propagation and its stability, but will also likely lead to flashing outbursts in emission.

The bending of the jet away from the mass-shedding donor star, investigated, for example, by \cite{perucho2008interaction} and subsequent papers, \cite{yoon2015global}, \cite{bosch2016effects}, and again confirmed in this paper, will lead to helical jet trajectories. In the most extreme cases, an observer will see the jet under different angles during an orbit of the system~\citep[see, e.g.,][]{horton-et-al:20}. Future works will lead to a more quantitative statement of this prediction. As a side note, we showed in Sect. \ref{sect:app_relat} that jets performed with a relativistic scheme display a higher propagation speed, even in the mildly relativistic case. Because the jet bending depends on the transverse momentum accumulated by the jet during its propagation, we would expect a weaker bending in relativistic simulations than in Newtonian simulations that are performed with the same parameters.

Another unknown is wind clumping. We know that winds from hot massive stars are clumped \citep[e.g.,][]{oskinova2012clumped}, with density contrasts $\langle\rho^2\rangle/\langle\rho\rangle^2\approx2$. \cite{poutanen2008superorbital} suggested that this is likely the case for Cygnus X-1 based on dips in the X-ray light curve \citep{2000MNRAS.311..861B, refId0, Hirsch2019}. It is unknown, however, where these clumps are formed: in the stellar atmosphere, or farther away from the star, for instance, by supersonic turbulence in the wind. Whether these clumps are small and dense, an intermittent density fluctuation of compressible turbulence, or a mix of both is unknown as well; see the discussion in \citet{2003IAUS..212..139W}. \cite{perucho20123d} suggested that even moderate wind clumping has strong effects on jet disruption, mass loading, bending, and likely energy dissipation. \cite{Delacita2017} suggested that the standing shocks introduced in the jet flow by its interaction with a clumpy wind would generate a higher apparent gamma-ray luminosity through inverse Compton scattering of the stellar photons, as well as efficient synchrotron cooling. In light of these results, we may expect that introducing clumpiness in the wind might significantly enhance the radiative cooling of the jet and modify its dynamics even more strongly.

\subsection{Wind driving and dynamics}
Velocity and density profiles in winds shed from hot massive stars in binaries are prone to a large uncertainty. This uncertainty pushed us to choose a grid of fixed wind parameters for the study presented in this paper and to make no attempts to model the acceleration of the wind. Winds from massive stars are driven by the radiative pressure on free electrons (to about one-third of the driving force) as well as the scattering of stellar photons in millions of UV and optical lines. This wind driving is expected to be vastly different in supergiants (as in Cygnus X-1) and WR stars (as in Cygnus X-3), in that the winds from WR stars are optically thick in the subsonic acceleration phase, whereas they are relatively optically thin in supergiant winds.

These winds are well understood for single stars, with the exception of the subsonic phase of WR winds; see the review by \citet{2000ARA&A..38..613K}. In binaries, the situation is more complex. \citet{2020A&A...634A..49H} presented an observation-based study of winds in HMXRBs and basically confirmed that the wind driving from stars in those binaries is the same as for single stars of the same type.  However, the situation is more complicated by the secondary radiation source, either another star or a compact object and its environment. This second source leads to an ionization structure within the wind that in contrast to the single-star situation does not fit the frequencies of the photons emitted by the star that produces the wind.

This in particular affects the region between the wind-shedding star and its companion. There, the driving by photon scattering in spectral lines can be suppressed and the wind acceleration may be weaker or even inhibited compared to the winds of a single star \citep{1990ApJ...365..321S, 1991ApJ...379..310S, 1994ApJ...435..756B}. Models that take this inhibition into account can be fit to observations \citep{2008ApJ...678.1237G}. The case of Cygnus X-1 was investigated through complex radiation hydrodynamics simulations by \citet{2012A&A...542A..42H} and \citet{2015A&A...579A.111K}, who suggested that strong inhibition of the wind speed might occur at the BH location. In this region, wind density and velocity profiles in binaries are very difficult to access observationally and are largely unknown. For Cygnus X-3, the situation is even more unknown because the wind is still optically thick at the BH location. Studies such as those by \cite{vilhu2021wind} find a strong wind-suppressing effect in particular in the extreme-UV region. However, we can firmly state that a wind is present in both systems, based on the modulation of the X-ray light curve over the binary orbit, for instance. This modulation is due to the different optical paths that X-ray photons have to travel through the wind and thus to the different attenuation of the X-ray light through absorption in the wind \citep{1981A&A...101..299B, poutanen2008superorbital, Grinberg2015}. Finally, the effect of the BH gravitational field must be considered, which may focus the stellar wind and modify the density and velocity profile that are encountered by the propagating jet, but might also modify the evolution of the jet cocoon at its base.

Another important open problem is the influence of magnetic fields on the wind because strong fields could substantially influence the mass-shedding process, for example, by a strong enhancement of the mass loss in the equatorial plane \citep{2006ApJ...640L.191U, 2008MNRAS.385...97U, 2016MNRAS.462.3672B}. A general discussion of magnetic fields of massive stars and how they influence their environment can be found in~\citet{2012SSRv..166..145W}.

\subsection{Nonthermal components and radiative processes emission}
Our losses are modeled as optically thin at every frequencies. This may not be the case, especially in UV, optical, and radio ranges at which our thermal electrons radiates. Reheating due to photon absorption \citep{Belmont2008} or synchrotron self-absorption as well as effects such as synchrotron self-Compton were not investigated and may have a strong impact on the jet temperature and therefore structure and dynamics. This may not hold for Cygnus X-1, for which cooling and its effects are moderate but it might play a strong role in Cygnus X-3 case, especially because winds from WR stars are optically thick over higher frequency bands than winds from O-type stars. Moreover, the temperatures attained in our simulations exceed the electron pair-production threshold and reach a regime in which proton distribution becomes relativistic, which will impact the radiative losses.

Another point we did not consider is the particle acceleration and the nonthermal processes: particles can be accelerated at magnetic reconnections \citep{2010MNRAS.408L..46G, 2009MNRAS.395L..29G, 2012MNRAS.419..573M, 2014A&A...570A.112M} and shocks \citep{Araudo2009, Bordas2009, Delacita2017} or by stochastic interaction with magnetized turbulence.  Shearing flow acceleration as also been invoked as potential injection process of relativistic particles \citep{Rieger2019}. An extensive review of these processes is given in \citet{2020LRCA....6....1M}. As all processes will significantly contribute to the nonthermal emission, they can inject a substantial fraction of the kinetic jet energy into nonthermal components and probably change the dynamics because part of the gas energy will be dissipated by the accelerated particles. Alternatively, the pressure imparted to these particles can directly modify the flow dynamics.

\subsection{Jet outburst dynamics and emerging structures}
The dynamics of the second (instability) phase is not well understood, and neither is the physical phenomenon that sets the fluctuation speeds. The physics of instability growth is deep and rich, and we may have overlooked some effects, in particular, concerning the nonlinear phase. Another limitation of this study is our grid resolution, which may not accurately capture all the shocks that form in such a system or the growth of small-amplitude modes, for which the impact was highlighted here. Furthermoer, the adiabatic index $\Gamma$ was chosen to be constant and equal to 5/3, where a variable index such as the method suggested in \cite{mignone2005piecewise} and \cite{mignone2007equation} would more accurately reflect the relativistic gas EoS \citep{synge1957relativistic} and lead to more realistic jet structures.

We included thermal radiative losses in our study of jets structure and dynamics. These effects were not included in similar works. No focus was placed on jet bending, but it is observable in the jets of Cygnus X-1 and Cygnus X-3, in agreement with similar studies such as \cite{yoon2015global}. Moreover, the jets are injected with no inclination relative to the orbital plane, in contrast to works such as \cite{dubus2010relativistic}, who suggested that the inclination angle for Cygnus X-3 lies between 20 and 80$^\circ$. Another strong point of this study is the simulation box size of $10^{14}$ cm, which is 33 times the orbital separation of Cygnus X-1 and close to 400 times that of Cygnus X-3. This allowed us to observe the outbreak and early dynamics details as well as the emergence of a steady "naked beam" structure for later timesteps that may be compared to similar works and translated into synthetic observational datas.

Last, our study is purely hydrodynamical and does not cover the potential stabilizing effects of the magnetic field as well as the development of MHD instabilities. The study of MHD jets is recent \citep[e.g.][]{mizuno2015recollimation, marti2016internal, mukherjee2020simulating} and the field will continue to grow as time passes. The influence of the stellar magnetic field may also lead to further modifications on the structure and dynamics of the jet.

\section{Summary and conclusions}\label{sect:conclusion}

We performed large-scale 3D relativistic hydrodynamical simulations of jet outbreak and early propagation in HMMQs. We added radiative cooling effects in the energy equation using a Maxwell-Jüttner distribution for the electrons. Two fiducial cases inspired by the HMMQs Cygnus X-1 and Cygnus X-3 were considered, along with parameter sensitivity studies. The jets of the two systems are in the same domain of dimensionless parameters, but with different instability growth and cooling timescales (about $\sim10^2$ shorter in the latter case), enabling us to better highlight the importance of these processes in jet structure and dynamics. In particular, on the timescales covered by our simulations, radiative losses, mainly free-free mechanism, play a relevant role for Cygnus X-3, but not for Cygnus X-1.

We investigated the impact of radiative losses on jet structures and dynamics and showed their strengthening effect on KHI growth, which was a focal point of our analysis. We identified three main dynamical phases: 1) an initial self-similar propagation in line with 1D momentum balance arguments, followed by 2) a modification of the inner jet structure and a phase of instability growth, and finally, 3) a turbulent cocoon and destabilized beam. We explored the effects of losses on the dynamics in the second phase through their modification of the KHI growth rate as well as the impact on the jet structure, and we examined the impact of the various parameters.

In Cygnus X-3, where radiative losses are observed on the timescales covered by our simulations, these losses are found to affect the volume ratio of the outer to inner cocoon. This means that simple postprocessing of simulations that do not take radiative cooling into account to study emissions from HMMQ jets must be met with caution. Likewise, the beam volume is found to obey a different power law in time depending on whether or not cooling is active. 

We find the sensitivities to jet power and wind parameters to be in line with the literature. We add to this picture by demonstrating that an increasing beam temperature results in a faster instability growth and a slower jet head velocity. In view of the wide variety of known HMMQs, but also given the difficult estimation of their system parameters, further and more extended parameter studies are clearly desirable in the future. Based on the results presented here, we would advocate that such studies not be limited to adiabatic simulations.

\begin{acknowledgements}
We acknowledge support from the French National Program for High Energies (PNHE). Simulations have been performed at the Pôle Scientifique de Modélisation Numérique (PSMN) hosted by the ENS de Lyon, and at the centers of the Grand Equipement National de Calcul Intensif (GENCI) under grant number A0090406960. We thank the referee for the helpful review and insights, which helped improve the paper significantly.
\end{acknowledgements}

%
%
\bibliographystyle{aa}
\bibliography{aanda}

\begin{appendix}
\section{Propagation and instabilities}\label{sect:propinstab}
\subsection{Model for jet propagation}\label{sect:propmodel}
A model for the propagation of a relativistic jet was derived, for instance, by \cite{marti1997morphology} and \cite{mizuta2004propagation}, who neglected multidimensional effects, but assumed 1D momentum balance between the beam and the ambient gas in the rest frame of the contact discontinuity at the head of the jet. In the case of an ambient medium with its own (nonrelativistic) flow speed $v_w$ , this gives
\begin{equation}
    S_b\left[\rho_b h_b \gamma_h^2\gamma_b^2(v_b-v_h)^2 + p_b\right] = S_w\left[\rho_w h_w \gamma_h^2(v_w-v_h)^2 + p_w\right],\label{eq:mombal1}
\end{equation}
where $\gamma$ are Lorentz factors, $S$ are cross-section areas, and $v$ are velocities. Subscripts $b$ refer to beam values, $w$ refers to the ambient medium, and $h$ to the interface at jet head. Solving Eq. \ref{eq:mombal1} for $v_h$, we obtain
\begin{equation}
    v_h=\dfrac{1}{A\eta^*-1}\sqrt{A\eta^*(v_b-v_w)^2-(A\eta^*-1)\dfrac{(AK-1)c_{s,w}^2}{\gamma_h^2 \Gamma}}\label{eq:mombal2},
\end{equation}
where $\eta^*\equiv\rho_b h_b \gamma_b^2/\rho_w h_w \gamma_w^2 = \xi_b/\xi_w$, $A$ is the cross-section ratio ($A\equiv S_b/S_w$), and $K$ is the pressure ratio defined in Sect. \ref{sect:previous}. This equation is consistent with the one derived by \cite{mizuta2004propagation} when taking $v_w=0$. The term including $K$ can be neglected as the sound speed in the stellar wind is much slower than the beam velocity. Assuming $A=1$, Eq. \ref{eq:mombal2} therefore becomes
\begin{equation}
    v_h=\dfrac{\eta^*v_b-v_w-\sqrt{\eta^*}(v_b-v_w)}{\eta^*-1}\label{eq:mombal3}.
\end{equation}

\subsection{Instabilities in relativistic jet flows}\label{sect:instab}
During the jet propagation, various hydrodynamical instabilities can be triggered and in the end perturb the beam, reducing the effective beam speed at the front shock and decelerating the jet. One of them is the KHI at the relativistic flow interface, which was extensively studied: \cite{turland1976instabilities}, \cite{blandford1976kelvin}, \cite{ferrari1978relativistic}, \cite{hardee1979configuration}, \cite{bodo1994kelvin}, \cite{hanasz1998kelvin}, \cite{hardee1998time}, \cite{hardee2001relativistic}, \cite{perucho2004stabilityI}, \cite{perucho2004stabilityII}, \cite{perucho2005nonlinear}, \cite{mizuno2007three}, \cite{rossi2008formation}, \cite{perucho2010stability}. 

Then, the Rayleigh-Taylor instability (RTI) can be triggered when a lighter fluid supports a heavier one against gravity, or equivalently, if the lighter fluid accelerates the heavier one (see \cite{norman1982structure}, \cite{allen1984rayleigh}, \cite{duffell2011tess} for nonrelativistic flows and \cite{matsumoto2017linear} for relativistic flows). This impacts the jet structures stability when the jet expands radially due to the centrifugal force, as studied by \cite{meliani2007transverse}, \cite{meliani2009decelerating}, \cite{millas2017rotation}, for example, or oscillates radially because of a pressure gradient, see \cite{matsumoto2013two}, \cite{toma2017rayleigh}.

Other instabilities include the Richtmyer-Meshkov instability (RMI); see \cite{nishihara2010richtmyer} for a review and \cite{matsumoto2013two} for a numerical study. The centrifugal instability (CFI) was studied in \cite{gourgouliatos2018relativistic}. Finally, we cite vortex formation and shedding at the contact discontinuity of jet head, as in \cite{norman1982structure}, \cite{scheck2002does}, or \cite{mizuta2004propagation}, for instance, which increases the cross-section of the jet head. This therefore diminishes the jet propagation speed. The vortices may also perturb the beam flow when they encounter it while propagating in the inner cocoon. \cite{norman1982structure} noted the importance of the beam Mach number in the vortex-shedding phenomenon, observing that lower Mach numbers produce higher vortices. This work will focus specifically on the KHI as other instabilities develop at the beam radius scale, meaning that most of their modes will not appear due to grid resolution limitations. 

\subsubsection{Kelvin-Helmholtz instability}\label{sect:KHI}
To numerically derive the linear growth time of the Kelvin-Helmholtz instability $t_{KHI}$, we chose to use the approach introduced in \cite{hanasz1996kelvin}, describing the jet with a 2D slab geometry. This choice was made because the solutions for the slab and cylindrical geometries behave in a similar way in a wide range of physical parameters, with only slight numerical differences \citep{ferrari1982magnetohydrodynamic}. In this analogy, symmetric and antisymmetric modes in the slab correspond to pinching and helical modes in the cylinder. Only high-order fluting modes do not have counterparts in slab jets.

\cite{hanasz1996kelvin} considered a core-sheet jet made of three layers: a beam with a relativistic flow, a cocoon with a nonrelativistic flow speed, and an ambient medium at rest. The transition layers at all interfaces are described in the vortex-sheet approximation. In the following, quantities with subscript $b$ refer to the beam, subscript $c$ refers to the cocoon, and $w$ refers to the ambient medium. We introduce the quantities $\eta_c\equiv\rho_b/\rho_c$ and $\eta_w\equiv\rho_c/\rho_w$ as the beam-cocoon and cocoon-ambient medium density contrast, respectively, and $R\equiv r_{c}/r_b$ as the radius ratio of the sheet and the core.

The deformations of the internal interface generate a sound wave that travels in the cocoon layer with the same frequency and wavenumber. As the maxima of the growth rate coincide with acoustic waves that travel an integer $n$ times their wavelength on their path back and forth in the cocoon, we chose to focus on these resonances to derive an estimate for the KHI growth time.

For a sound wave traveling between the internal and external interfaces with a wavelength $\lambda_c=2\pi/\sqrt{k_x^2+k_z^2}$ and a propagation angle $\alpha$ defined by $\tan\alpha=k_x/k_z$, the distance traveled by the wave between the interfaces is $L=(R-1)/\cos\alpha$. From the full derivation of the dispersion relation, we have in the cocoon
\begin{align}
    k_z &= \left(\frac{\omega_0^2}{\eta_c\Gamma}-k_x^2\right)^{1/2},\\
    \omega_0 &= \omega - \sqrt{\eta_c\Gamma}M_ck_x.
\end{align}
Substituting these expressions in the resonance condition $2L=n\lambda_c$, we obtain the following expression:
\begin{equation}
\begin{split}
&\left[(R-1)-n\pi\left(\frac{\omega^2}{\eta_c\Gamma} + (M_c-1)k_x^2 + 2\omega k_x\frac{M_c}{\sqrt{\eta_c\Gamma}} \right)^{1/2} \right]\\ &\quad\quad\quad\quad\times\left(\frac{\omega}{\sqrt{\eta_c\Gamma}} - M_ck_x \right) = 0.
\end{split}
\end{equation}
This equation can then be solved using a Newton-Raphson method and leads to the calculation of the wavenumber $k_x$ maximizing $\omega$ with the densities and radii as parameters.  The linear growth time for the KHI is obtained from the corresponding $\omega$ by taking $t_{KHI}=\omega^{-1}$. To derive the growth time from our simulations, densities were measured as a volume-averaged value over each jet zone (more on this method in Sect. \ref{sect:postproc}), while the beam and inner cocoon radius were derived from the respective measured volume and length in the propagating direction of the zones by approximating them as coaxial cylinders. These timescales correspond to the "linear" growth times, whereas the observed growth rate in our simulations will be significantly higher due to nonlinear effects. They are still of interest when we compare them for different runs because the relation between different linear growth time is the same as for the observed runs.

\subsubsection{Rayleigh-Taylor instability}
Neglecting the effects of interface curvature and assuming the perturbation of physical variables have the Wentzel-Kramers-Brillouin spatial and temporal dependence in $\exp\left[i(ks-\omega t)\right],$ where $s$ is the local coordinate tangent to the interface and perpendicular to jet propagation, we can derive the linear growth time using the dispersion relation from \cite{matsumoto2017linear},
\begin{equation}
    \omega^2 = -g k \dfrac{\gamma_b^2\rho_bh_b - \rho_{ic}h_{ic}+\Gamma\gamma_b^2 p_0/c^2}{\gamma_b^2\rho_bh_b + \rho_{ic}h_{ic}}.\label{eq:RTI_disp}
\end{equation}
We take $\gamma_{ic} = 1$ (which is verified in all our runs), and $g$ is the acceleration of the surface assuming that the amplitude of the oscillation is roughly equal to the beam radius and $p_0$ is the pressure at the interface, we take $p_0 = (p_b+p_{ic})/2$ as a first approximation. We deduce from Eqs. \ref{eq:RTI_disp} that the RTI timescale $t_{RTI} = \text{Im }\omega^{-1}$ is proportional to the square root of the wavelength $\lambda = k^{-1}$, meaning that the temporal growth of the shorter-wavelength modes is faster. However, when the instability to wavelengths smaller than the jet radius and in a numerical grid is neglected, only the modes with wavelengths of several cell length will matter. This restricts the growth of the RTI to only a few possible wavelengths. This explains the large difference to the growth time of the KHI: we find $t_{KHI}$ to be mainly in the range $10^{1-2}$s, while $t_{RTI}$ is found to have values of a few $10^4$ to $10^5$s using this derivation.

\section{Radiative processes}\label{sect:radprocA}
A thermal plasma distribution of the electrons can be written in a general way using a Maxwell-J\"uttner distribution of the electrons in the plasma rest frame from \cite{juttner1911maxwellsche}, as in \cite{wardzinski2000thermal},
\begin{equation}
n_e(\gamma)= {\frac{n_e}{\Theta}} \frac{\gamma_e (\gamma_e^2 -1)^{1/2}}{K_2(\Theta^{-1})} \exp(-\gamma_e/\Theta) \label{eq:MaxJutt} ,
\end{equation}
where $K_2$ is the modified Bessel function of the second kind of order 2 and $\Theta= k_B T/m_e c^2$ is the normalized temperature. This formulation was adapted to our model because temperatures can exceed $10^{10}$ K behind shocks. The proton distribution can at first be assumed to be nonrelativistic at the temperatures reached by our simulations as $\Theta_p = k_BT/m_pc^2\ll1$ for $T<10^{12}$ K, where the Maxwell-J\"uttner and Boltzmann distributions coincide. An improved model should include a relativistic treatment of protons and $e^\pm$ pair production because some shock zones may approach temperatures at whic either protons start to become relativistic or where pair creation starts to be effective.

The following expressions were derived in the plasma rest frame, but the calculated volumic power loss is invariant with frame because the volume dilatation is compensated for by the time dilatation of the boost. The numerical approximation of the Bessel K function used in A-MaZe is detailed in Appendix \ref{sect:BessK} .

\subsection{Free-free radiation}
\begin{figure}
    \centering
    \includegraphics[width=\hsize]{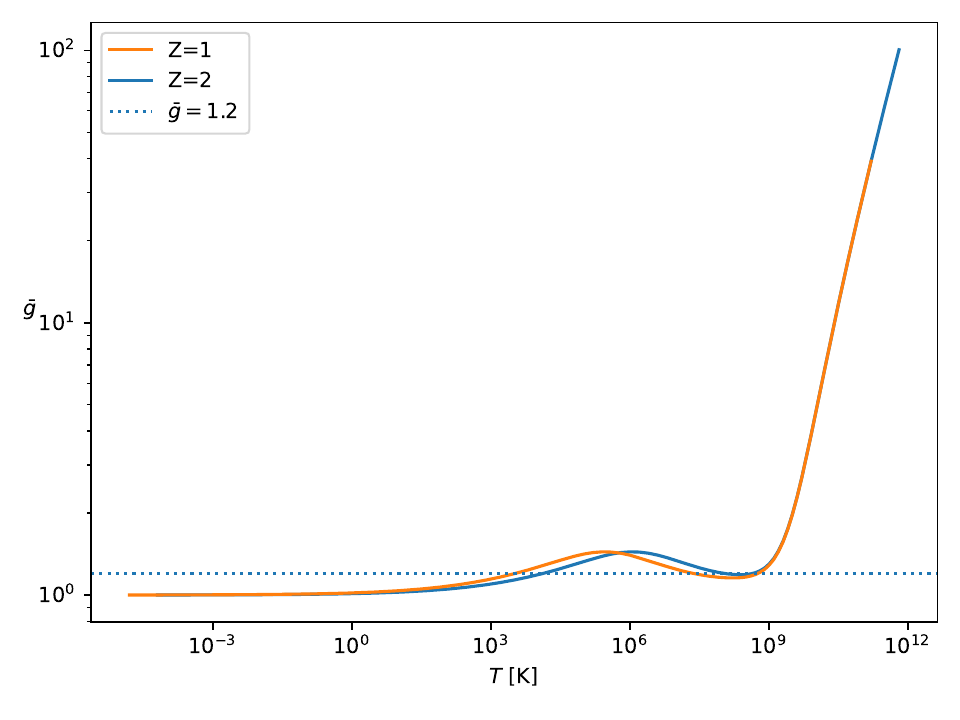}
    \caption{Gaunt factor for hydrogen and helium across the equivalent tabulated range of temperature covered in \cite{van2015accurate} and used in our Cygnus X-3 runs. Above $10^9$ K, $\bar{g}\propto T^{\sim3/4}$. The dotted line shows the mean value $\bar{g}=1.2$ used for the Cygnus X/-1 runs. }
    \label{fig:gaunts}
\end{figure}

Free-free radiation is emitted as an electron is accelerated by the Coulomb field of an ion. The expression of the volumic power loss due to the free-free mechanism of electrons in a hydrogen - helium plasma, including relativistic corrections, can be found in \cite{rybicki2008radiative}, 
\begin{equation}\label{eq:ff}
    P_{ff} = 1.4\cdot10^{-27}T^{1/2}n_e \left(n_H \Bar{g}_{H}+ Z_{He}^2 n_{He} \Bar{g}_{He}\right)(1+4.4\cdot10^{-10}T).
\end{equation}
This expression shows a dependence on $\rho^2$ and $T^{1/2}$ at classical electron temperatures, which becomes $T^{3/2}$ at high temperatures due to relativistic corrections. \cite{rybicki2008radiative} suggested $\Bar{g}_H = \Bar{g}_{He} = 1.2$ as a good numerical approximation for the frequency-averaged Gaunt factor in all temperature ranges, which is the value we used for the Cygnus X-1 runs. This approximation is good up to $T\sim10^9$ K, but a better approximation is needed because the temperature in the jet can become relativistic. A better evaluation of this term was given in \cite{van2015accurate}, who provided the values for $\Bar{g}_H$ and $\Bar{g}_{He}$ as functions of a parameter $\propto T^{-1}$. An illustration of the coefficients is given in Fig. \ref{fig:gaunts}. These functions were implemented in the Cygnus X-3 runs.

\subsection{Synchrotron}
\cite{ghisellini2013radiative} derived the synchrotron power emitted by a single electron with a Lorentz factor $\gamma_e=(1-\beta_e^2)^{-1/2}$ and pitch angle $\theta$ (the angle between its velocity and magnetic field lines) in the flow frame,
\begin{equation}
    P_e(\gamma_e,\theta) = 2\sigma_T c \gamma_e^2\beta_e^2 \sin\theta\ U_B \ ,
\end{equation}
with $\sigma_T$ the Thomson scattering cross-section and $U_B = B^2/8\pi$ the magnetic field energy density. This power can be averaged over the pitch angle $\theta$ because the electron distribution is assumed here to be isotropic,
\begin{equation}
    P_e(\gamma_e) = \frac{4}{3}\sigma_T c\gamma_e^2\beta_e^2 U_B.
\end{equation}
The volumic power loss is derived by integrating over the electron distribution,
\begin{align}
    P_{syn} &= \int P_e(\gamma)n_e(\gamma)d\gamma \nonumber\\
    &= \frac{4}{3}\sigma_T c U_B \frac{n_e}{\Theta K_2(\Theta^{-1})} \int_1^\infty \gamma(\gamma^2 -1)^{3/2}e^{-\gamma/\Theta}d\gamma \nonumber\\
    &= 4 \sigma_T c n_e \Theta \frac{K_3(\Theta^{-1})}{K_2(\Theta^{-1})} U_B,\label{eq:syn}
\end{align}
with $K_3$ the modified Bessel function of the second kind of order 3. In our simulations, the magnetic energy density is the sum of the contributions from the jet and the star, weighted by the tracer $J$.

These two contributions were modeled differently in the Cygnus X-1 and Cygnus X-3 runs: the magnetic field was chosen to be constant in Cygnus X-1 as a first attempt to model losses, at the risk of overestimating synchrotron losses in the long run. Seeing no impact of radiative losses until late simulation times in the Cygnus X-1 runs, we chose to launch runs based on Cygnus X-3 and updated our assumptions on the magnetic field. Because many authors such as \cite{perucho2019propagation} suggested that the jet magnetic field structure is presumably toroidal, we chose a linear decrease with distance for the jet inner field to reflect this assumption. The stellar magnetic field was assumed to be a dipole, decreasing as $r^{-3}$, with $r$ the distance to the stellar center. This assumption does not take effects such as increased magnetic field downstream of shocks in consideration and may cause us to underestimate the synchrotron losses in the beam and inner cocoon. A better treatment of synchrotron losses would require magnetohydrodynamical simulations, which are beyond the scope of this study.

\subsection{Inverse Compton scattering}
Following \cite{ghisellini2013radiative}, the radiated power per electron in the flow frame is
\begin{align}
    P_{ph}(\gamma_e,\psi) &= \sigma_T c \gamma_e^2(1-\beta_e\cos\psi)^2 U_{ph} \ ,
\end{align}
with $\psi$ the incident photon angle and $U_{ph}$ the radiative energy density in the flow frame. In this frame, the electron distribution is assumed to be isotropic, therefore $(1-\beta_e\cos\psi)^2$ can be averaged over the solid angle, from which we obtain $1+\beta_e^2/3$. The power loss of a single electron is then
\begin{align}
    P_e(\gamma_e) &= \langle P_{ph}(\gamma_e,\psi) \rangle - \sigma_T c U_{ph} \nonumber\\
    &= \frac{4}{3}\sigma_T c \gamma_e^2\beta_e^2 U_{ph}.
\end{align}
The volumic power loss in the fluid frame is obtained by integrating over the electron distribution, which yields
\begin{equation}
    P_{IC} = 4 \sigma_T c n_e \Theta \frac{K_3(\Theta^{-1})}{K_2(\Theta^{-1})} U_{ph}.\label{eq:ic}
\end{equation}
Considering the star as the sole source of seed photons for inverse Compton scattering, we derive the radiative energy density in the rest frame of the flow moving with a speed $v_j = \beta_j c$. When we define $\theta$ as the angle between the photon direction and the flow direction in the star rest frame, the radiative energy density is
\begin{equation}
     U_{ph} = \gamma_j (1-\beta_j\cos\theta)\frac{\sigma {T^\star}^4}{\pi}\left(\frac{R^\star}{r}\right)^2,
\end{equation}
where $r$ is the distance to the stellar center in the stellar rest frame. Synchrotron and inverse Compton cooling follow the same law, and their ratio is equal to the ratio of the magnetic and stellar photon energy density, $U_B$ and $U_{ph}$, respectively.

\subsection{Line and recombination cooling}
This term accounts for the collisional excitation of resonance lines and dielectronic recombination, where an ion captures an electron into a high-energy level and then decays to the ground state. We assumed solar photospheric abundances and the thermodynamical equilibrium of the plasma (Saha equilibrium), although in reality, the recombination may be delayed and may not correspond to the actual temperature of the plasma. This term follows the law
\begin{equation}
    P_{line} = \sum_i n_e n_{ion, i} 10^{\Lambda_i(T)},
\end{equation}
with i the different ion species. To facilitate the calculations, the various ions were taken into account in a single parameter $\Lambda(T)$ from \cite{cook1989effect}, such that we can take $P_{line} = n_e^2 10^{\Lambda(T)}$. This parameterization was then extended in temperature range and implemented numerically in \cite{walder1996radiative} and subsequent works. We chose an upper temperature of $10^{7.7}$ K for this process, which corresponds to the recombination of fully ionized iron and the Fe-$\alpha$ line. This very efficient process is only effective in the coolest and most external parts of the cocoon.

\subsection{Scalings of radiative processes}\label{sect:domRad}
The scalings of the various radiative losses with density, temperature, and distance to the star are listed in Table \ref{tab:powloss}. For free-free losses, the dependence of the Gaunt factors on temperature modifies the high-temperature scaling from $T^{3/2}$ in the Cygnus X-1 runs to $T^2$ in the Cygnus X-3 runs. For the synchrotron and inverse Compton losses, the term $K_3/K_2(\Theta^{-1})$ is constant at low temperatures and proportional to $T$ at relativistic temperatures, which explains the evolution of the scaling with temperature from a linear to a square power-law for these two processes. Last, the line and recombination losses have a non-power-law dependence on T. At low temperatures, which are found only in the outer cocoon, the main cooling process is line recombination. At higher temperatures, free-free losses take over. At the highest temperatures, the cooling is dominated by synchrotron or inverse Compton processes, depending on our choice for the magnetic field intensity.

\begin{table}
    \centering
    \begin{tabular}{|c||c|c|c||c|c|c|}
    \hline
    setup&\multicolumn{3}{c||}{Cygnus X-1}&\multicolumn{3}{c|}{Cygnus X-3}\\\hline
    variable & $\rho$ & $T$ & $r$ & $\rho$ & $T$ & $r$ \\\hline
    $P_{ff}$ & 2 & 1/2$\rightarrow$3/2 & 0 & 2 & 1/2$\rightarrow$2 & 0 \\
    $P_{syn}$ & 1 & 1$\rightarrow$2 & 0 & 1 & 1$\rightarrow$2 & -6 \\
    $P_{ic}$ & 1 & 1$\rightarrow$2 & -2 & 1 & 1$\rightarrow$2 & -2 \\
    $P_{line}$ & 2 & / & 0 & 2 & / & 0 \\\hline
    \end{tabular}
    \caption{Power-law exponent of the main variables in the loss term. The slash indicates a non-power-law scaling.} 
    \label{tab:powloss}
\end{table}

\subsection{Cooling time}\label{sect:tc}
The cooling time in the observer's frame of a fluid particle with rest frame temperature $T$ and Lorentz factor $\gamma$ is defined as $t_{cool} = \gamma T/\Dot{T}$, where the dot marks the derivation with respect to the proper time of the fluid. For a perfect gas, $T = p/\mathcal{R}\rho,$ with $\mathcal{R}$ the gas constant divided by the molar mass of the fluid. Therefore
\begin{equation}
    t_{cool} = \gamma\dfrac{p}{\Dot{p}+p\frac{\Dot{\rho}}{\rho}} = \gamma\left(\frac{\Dot{p}}{p}+\frac{\Dot{\rho}}{\rho}\right)^{-1},
\end{equation}
with all thermodynamic quantities measured in the comoving frame of the flow. Two extreme cases can be considered: the isobaric case ($\Dot{p}=0$), where $t_{c,p} = \gamma\rho/\Dot{\rho,}$ and the isochoric case ($\Dot{\rho}=0$), where $t_{c,\rho} = \gamma p/\Dot{p}$. From the definitions given in Sect. \ref{sect:srhd},
\begin{equation}
    \Dot{\tau} = 2\gamma(\rho c^2+\Gamma_1p)\Dot{\gamma}+(\gamma^2\Gamma_1-1)\Dot{p} + \gamma^2c^2\Dot{\rho},
\end{equation}
with $\Gamma_1=\Gamma/(\Gamma-1)$. Then, using $\dfrac{d\tau}{dt}=\gamma^{-1}\Dot{\tau}=P_{rad}$ and considering  $\Dot{\gamma}\ll\Dot{p},\Dot{\rho}$ as an approximation in the weakly relativistic case, we can approximate these timescales as
\begin{align}
    t_{c,p} &= \frac{\gamma^2\rho c^2}{P_{rad}},\\
    t_{c,\rho} &= \frac{(\gamma^2\Gamma_1-1)p}{P_{rad}}.
\end{align}
Assuming $\gamma^2 = 1$ to approximate these cooling times, $t_{c,p} \propto 10^{21}\rho/P_{rad}$ and $t_{c,\rho} \propto 1.5 p/P_{rad}$. In all the cases considered in the article, the isochoric cooling time is the shortest by about two orders of magnitude.

\section{Detailed numerical methods}\label{sect:det_num}
We detail here the numerical methods we used to perform the simulations described in this paper. We use the hydrodynamical framework of the A-MaZe simulation toolkit as described in \citet{2019A&A...630A.129P}, on a Cartesian static mesh and without the well-balanced option.

\subsection{Integration scheme}

Semidiscretization of Eq.~\ref{Eq:BalanceLaw} in space results in
%

\begin{eqnarray}
\label{Eq:MOL}
\frac{\partial{\bf U}_{i,j,k}}{\partial t} 
& + & \frac{{\bf F}_{i+1/2,j,k}-{\bf F}_{i-1/2,j,k}}{dx} + \nonumber \\
&   & \frac{{\bf G}_{i,j+1/2,k}-{\bf G}_{i,j-1/2,k}}{dy} + \nonumber \\
&   & \frac{{\bf H}_{i,j,k+1/2}-{\bf H}_{i,j,k-1/2}}{dz}={\bf \Psi}_{i,j,k}.
\end{eqnarray}

Here, $dx$, $dy$, and $dz$ represent the spatial discretization in the $x$-, $y$-, and $z$-direction, and ${\bf U}_{i,j,k}$ is the vector of the discrete conserved variables at cell centers $(i,j,k) \in (1,\ldots, N_{x}, 1,\ldots, N_{y}, 1,\ldots, N_{z})$, with $N_{x}, N_{y}$, and $N_{z}$ the number of cells in the $x-$, $y-$, and $z-$direction of the computational space. Half indices denote cell faces. ${\bf F}_{i\pm1/2,j,k}$, ${\bf G}_{i,j\pm1/2,k}$, and ${\bf H}_{i,j,k\pm1/2}$ denote the fluxes through the cell faces in the x-, y-, and z-direction. ${\bf \Psi}_{i,j,k}$ represents the source terms that are also evaluated at the cell centers.
  
We performed a time integration of the $N_x\times N_y\times N_z$-D system of ordinary differential equations, Eq.~\ref{Eq:MOL} with a first-order Runge-Kutta method (forward Euler method), although A-MaZe also offers strong stability-preserving (SSP) higher-order integration schemes~\citep{1988JCoPh..77..439S, 2001SIAMR..43...89G}. 
  
We used a simple central scheme to evaluate the fluxes, as detailed here for the the flux through the right x-interface of cell $(i,j,k)$,
\begin{eqnarray}
    {\bf F}_{i+1/2,j,k} = & &\frac{ F(U^{L}_{i+1/2,j,k}) +  F(U^{R}_{i+1/2,j,k})}{2} - \nonumber \\  & & - \frac{\lambda_{max}}{2}  \left(U^{R}_{i+1/2,j,k} -  U^{L}_{i+1/2,j,k} \right).
\end{eqnarray}
$U^{L}_{i+1/2,j,k}, U^{R}_{i+1/2,j,k}$ are the limited reconstructed variable values to the left and right of the cell interface ${i+1/2,j,k}$. We used linear reconstruction and minmod limiters. $\lambda_{max}$ is the highest characteristic speed. This integrator is relatively diffuse, but easy to implement for any hyperbolic system of equations and for an arbitrary EoS.

\subsection{Inversion scheme}\label{sect:app_invscheme}
To solve these equations for the conservative variables $(D,S^j,\tau)$, primitive variables $(\rho,v^j,p)$ are also necessary to compute the fluxes $\mathcal{F}^i$. The following system is obtained from the definition of the conservative variables:
\begin{align}
    \rho &= D/\gamma,\\
    v^j &= S^j/\xi,\\
    p &= \xi - \tau, \label{eq:p}
\end{align}
where $\xi=\gamma^2\rho h$ needs to be determined to derive the primitive variables. \cite{del2002efficient} suggested a method adapted to the polytropic EoS. With Eqn. \ref{eq:p} and $\Gamma_1 \equiv \Gamma/(\Gamma-1)$, we obtain
\begin{equation}
    \xi = \dfrac{\gamma^2\Gamma_1\tau-\gamma Dc^2}{\gamma^2\Gamma_1-1}.\label{eq:xi1}
\end{equation}
Using the definitions of the Lorentz factor and $\Vec{S}$,
\begin{equation}
    \xi^2 = \frac{S^2}{c^2(1-\gamma^{-2})}.\label{eq:xi2}
\end{equation}
The two expressions for $\xi$ are combined to obtain the final equation for $\gamma$,
\begin{equation}
\left(\dfrac{\gamma^2\Gamma_1\tau-\gamma Dc^2}{\gamma^2\Gamma_1-1}\right)^2 c^2 (1-\gamma^{-2}) - S^2 =0,\label{eq:inversion}
\end{equation}
which is solved numerically using the Brent method \citep{brent1973algorithm}. Primitive variables were then computed using the formulas
\begin{align*}
    \rho &= D/\gamma,\\
    \xi &= \dfrac{\gamma^2 \Gamma_1 \tau - \gamma Dc^2}{\gamma^2\Gamma_1 -1},\\
    v^j &= S^j/\xi,\\
    p &= \gamma^{-2}\Gamma_1^{-1}(\xi-\gamma Dc^2).
\end{align*}
This method is quite efficient, but is only valid for a constant $\Gamma$-law EoS. \cite{mignone2007equation} suggested a discussion of the validity of a constant $\Gamma$-law EoS and an inversion method suitable for all EoS, but we find that the method described above is suitable in our case.

\begin{figure*}[tbp]
    \centering
    \includegraphics[width=1\textwidth]{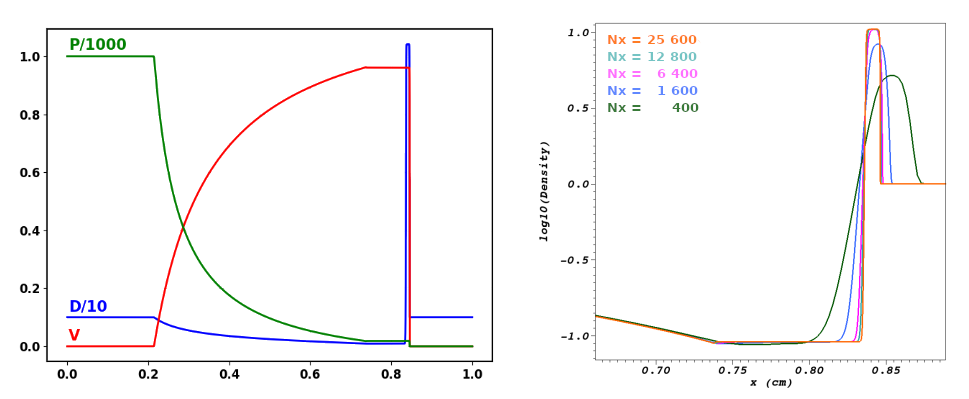}
    \caption{Solution of the relativistic blast problem with a Lorentz factor $\gamma=6$ shock detailed in the text at time $t=0.35$. \textbf{Left:} Pressure (P), density (D), and velocity (V) as computed on a very fine mesh (25600 cells) showing the shock, the contact discontinuity, and the rarefaction fan. \textbf{Right:} Zoom into the thin high-density layer between the shock and the contact discontinuity. This feature is the most difficult to resolve. We show the solutions based on different resolutions: 400 cells (black), 1600 cells (blue), 6400 cells (pink), 12800 cells (green), and 25600 cells (orange).}
    \label{fig:RP-Test-case}
\end{figure*}

\begin{figure}[tbp]
    \centering
    \includegraphics[width=0.48\textwidth]{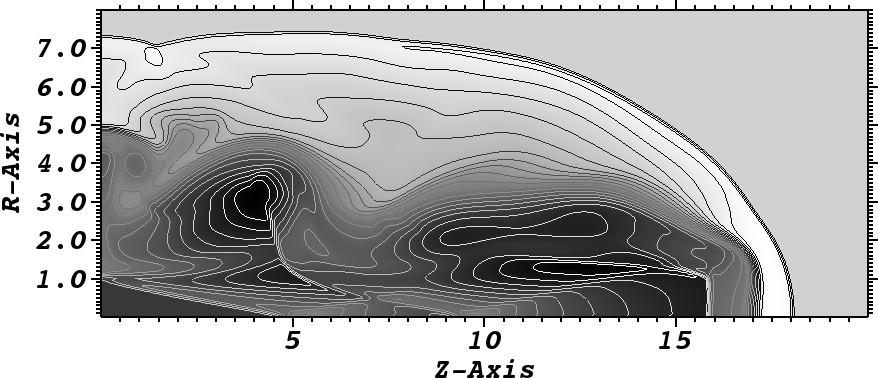}
    \caption{$\gamma = 7.1$ (Mach $\approx 17.9$) jet simulated in \citet{del2002efficient} and reproduced by the scheme used for this work. The image shows time=40 and can be directly compared with the last panel in Fig.9 of \citet{del2002efficient}. }
    \label{fig:Echo-Test-case}
\end{figure}

\subsection{Benchmark for the scheme}
Central schemes similar the one used in this paper have been widely used to perform (magneto-)hydrodynamical simulations \citep[e.g.,][]{del2002efficient, 2008CoPhC.179..617V, del2007echo}. These schemes are easy to implement and very robust, but are relatively diffuse (see, e.g., \citet{1996JCoPh.128...82T} for a discussion). As these schemes are not based on (even partial) characteristic decomposition, contact interfaces in particular are smeared out relatively strongly. This has consequences for the growth of instabilities along these interfaces.

Central schemes have become more popular again because more sophisticated Riemann solvers -- in particular exact solvers -- are very CPU costly. Moreover, they are also not really adapted to the situation when more complex physics is involved in addition to (M)HD. Flows that include radiation, gravity, and particles show a different wave pattern, and waves have different velocities than pure (M)HD waves.
  
There are two ways to overcome the large diffusivity of central schemes: 1) higher-order spatial reconstruction schemes as proposed in \citet{del2007echo}, for example, or 2) meshes with a finer spatial discretization. This second approach was chosen for this work, where we used concentrated fine meshes along the beam of the jet where the instabilities develop. Ideally, the two approaches and the adaptive mesh algorithm implemented in A-MaZe may be combined for a more relevant mesh refinement.

\subsubsection{Basic tests of the adiabatic scheme}

The central scheme has been used by the authors for other work \citep{2004A&A...414..559F}. We tested the implementation of SR by performing about 20 tests as proposed in the literature and found that we can reproduce these results well. Here, we only show two examples. The first is the relativistic blast wave problem as originally proposed by~\citet{1998JCoPh.146...58D}. This Riemann problem is defined by setting the state to the left or right of the original interface located at 0.5 to $(\rho, v, p)_L = (1, 0, 1000)$ and $(\rho, v, p)_R = (1, 0, 0.01)$, resulting in a $\gamma=6$ blast shock propagating to the right and a strong rarefaction fan propagating to the left. The solution at $t=0.35$ on a very fine mesh of 25600 cells is shown in the left panel of Fig.~\ref{fig:RP-Test-case}. The problem is tough and demonstrates why relativistic hydrodynamics is a numerical challenge.

\citet{1998JCoPh.146...58D} presented a solution based on a third-order scheme combined with the Marquina solver, which usses the full spectral decomposition. \citet{del2002efficient} presented two solutions of the same problem based on a mesh of 400 cells. The first solution was computed with a third-order scheme based on the Harten, Lax, van Leer (HLL) solver (which uses only a part of the spectral information) and a monotized central (MC) limiter (their third-order convex essentially non-oscillatory - or CENO3 - scheme). The second solution is based on the same method as used in this paper, the second-order Lax-Friedrichs-scheme and minmod limiters. The hard part to compute is the thin high-density shell between the shock wave and the contact interface. These shells are typical for relativistic flows. In Fig.~\ref{fig:RP-Test-case}, at $t=0.35$, it is located between $x=0.84$ and $x=0.85$, where the density jumps by about two orders of magnitude in the shock wave and by three orders of magnitude in the contact interface. Most of the mass is concentrated within a region covering only about 1\% of the domain.

Based on a discretization of 400 cells, none of the described schemes resolved the shell: The third-order schemes of~\citet{1998JCoPh.146...58D} and~\citet{del2002efficient} reached a density of about~$7.3$ in 1-2 cells, and the second-order LF-scheme in~\citet{del2002efficient} reached a density of about~$6.5$ over 1-2 cells. The correct value is about $10.5$, however. The convergence of our scheme to the correct density value is demonstrated in the right panel of Fig.~\ref{fig:RP-Test-case}. With 400 cells, our scheme is in line with that of~\citet{del2002efficient}. When 1600 cells are used, the density peaks at about $9.0$ in 1-2 cells and at almost the correct value when 6400 cells are used. With 12800 cells, the density peak is well resolved, but the contact interface is still somewhat smeared out. The simulation using 25600 cells fully resolves the thin shell with some tens of cells, and the transition to the contact is quite sharp. We note that the computational costs for our scheme for 1600 cells is probably not (much) more than using a third-order scheme and a Riemann solver based on spectral decomposition of 400 cells. This demonstrates that a strategy based on fine meshes and a simple solver can be efficient. Admittedly, data files are more heavy than those produced on a 400 cell mesh, however.

The second test is the jet test-case proposed in \citet{del2002efficient}\,: In cylindrical geometry, a $\gamma = 7.1$ jet is launched into a uniform environment with a low pressure, corresponding to a relativistic Mach number of about 17.9. This test is harder to simulate than the jets presented in this paper. Twenty cells covered the beam width, and the mesh in the domain was 160x400 in radial- and z-direction, respectively. Comparing the result obtained with our code (see Fig.~\ref{fig:Echo-Test-case}) with Fig.~9 of \citet{del2002efficient}, we observe an excellent agreement in the position of the front bow-shock, the position of the Mach stem at the end of the beam, the position of the cross-shocks in the beam, and the general shape of the cocoon. In our case, the interface between inner and outer cocoon is more smeared out. The smaller modes in the instability developing along this interface are less resolved than in \citet{del2002efficient}. This discrepancy is natural because \citet{del2002efficient} used the more accurate CENO3 scheme, while our result is based on the second order in the space LF method. However, this drawback can be overcome by using a finer mesh (not shown).

\subsubsection{Uncertainty for simulations of turbulent and cooling flows}
\begin{figure}[tbp]
    \centering
    \includegraphics[width=0.48\textwidth]{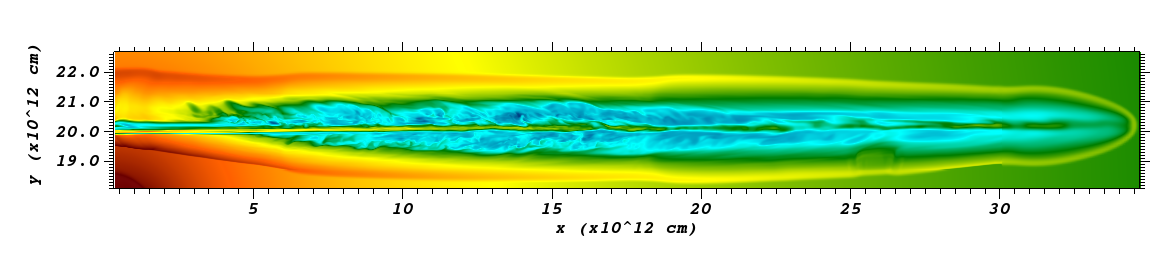}
    \caption{Simulation of the fiducial case of CygX1 on a mesh that is twice
      as fine as the simulation presented in
      Fig.~\ref{fig:CygX1_jetevol}. We show the simulation at
      $6000$~s, corresponding to the middle panel of
      Fig.~\ref{fig:CygX1_jetevol}.}
    \label{fig:CygX-1_Fine}
\end{figure}

The exactness of the simulation presented in this paper is harder to estimate. The flows are turbulent, and cooling introduces more instabilities, waves, and interfaces. The turbulent region of the cocoon has no fixed boundary, but is connected by shocks and material interfaces to the environment. Interior turbulent fluctuations will impact the shape of the interfaces and, inversely, the dynamics of the interfaces will act on the interior turbulence. Based on these arguments, we cannot expect to find a converged solution in the sense demonstrated in Fig.~\ref{fig:RP-Test-case}. We have to trust the general correctness of the scheme and have to give some reasons why the presented solutions are close to correct. A rigorous error analysis based on statistical analysis of many simulations that differ slightly in their initial conditions would be desirable, but is sophisticated, complex, and computationally expensive and thus beyond reach for this study. A step in this direction, nevertheless, is presented in Fig.~\ref{fig:CygX-1_Fine}. We list in the following some points that shed some light on the uncertainties.

\paragraph{\textbf{1. Turbulence:}}
Reynolds numbers are too high to resolve the turbulent cascade with any numerical scheme. Moreover, ideal hydrodynamics does not treat diffusion explicitly.  A numerical scheme implicitly introduces a certain diffusion \citep{hirsch2006numerical, leveque2002finite}, however, which is much higher in astrophysical rarefied flows than the physical diffusion. However, as pointed out by \citet{1992FlDyR..10..199B} and further explored by \citet{1992ThCFD...4...13P} and \citet{1994ApJS...93..309P}, finite-volume methods such as ours cut the turbulent cascade in a way that does not lead to an energy pile-up or -sink at the numerical diffusion scale, thus cutting the cascade correctly, at least to first order. This approach is called monotone integrated large-eddy simulation  (MILES). A more rigorous study of the MILES approach is given in~\citet{1999JCoPh.153..273G}, for example, and a summary of the idea and more references can be found in~\citet{2006A&A...459....1F}. These studies show that turbulent flows are relatively well captured by finite-volume methods and do not introduce large errors.

\paragraph{\textbf{2. Cooling:}}
Radiative shocks are prone to an overstability whenever the slope of the cooling law is sufficiently shallow or negative. For a radiative cooling parameterized as a function of density and temperature $\dot{\tau}(\rho, T) = \rho^2 \Lambda(T)$ with $\Lambda(T) = \Lambda_0 T^{\beta}$, which applies for free-free and line cooling, \citet{1982ApJ...261..543C} and \citet{1986ApJ...304..154B} have shown that the overstability is present whenever $\beta \lapprox 0.4$ (fundamental mode), and $\beta \lapprox 0.8$ (first-overtone mode). We have shown \citep{walder1996radiative} that the presence and amplitude of the overstable modes in a numerical study critically depend on the numerical resolution because smeared-out interfaces radiate more than better resolved interfaces. The resolution we chose for the simulations is sufficient to resolve the overstability (not shown). We add two remarks, however. First, the numerical model we used does not include mass diffusion and, in particular, heat diffusion, which physically determines the smearing of the interface. It is thus not clear whether we under- or over-estimate this particular effect. Second, radiative multidimensional shocks can generate and drive turbulence \citep{1998A&A...330L..21W} and turbulent thin shells~\citep{2006A&A...459....1F, 2014A&A...562A.112F}.

\paragraph{\textbf{3. Resolution comparison:}}
The fiducial case of CygX1 was simulated on a mesh twice finer than the generic mesh up to about 10'000~s. The snapshot at 6000~s is shown in Fig.~\ref{fig:CygX-1_Fine}. This can be compared to the snapshot of the generic case shown in the middle panel of Fig.~\ref{fig:CygX1_jetevol}. The comparison shows that the instability sets in at about the same time in both simulations. However, in the simulation on the finer mesh, the jets propagate about 10\%\ faster than in the simulation on the generic mesh. We also observe similar effects, of the same order, in 1D test simulations. This is expected on the basis of the arguments given in the point above. Better resolving the contact interface at the head of the jet will reduce cooling there, leaving slightly more energy to push the bow shock to a larger distance. An error of 10\%\ is quite a good result for most large-scale fluid-dynamical simulations. We stress again that the correct jet speed depends on the physical diffusion.

To even improve the confidence in the solutions presented in this paper, we ran a 2D resolution study, using the generic and the finer mesh in the 3D case. This study also covered the turbulent phase. Again we find that the essential features of the jet (number and location of the cross shocks, cocoon shape, and time at which the instability and the turbulence set in) are independent of the resolution. However, the 2D simulations cannot directly be compared to the 3D simulations because the character of the turbulence is different in 2D and 3D.
  
\subsection{Necessity of relativistic simulations}\label{sect:app_relat}
\begin{figure}[tbp]
\centering
\includegraphics[width=.5\textwidth]{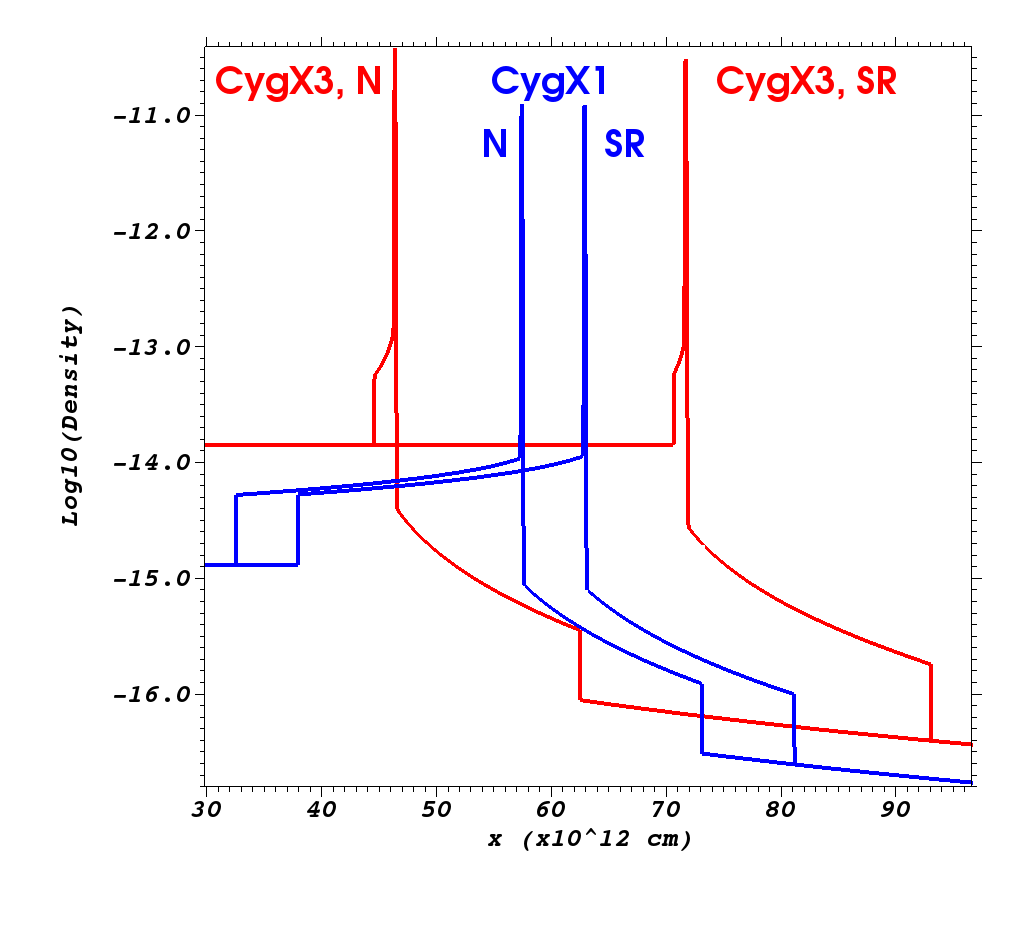}
\caption{Comparison in density profiles of 1D simulations for parameters similar to the fiducial cases. We show in the observer frame the comparison between a Newtonian and a relativistic simulation for CygX1 at 30 000~s (blue) and for CygX3 for 6 500~s (red). The simulations include all cooling terms. The pattern with forward and reverse shock and a thin layer of cooled gas to the left of the contact interface between beam and environment material is similar for all simulations. The mesh consists of 12800 cells, about as many as the mesh covering the beam in the 3D simulations presented in the paper.}
\label{fig:SRvN}
\end{figure}

Because computational costs are considerably higher for a relativistic simulation, it might be questioned whether it is necessary to perform relativistic simulations to obtain correct solutions for the mildly relativistic problems presented in this paper, with $\gamma_b \approx 1.06$ for CygX1 and $\gamma_b \approx 1.51$ for CygX3. However, even these small Lorentz factors lead to a significant difference in the jet propagation between a relativistic and a Newtonian simulation. This is illustrated in Fig.~\ref{fig:SRvN}, which shows 1D simulations at (observer) time $t=6500$~s of the jet propagation of CygX3, including all cooling terms. The jet head is located in the thin high-density shell. In the Newtonian case, this shell in the observers frame is located at about $x=45\cdot 10^{12}$~cm. The shell in the relativistic case is located at $x=71\cdot 10^{12}$~cm. Both simulations used 12800 cells. Thus, the relativistic jet head propagates about 1/3 faster than the Newtonian jet head. This can be explained on the basis of Eq.~\ref{Eq:Relate_Mu_MuStar}: the ratio of $\eta^*$ and $\eta$ for CygX3 (see Table~\ref{tab:CygX3params}) is about $2.3$, resulting in a difference in the jet-propagation speed of about 45\%. The difference for CygX1 is smaller, but still about 5\%. In 1D, the shocked beam will cool down, in contrast to the multidimensional simulations, in which the beam is regularly reheated by the cross shocks.

\subsection{Impact of the adiabatic index choice on jet propagation}\label{sect:app_Gamma}
\begin{figure}
    \centering
    \includegraphics[width=\hsize]{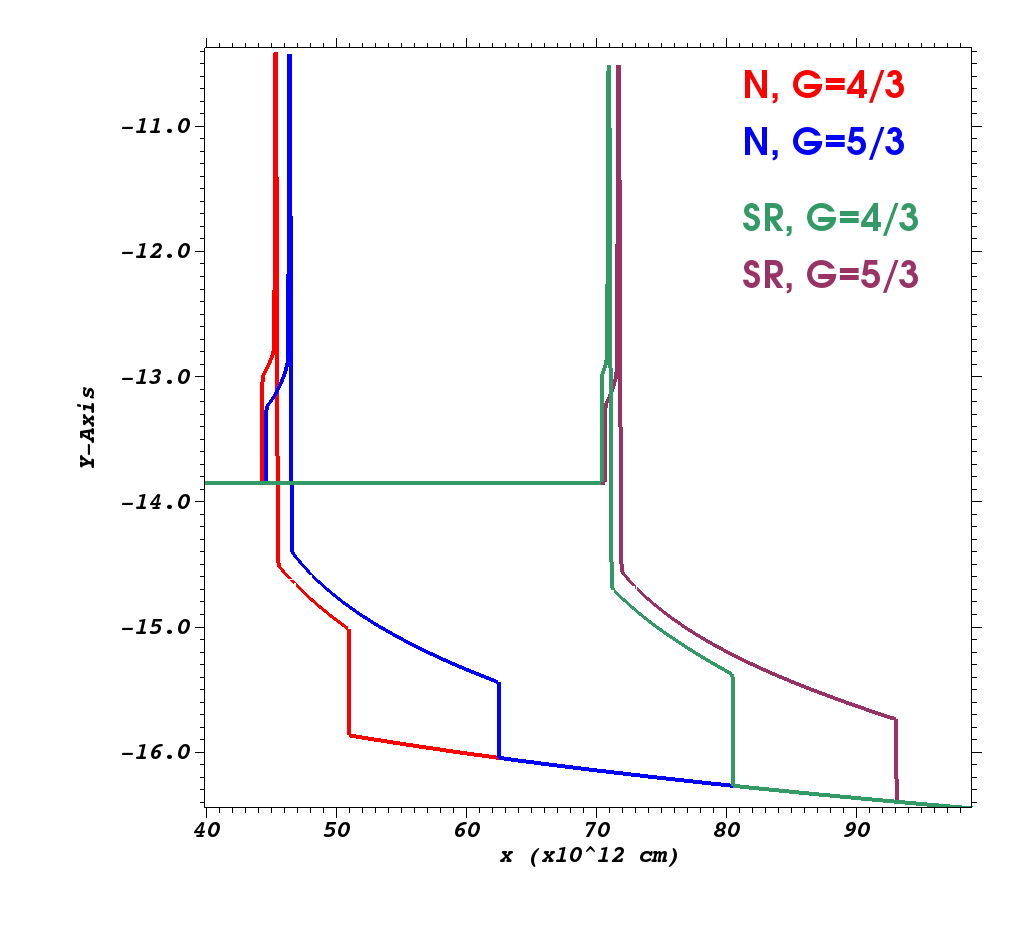}
    \caption{Comparison in density profiles of 1D simulations similar to the fiducial case CygX3. Everything is the same as in Fig. \ref{fig:SRvN}, with the exception of the adiabatic index $\Gamma$. Little effects are observable on the jet head (the thin high-density shell) propagation when $\Gamma$ is varied, but choosing $\Gamma=5/3$ results in a more advanced front shock by 25\% in the Newtonian case and 16\% in the relativistic one. This is caused by a stronger cooling of the shocked gas due to higher densities.}
    \label{fig:GammaSens}
\end{figure}

Because the simulated flows are relativistic, the choice of using a constant adiabatic index of value 5/3 may be discussed. We show in Fig. \ref{fig:GammaSens} a 1D test similar to the one presented in Sect. \ref{sect:app_relat}, comparing jets with $\Gamma=5/3$ and $\Gamma=4/3$ with the relativistic and the Newtonian solver. Changing the adiabatic index has an almost negligible effect on the jet head (the thin high-density shell) propagation, but jets with $\Gamma=5/3$ present a more advanced front shock by 25\% in the Newtonian case, and it is more advanced by 16\% with the relativistic solver. This is caused by the higher post-shock densities for $\Gamma=4/3,$ which result in a stronger cooling of the shocked gas. The situation is different in 2 and 3D, however, because \cite{mignone2007equation} reported that jets with a smaller adiabatic index propagate faster. This is due, as mentioned Sect. \ref{sect:previous} \citep[following][]{marti1997morphology}, to the first recollimation shock in the beam, which is strong enough to reaccelerate the beam flow at $\Gamma=4/3$.

\subsection{Numerical approximation of Bessel K functions}\label{sect:BessK}
\begin{figure}
    \centering
    \includegraphics[width=\hsize]{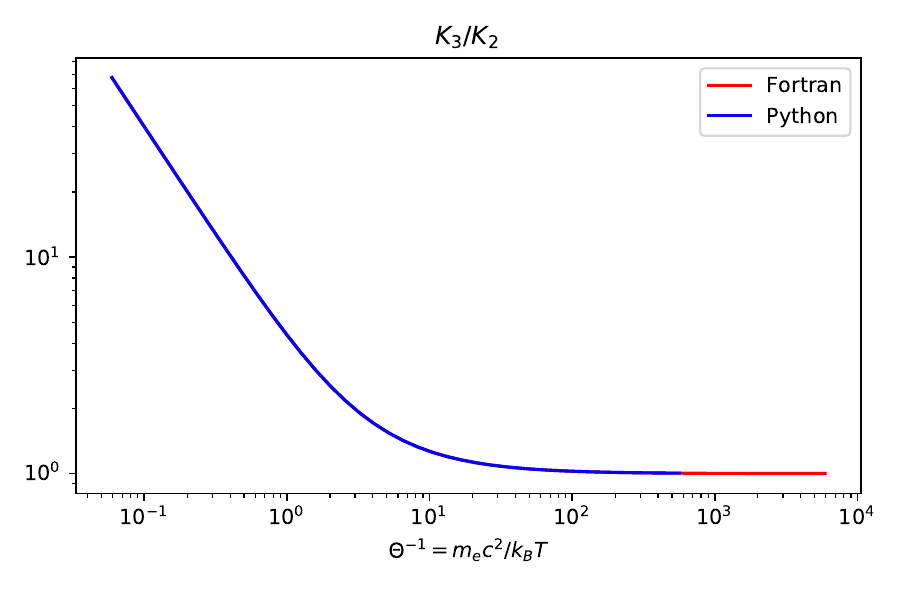}
    \caption{Comparison of our Fortran method for the ratio $K_3/K_2$ with the simple division of both terms using the functions defined in Python package SciPy. Our Fortran method avoids underflows for high values of $\Theta^{-1}$ with a relative error that never exceeds $10^{-8.}$}
    \label{fig:K3ovK2}
\end{figure}

Equations \ref{eq:MaxJutt} and therefore Eqns. \ref{eq:syn} and \ref{eq:ic} use the modified Bessel function of the second kind (also called Bessel K function or Macdonald function), especially the ratio $K_3/K_2$. A Fortran 90 implementation of this function by \cite{moreau2005numerical} was ported to A-MaZe, but as both functions tend to zero at low temperature, a simple division of $K_3(\Theta^{-1})$ by $K_2(\Theta^{-1})$ caused underflows during calculations. We therefore modified the method to derive the ratio directly. Figure \ref{fig:K3ovK2} compares our Fortran method with the built-in Bessel K functions from the SciPy package and shows the stability of our method over the whole temperature range compared to a simple division.

\section{Simulations setups}
Tables \ref{tab:CygX1} and \ref{tab:CygX3} show the input parameters for each run in CGS units. Tables \ref{tab:CygX1params} and \ref{tab:CygX3params} show the same inputs in terms of the a-dimensional parameters introduced Sect. \ref{sect:previous}.
\begin{table*}    
\centering
\small
\begin{tabular}{|l||CCCCCC|CCCC|} 
\hline     
 & \multicolumn{6}{c|}{Jet parameters} & \multicolumn{4}{c|}{Star parameters} \\
 setup name & L_j\text{ (erg s}^{-1}) & \rho_j\text{ (g cm}^{-3}) & v_j\text{ (cm s}^{-1}) & r_0\text{ (cm)} & T_j\text{ (K)} & B_j\text{ (G)} & \Dot{M}^\star & v_\infty\text{ (km s}^{-1}) & T^\star\text{ (K)} & B^\star\text{ (G)}\\
\hline                    
CygX1 & 5.1\cdot10^{36} & 1.3\cdot10^{-15} & 1\cdot10^{10} & 5\cdot10^{10} & 10^8 & 10. & 3\cdot10^{-6} & 1000 & 3\cdot10^4 & 10. \\
CygX1\_noLoss & 5.1\cdot10^{36} & 1.3\cdot10^{-15} & 1\cdot10^{10} & 5\cdot10^{10} & 10^8 & 0. & 3\cdot10^{-6} & 1000 & 3\cdot10^4 & 0.\\
CygX1\_wind & 5.1\cdot10^{36} & 1.3\cdot10^{-15} & 1\cdot10^{10} & 5\cdot10^{10} & 10^8 & 10. & 3\cdot10^{-6} & 1500 & 3\cdot10^4 & 10. \\
CygX1\_mP & 5.1\cdot10^{35} & 1.3\cdot10^{-16} & 1\cdot10^{10} & 5\cdot10^{10} & 10^8 & 10. & 3\cdot10^{-6} & 1000 & 3\cdot10^4 & 10. \\
CygX1\_T7 & 5.1\cdot10^{36} & 1.3\cdot10^{-15} & 1\cdot10^{10} & 5\cdot10^{10} & 10^7 & 10. & 3\cdot10^{-6} & 1000 & 3\cdot10^4 & 10. \\
CygX1\_T9 & 5.1\cdot10^{36} & 1.3\cdot10^{-15} & 1\cdot10^{10} & 5\cdot10^{10} & 10^9 & 10. & 3\cdot10^{-6} & 1000 & 3\cdot10^4 & 10. \\
\hline
\end{tabular}
\small
\caption{Simulation parameters of the runs based on Cygnus X-1. The mass-loss rate $\Dot{M}^\star$ is given in units of $M_\odot$ yr$^{-1.}$}    
\label{tab:CygX1}
\end{table*}

\begin{table*}
    \centering
    \begin{tabular}{|l||CCCCCC|}
        \hline
        setup name & \beta_j & M_j & \mathcal{M}_j & \eta & \eta^* & K \\
        \hline
        CygX1 & 0.334 & 67 & 71 & 0.077 & 0.087 & 258\\
        CygX1\_noLoss & 0.334 & 67 & 71 & 0.077 & 0.087 & 258\\
        CygX1\_wind & 0.334 & 67 & 71 & 0.116 & 0.130 & 387\\
        CygX1\_mP & 0.334 & 67 & 71 & 0.008 & 0.009 & 26\\
        CygX1\_T7 & 0.334 & 211 & 224 & 0.077 & 0.087 & 26\\
        CygX1\_T9 & 0.334 & 21 & 22 & 0.077 & 0.087 & 2577\\\hline
    \end{tabular}
    \caption{Dimensionless parameters of the Cygnus X-1 runs.}
    \label{tab:CygX1params}
\end{table*}

\begin{table*}
\centering
\small
\begin{tabular}{|l||CCCCCC|CCCCC|} 
\hline     
 & \multicolumn{6}{c|}{Jet parameters} & \multicolumn{4}{c|}{Star parameters} \\
 setup name & L_j\text{ (erg s}^{-1}) & \rho_j\text{ (g cm}^{-3}) & v_j\text{ (cm s}^{-1}) & r_0\text{ (cm)} & T_j\text{ (K)} & B_j\text{ (G)} & \Dot{M}^\star & v_\infty\text{ (km s}^{-1}) & T^\star\text{ (K)} & B^\star\text{ (G)}\\
\hline
CygX3 & 10^{38} & 1.4\cdot10^{-14} & 2.25\cdot10^{10} & 2\cdot10^{10} & 10^8 & 10 & 10^{-5} & 1500 & 8\cdot10^4 & 100 \\
CygX3\_noLoss & 10^{38} & 1.4\cdot10^{-14} & 2.25\cdot10^{10} & 2\cdot10^{10} & 10^8 & 10 & 10^{-5} & 1500 & 8\cdot10^4 & 100 \\
CygX3\_mW & 10^{38} & 1.4\cdot10^{-14} & 2.25\cdot10^{10} & 2\cdot10^{10} & 10^8 & 10 & 7.5\cdot10^{-6} & 1000 & 8\cdot10^4 & 100 \\
CygX3\_mP & 5.0\cdot10^{37} & 7\cdot10^{-15} & 2.25\cdot10^{10} & 2\cdot10^{10} & 10^8 & 10 & 10^{-5} & 1500 & 8\cdot10^4 & 100 \\
CygX3\_mPmW & 5.0\cdot10^{37} & 7\cdot10^{-15} & 2.25\cdot10^{10} & 2\cdot10^{10} & 10^8 & 10 & 7.5\cdot10^{-6} & 1000 & 8\cdot10^4 & 100 \\
CygX3\_mPmmW & 5.0\cdot10^{37} & 7\cdot10^{-15} & 2.25\cdot10^{10} & 2\cdot10^{10} & 10^8 & 10 & 7.5\cdot10^{-6} & 750 & 8\cdot10^4 & 100 \\\hline
\end{tabular}
\small
\caption{Simulation parameters of the runs based on Cygnus X-3. The mass-loss rate $\Dot{M}^\star$ is given in units of $M_\odot$ yr$^{-1.}$}
\label{tab:CygX3} 
\end{table*}

\begin{table*}[ht!]
    \centering
    \begin{tabular}{|l||CCCCCC|}
        \hline
        setup name & \beta_j & M_j & \mathcal{M}_j & \eta & \eta^* & K \\
        \hline
        CygX3 & 0.75 & 150 & 228 & 0.0028 & 0.0065 & 3.5\\
        CygX3\_noLoss & 0.75 & 150 & 228 & 0.0028 & 0.0065 & 3.5\\
        CygX3\_mW & 0.75 & 150 & 228 & 0.0025 & 0.0058 & 3.1\\
        CygX3\_mP & 0.75 & 150 & 228 & 0.0014 & 0.0032 & 1.8\\
        CygX3\_mPmW & 0.75 & 150 & 228 & 0.0013 & 0.0029 & 1.6\\
        CygX3\_mPmmW & 0.75 & 150 & 228 & 0.0009 & 0.0022 & 1.2\\\hline
    \end{tabular}
    \caption{Dimensionless parameters of the Cygnus X-3 runs.}
    \label{tab:CygX3params}
\end{table*}

\section{Instability growth phase}
\begin{figure*}
    \centering
    \includegraphics[width=\hsize]{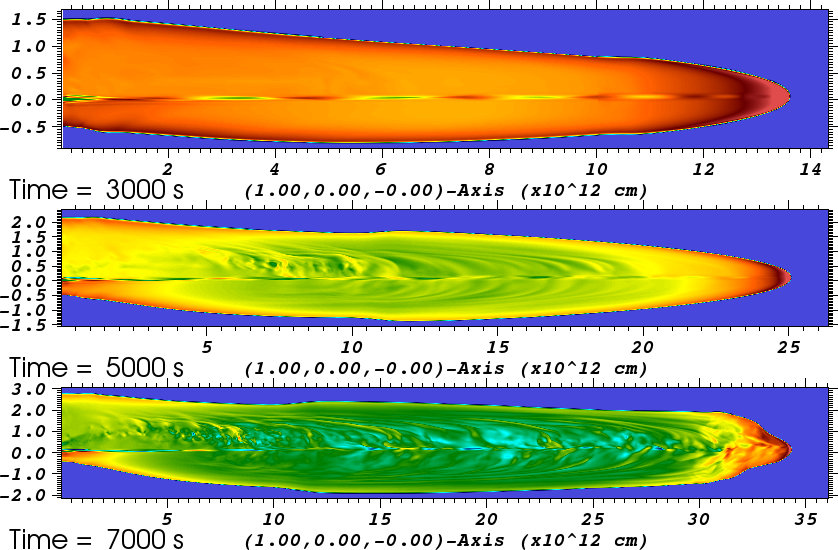}
    \caption{Pressure slices during the instability growth phase of the fiducial Cygnus X-1 run CygX1. The color scale is fixed from 1 (blue) to 1000 Ba (red) to better highlight the beam structure. The beam shows alternating over- and underpressured zones whose number has risen at the 5000 s mark. The inner cocoon shows ripple-like structures alternating on either side of the beam with increasing intensity as the jet evolves.}
    \label{fig:CygX1_instgr_p}
\end{figure*}

\begin{figure*}
    \centering
    \includegraphics[width=\hsize]{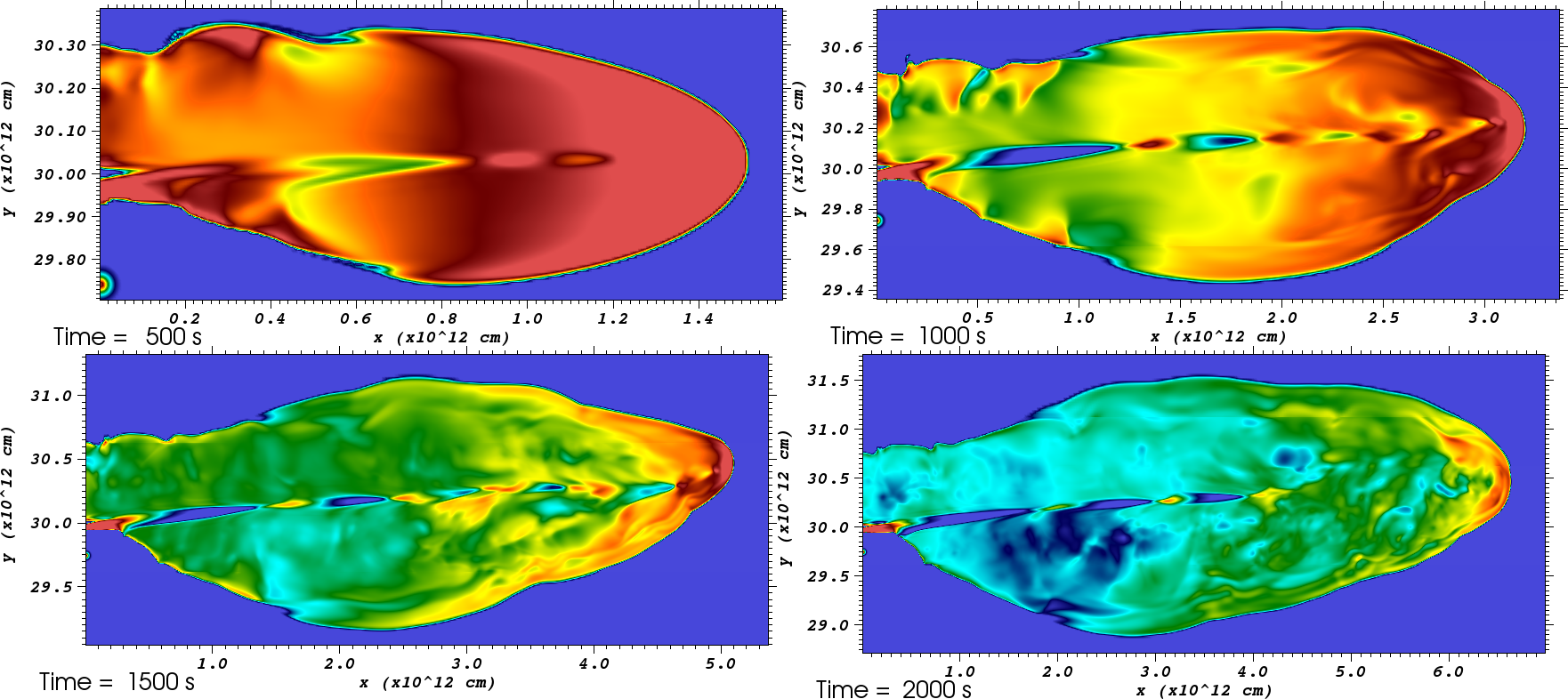}
    \caption{Pressure slices during the instability growth phase of the fiducial Cygnus X-3 run CygX3. The color scale is fixed from $10^3$ (blue) to $10^5$ Ba (red) to better highlight the beam structure.}
    \label{fig:CygX3_instgr_p}
\end{figure*}

Figures \ref{fig:CygX1_instgr_p} and \ref{fig:CygX3_instgr_p} show pressure slices of the runs CygX1 and CygX3 runs, respectively. They highlight the internal structure of the beam with alternating over- and under-pressured regions compared to the fluid in the surrounding cocoon.

\end{appendix}
\end{document}